\documentclass[11pt]{article}
\pdfoutput=1
\usepackage[margin=1.0in]{geometry}
\usepackage{amsmath,amssymb,graphicx,bbm}
\usepackage{hyperref}
\usepackage{slashed}

\usepackage[utf8]{inputenc}

\usepackage[T1]{fontenc}
\usepackage{mathpazo}
\usepackage[font=small,labelfont=bf]{caption}
\usepackage{cancel}
\usepackage{cite}
\usepackage{multirow}

\newcommand{\be}{\begin{equation}}
\newcommand{\ee}{\end{equation}}

\usepackage{afterpage}

\usepackage[table,usenames,dvipsnames]{xcolor}
\usepackage{pdfpages}
\usepackage{framed}
\usepackage{afterpage}

\usepackage{tikz}
\usetikzlibrary{trees}
\usetikzlibrary{decorations.pathmorphing}
\usetikzlibrary{decorations.markings}
\tikzset{
    photon/.style={decorate, decoration={snake}, draw=black},
    electron/.style={draw=black, postaction={decorate},
        decoration={markings,mark=at position .55 with {\arrow[draw=black]{>}}}},
    scalar/.style={draw=black,dashed},
    gluon/.style={decorate, draw=black,
        decoration={coil,amplitude=4pt, segment length=5pt}}
}

\title{\bf Opening the black box of neural nets:\\ case studies in stop/top discrimination}
\author{Thomas Roxlo and Matthew Reece\\
{\small Department of Physics, Harvard University, Cambridge, MA, 02138}}

\begin{document}
\maketitle

\begin{abstract}
We introduce techniques for exploring the functionality of a neural network and extracting simple, human-readable approximations to its performance. By performing gradient ascent on the input space of the network, we are able to produce large populations of artificial events which strongly excite a given classifier. By studying the populations of these events, we then directly produce what are essentially contour maps of the network's classification function. Combined with a suite of tools for
identifying the input dimensions deemed most important by the network, we can utilize these maps to efficiently interpret the dominant criteria by which the network makes its classification.

As a test case, we study networks trained to discriminate supersymmetric stop production in the dilepton channel from Standard Model backgrounds. In the case of a heavy stop decaying to a light neutralino, we find individual neurons with large mutual information with $m_{T2}^{\ell\ell}$, a human-designed variable for optimizing the analysis. The network selects events with significant missing $p_T$ oriented azimuthally away from both leptons, efficiently rejecting $t\overline{t}$ background. In the case of a light stop with three-body decays to $Wb{\widetilde \chi}$ and little phase space, we find neurons that smoothly interpolate between a similar top-rejection strategy and an ISR-tagging strategy allowing for more missing momentum. We also find that a neural network trained on a stealth stop parameter point learns novel angular correlations.
\end{abstract}

\section{Introduction}

The Large Hadron Collider (LHC) had a stunning success with the discovery of the Higgs boson, but so far all of the LHC's measurements appear to be consistent with the Standard Model (SM). This suggests that new physics, if it exists at the TeV scale, may be hidden in vast samples of superficially similar background events. To extract such physics, we will need an increasingly precise understanding of what the Standard Model predicts, together with powerful statistical tools for
searching for deviations from SM predictions. Such thinking has spurred increased use of the tools of Machine Learning, such as Deep Neural Networks (DNNs). In recent years, DNNs and related forms of so-called Deep Learning \cite{lecun2015deep,Goodfellow-et-al-2016} have shown great utility in a variety of applications previously the sole domain of biological brains, such as speech recognition \cite{amodei2016deep}, driving \cite{SelfDriving:2016}, and playing the game of Go
\cite{silver2016mastering}. Machine Learning techniques have increasingly been applied to particle physics problems, including (but not limited to) tagging boosted $W$ bosons \cite{Cogan:2014oua,deOliveira:2015xxd} or top quarks \cite{Kasieczka:2017nvn,Almeida:2015jua,Macaluso:2018tck}, reducing sensitivity to systematic uncertainties \cite{Louppe:2016ylz}, discriminating between quark and gluon jets \cite{Komiske:2016rsd,Metodiev:2017vrx}, mitigating pileup \cite{Komiske:2017ubm}, and distinguishing supersymmetric events from SM backgrounds \cite{Cohen:2017exh}. Further references may be found in the recent review \cite{Larkoski:2017jix}.

However, machine learning solutions in general and neural networks in particular often have the problem of being relatively opaque in their operation. That is, while one can train a DNN to recognize almost any pattern and seamlessly categorize datasets based on learned observables, the details of its learning can be hard to extricate---in a very real sense, a DNN often acts as a ``black box,'' stubbornly resisting attempts to peer inside. In the case of collider analysis, for example, one could easily imagine training a multi-level perceptron to distinguish real events from events simulated with programs such as Pythia \cite{Sjostrand:2014zea} or Herwig \cite{Bellm:2015jjp}. If the classifier was then able to distinguish data from simulation, and even perhaps classify certain events as ``very data-y,'' this could be taken as evidence that there might possibly be exotic physics at play in these special events.

Unfortunately, as written such a classifier would probably be of limited usefulness in terms of providing insight into new physics. The reason is that, even upon finding out that the DNN considered a certain subset of events to deviate significantly from the simulation distribution, a researcher would have little idea why this was so, unless it was obvious from a direct examination of the events. Thus, in order to interpret the dictate of the DNN, the researcher would have to perform much of the work the DNN was designed to replace. Given that the unavoidable limitations of the available simulations would no doubt result in a number of false ``signals,'' this technique may not save much effort in the end. In particular, one would want to be sure that the DNN was really tagging physical differences and not systematic failures of Pythia or Herwig in modeling complex physics like hadronization. It is possible that the use of generative adversarial networks \cite{2014arXiv1406.2661G} could help to reduce the dependence of the classifier on internal Pythia nuisance parameters (cf.~\cite{Louppe:2016ylz}). The use of data planing may also help to extract insight into which physical features a neural network finds most useful \cite{Chang:2017kvc}. Nonetheless, it is fair to say that the black box nature of neural networks remains a concern for many working particle physicists.

Here we propose a method to improve this situation and ``open the black box'' by constructing large numbers of artificial inputs which strongly activate the classifier of the neural network, and then isolating the important features the network is using to perform its classification. To illustrate this method, we will apply neural nets to a well-studied physics example: discriminating top quarks from their supersymmetric partner particles, scalar top quarks or ``stops.'' The reason is that this
is a well-studied problem. We can thus compare the variables physicists have chosen to those the machine is using, and determine if the machine is learning qualitatively new things or pursuing similar strategies. We find that, although the network is usually classifying similar events to physicists, it is often doing so in subtly different ways that betray its data-driven, rather than physics-driven, approach to the problem.

\subsection{Stops as a Test Case}

As our test case for understanding how well neural nets can help us to understand the physical differences between new physics and Standard Model phenomena, we will focus on the stop/neutralino sector of supersymmetry. Despite increasingly strong experimental constraints, which often push into at least moderately fine-tuned regions of parameter space, the general framework of supersymmetry remains one of the most compelling possibilities for a natural theory of the electroweak scale. The measured gauge couplings unify if superpartners lie near the weak scale, which is a strong empirical motivation for the study of supersymmetry. It has long been appreciated that certain particles in the supersymmetric spectrum play a more central role for naturalness than others, beginning with higgsinos at tree level \cite{Barbieri:1987fn} and subsequently stops and gluinos at one and two loops \cite{Dimopoulos:1995mi, Pomarol:1995xc, Cohen:1996vb}. This led to various suggestions for focusing on these particular particles in experimental searches. Such searches use ``simplified models'' that include only the particles most essential for naturalness---and for the associated experimental signal \cite{Meade:2006dw, Kitano:2006gv, Perelstein:2007nx, Kats:2011it, Kats:2011qh, Essig:2011qg, Brust:2011tb, Papucci:2011wy, Kribs:2013lua}. Independent of naturalness arguments, early studies of SUSY phenomenology often focused on stops as a key experimental target since large renormalization group running and mixing effects often make them one of the lightest scalar superpartners in simple models of SUSY breaking \cite{Ellis:1983ed, Hikasa:1987db, Drees:1990te, Baer:1991cb}.

Apart from its strong intrinsic motivation for electroweak naturalness, a compelling reason to focus on the stop/neutralino simplified model in our work is that it has already been thoroughly explored by theorists. The parameter space contains regions spanning a range of different kinematics, which require different experimental search strategies. This makes it a good test case for the application of Machine Learning techniques. If we find that a neural net does significantly better than previously proposed strategies, this is significant, since ample thought has already gone into identifying good strategies. On the other hand, we expect that the neural net will often use strategies similar to those devised by humans; in this case, our prior familiarity with these strategies will help us to analyze how the neural net is operating. In the future it might be very interesting to apply neural nets to previously unexplored signal/background discrimination problems, but for our current purposes we believe it is best to stick to well-trod ground.

\begin{figure}[!h]
\begin{center}
    \includegraphics[width=0.7\textwidth]{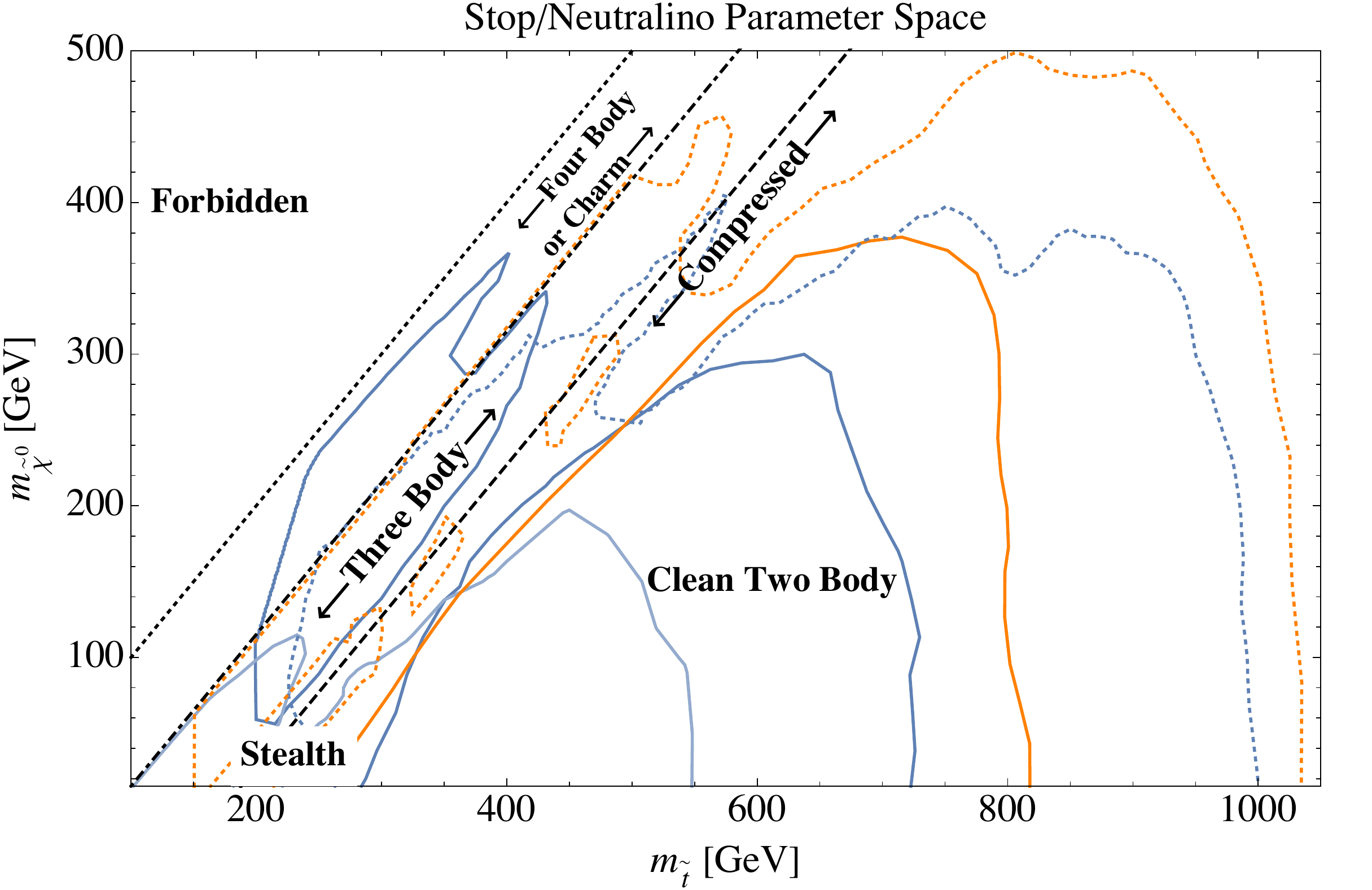}
\end{center}
\caption{Parameter space of a stop/neutralino simplified model with $\widetilde{t} \to t^{(*)} \widetilde{\chi}_0$, with regions of distinct kinematics (explained in the text) labeled. Selected exclusion curves are shown from CMS (orange) \cite{Sirunyan:2017wif, Sirunyan:2017leh} and ATLAS (blue) \cite{Aad:2015pfx, Aaboud:2017nfd,Aaboud:2017ayj}. Solid exclusion curves come from studies of dilepton final states (as studied in this paper), while dotted exclusion curves come from studies of hadronic final states. The lighter blue curve is a result from 8 TeV data while the darker blue and orange curves are from 13 TeV data.}
\label{fig:stopmap}
\end{figure}

In a general model, stops can have a variety of decays, for instance ${\widetilde t} \to t {\widetilde \chi}^0$ (with the top quark possibly off-shell) and ${\widetilde t} \to b {\widetilde \chi}^+$. In this paper we will focus on the first case, arising for example if the lightest neutralino is a bino (or, with some phenomenological differences, if the stop is an NLSP decaying to a gravitino). The parameter space for a stop--neutralino simplified model is illustrated in Fig.~\ref{fig:stopmap}. If $m_{\widetilde t} \gg m_{{\widetilde \chi}^0_1}$, we have simple events with large missing transverse momentum carried away by the neutralinos. This region is labeled ``Clean Two Body'' and, as the figure illustrates, experiments have already set significant constraints. The experimental search becomes increasingly challenging when the decays have small phase space. These occur near the three diagonal black lines in the plot: a dashed line where $m_{\widetilde t} = m_t + m_{{\widetilde \chi}^0_1}$, a dot-dashed line where $m_{\widetilde t} = m_W + m_b + m_{{\widetilde \chi}^0_1}$, and a dotted line where $m_{\widetilde t} = m_{{\widetilde \chi}^0_1}$. Near the first dashed line, we have a region of ``Compressed'' two-body decays. In the rest frame of the decaying stop, the top and neutralino have low momentum. In the lab frame, they inherit momentum from the top, but the overall missing transverse momentum in the event tends to be small. As mass splittings become increasingly small, we find the region labeled ``Three Body'' where the dominant decay is ${\widetilde t} \to W^+ b {\widetilde \chi}^0_1$ and the region labeled ``Four Body or Charm'' where either the decay ${\widetilde t} \to f {\overline f}' b {\widetilde \chi}^0_1$ (through an off-shell $W^* \to f {\overline f}'$) or the decay ${\widetilde t} \to c {\widetilde \chi}^0_1$ dominates, depending on the amount of flavor violation in the model. Finally, the corner of the compressed region where $m_{{\widetilde \chi}^0_1} \ll m_{\widetilde t} \approx m_t$ is labeled ``Stealth.'' In this region, the kinematics of the top decay products are almost identical to Standard Model $t{\overline t}$ production, as the neutralino carries very little momentum even in the presence of initial state radiation (ISR). Finally, we label the region with $m_{{\widetilde \chi}^0_1} > m_{\widetilde t}$ ``Forbidden,'' not because it is theoretically impossible but because it is incompatible with the decay we are choosing to study.

A number of studies of how to distinguish signal and background have been carried out for stops; a sampling includes \cite{Cao:2012rz, Kilic:2012kw, Buckley:2013lpa, Belanger:2015vwa, Baer:2016bwh}. Studies focused on the 4-body or charm region of very light stops include \cite{Boehm:1999tr, Das:2001kd, Muhlleitner:2011ww, Krizka:2012ah, Ferretti:2015dea}. In the compressed region, missing momentum is typically very small unless the stops are recoiling against energetic ISR jets (see e.g.~Fig.~1 of \cite{Alwall:2008ve} for a clear illustration, and \cite{Martin:2007gf, Baer:2007uz, Martin:2007hn} for general discussion of compressed supersymmetry). Searches in this region can make use of the ISR jet, and in particular the ratio of its $p_T$ to missing transverse momentum \cite{Hagiwara:2013tva, An:2015uwa, Macaluso:2015wja}. Other studies of the compressed stop region include \cite{Dutta:2013gga,Cheng:2016mcw,Cheng:2017dxe}. The case where 3-body decays dominate, dubbed the ``W corridor,'' has been recently studied in \cite{Cheng:2017koe}. The existence of the stealth region as a kinematically distinct case was emphasized in \cite{Fan:2011yu, Kats:2011it, Fan:2012jf}. It has been studied using spin correlations \cite{Han:2012fw, Aad:2014mfk} and the $t {\overline t}$ cross section \cite{Czakon:2014fka, Eifert:2014kea, Cohen:2018arg}.

In this paper we will focus on studies where both stops decay semileptonically. Recent searches for stops in the 2-lepton channel have been performed at ATLAS \cite{Aaboud:2017nfd} and at CMS \cite{Sirunyan:2017leh}. From these papers we learn that dominant backgrounds for these searches include $t{\overline t}$, including in association with a $Z$, $W$, or $H$ boson; single top production ($tW$ and $tbZ$); diboson events, including $WW$, $WZ$, and $ZZ$ events with leptonic decays; and triboson production ($WWZ$). In our study we simulate all of these background processes as well as signal events, and train neural nets to discriminate between them for different regions of signal parameter space. Including some mild generator level cuts (the most stringent being $p_T^{\rm miss} > 50$ GeV), the background we simulate is dominantly composed of $t{\bar t}$ events, with order 10\% contributions from $tW$ and $WW$. The other processes contribute at lower rates, but are nonetheless included. We scale the backgrounds to state-of-the-art theory calculations of inclusive cross sections using results from \cite{Czakon:2011xx,Kidonakis:2016sjf,Gehrmann:2014fva,Grazzini:2016ctr,Aaboud:2017qkn,deFlorian:2016spz,Anastasiou:2016cez,Lazopoulos:2008de,Campbell:2012dh,Garzelli:2012bn,Nhung:2013jta,Cascioli:2014yka,Grazzini:2015hta,Grazzini:2016swo}. We scale the signals to rates from \cite{Borschensky:2014cia}.

\subsection{Summary of our approach}

The outline of this paper is as follows. We explain our methods of unpacking the behavior of neural networks in \S\ref{sec:methods}, using some toy problems to visualize network performance outside the context of physics. Our first physics case study is in \S\ref{sec:uncompressed}, where we study a parameter point with clean (uncompressed) two-body decays, $m_{\widetilde t} = 750$ GeV and $m_{\widetilde \chi} = 1$ GeV. We find that, although the network's performance is comparable or
superior to discrimination on the variable $m_{T2}^{\ell\ell}$, it appears to be learning a less complicated pattern. In \S\ref{sec:lepmetphi} we elaborate on how the neural network's strategy of selecting events with two leptons pointing azimuthally opposite the missing momentum can efficiently reject $t\overline{t}$ background. In \S\ref{sec:threebody} we move on to discussing a parameter point with $m_{\widetilde t} = 350$ GeV and $m_{\widetilde \chi} = 200$ GeV, with three-body decays ${\widetilde t} \to W^+ b {\widetilde{\chi}}$ and relatively little phase space available to the daughter particles. We find a neuron that, over some range of missing $p_T$, exploits a similar strategy to that we saw for the uncompressed point, but which increasingly allows the leptons to point at more moderate angles to missing $p_T$ when the missing $p_T$ is very large. We interpret this as an event selection that smoothly pivots from a pure top-rejection strategy at moderate $p_T^{\text{miss}}$ to an ISR-tagging strategy at large $p_T^{\text{miss}}$. In \S\ref{sec:stealth}, we study a ``stealthy'' parameter point where $m_{\widetilde t} = 185$ GeV and $m_{\widetilde \chi} = 5$ GeV. The neural network learns certain angular correlations associated with the difference between scalar and fermion production known in previous literature, but also displays more complicated patterns not previously known. In \S\ref{sec:conclusions}, we conclude.

\section{Methods and tools} \label{sec:methods}

\subsection{Simulations} \label{sec:simulation}


Signal and SM background distributions were simulated at tree level using the Madgraph \cite{Alwall:2014hca} (v5.2.5.4) event generator with the CTEQ6L1 PDF. Parton showering and hadronization were simulated in Pythia \cite{Sjostrand:2014zea} (v8.2.26), and the events were then passed through the Delphes detector simulation with CMS detector model \cite{Ovyn:2009tx, deFavereau:2013fsa} (v3.4.1). 

The background distribution was constructed by combining the dilepton decay channels of the following processes: $t\bar{t}$, $tW$, $t\bar{t} + W/Z/H$, $tbZ$, $ZZ$, $WW$, $WZ$, and $WWZ$. The signal distribution was composed of dilepton decays of ${\widetilde t}_1{\widetilde t}_1^*$ pairs with varying stop and neutralino masses; for mass pairs near the compressed region we also included events with an extra ISR jet through jet matching \cite{Alwall:2007fs,Alwall:2008qv}.

Generator-level cuts were applied to both signal and background to require the events to satisfy the following criteria: at least 2 electrons or muons above 15 GeV but not 3 over 20 GeV, at least one $b$-jet over 20 GeV, and at least 50 GeV of missing transverse momentum. After the detector simulation, the event preselection criteria of the CMS dilepton search \cite[Table 1]{Sirunyan:2017leh} were applied. These cuts ensure that the missing $p_T$ is not aligned with one of the leading jet $p_T$s, that the dilepton invariant mass is above 20 GeV, and that the two leptons do not reconstruct a $Z$ boson.

The event data was condensed into a set of 19 real variables which hopefully encompass the relevant kinematic information; these variables were then passed to the neural network for training. They were composed of kinematic observables related to the following particles: the highest $p_T$ lepton ($\ell_1$), the second-highest $p_T$ lepton ($\ell_2$), and the two highest $p_T$ $b$-tagged jets. If there were less than 2 $b$-tagged jets in an event, the highest $p_T$ non-$b$-tagged jets were used instead. In addition, following \cite[\S 5]{Sirunyan:2017leh}, the two chosen jets were paired with the leptons by selecting the pairing which minimizes the maximum invariant mass of the pairs ($\ell_1$, $j_1$) and ($\ell_2$, $j_2$). Notice that with this choice, the jets $j_1$ and $j_2$ that the neural network sees are not necessarily the highest-$p_T$ jets in the event. Finally, the missing momentum, defined as minus the sum of the momenta of all visible particles in the event, was taken as an additional ``particle.''

With the particles defined, the 19 kinematic variables fed into the neural network were as follows (all variables were scaled by their mean and standard deviation over the training set):
\begin{itemize}
    \item The natural logarithm of $p_T^{\text{miss}}$ and the $p_T$ of all 4 particles.
    \item The pseudorapidity of the 4 particles. The sign of this variable was taken relative to $\eta^{\ell_1}$; that is, we took $\eta^{\ell_1} \rightarrow |\eta^{\ell_1}|$ and $\eta^{i} \rightarrow \eta^{i}\frac{\eta^{\ell_1}}{|\eta^{\ell_1}|}$ for $i \ne \ell_1$. The $\eta$ value of the missing energy was not considered because it is mostly dependent on longitudinal boosts of the event center of mass, which are not physically relevant.
    \item The magnitudes and signs of the azimuthal angles of the 4 particles. These $\phi$ values were defined relative to the $\phi$ of the missing energy and range over $0 \leq \phi \leq \pi$; a separate sign variable was included which is $1$ if the difference was defined in the counter-clockwise direction and $-1$ if in the clockwise direction. The form of these variables are designed to avoid periodicity in the inputs to the neural network.
    \item The $b$-tag status of the two jets: $1$ for $b$-tagged and $-1$ for non-$b$-tagged.
\end{itemize}

\subsection{Neural net design} \label{sec:nndesign}

For the purposes of this paper, we will confine ourselves to simple feed-forward multilayer perceptrons composed of a series of dense layers with relatively small numbers of nodes. Such networks are sufficient for our projected use case: modelling discrimination variables which are complicated non-linear functions of a only a few relevant physics variables. However, we expect that most of the techniques we use can be generalized to more complicated architectures.

The networks used for the stop/top discrimination study were composed of an input layer, 11 hidden layers, and an output layer. The hidden layers had, in order, the following numbers of nodes: 128, 64, 32, 16, 16, 8, 8, 4, 4, 2, and 2. All hidden layers used leaky ReLU transfer functions \cite{Maas2013RectifierNI} with an inactive gradient of $0.1$; the output layer used a $\tanh$ activation function. The Adam optimizer \cite{Kingma:2014} was used to improve convergence time. A small L1 regularization term was applied to all layers. In training, we held out 20\% of the data to use as a test set.

\subsection{Mutual information}

The first thing we would like to do to get a handle on the internal behavior of our networks is to compare the outputs of the hidden neurons --- representing intermediate, atomic terms that are combined together in the overall computation --- to the various variables that physicists have come up with over the years to solve the problem the network is tackling. To do this, we compute the normalized \emph{mutual information}, defined as
\be
I(X; Y) = \frac{1}{\sqrt{H(X)H(Y)}} \sum_y \sum_x p(x,y) \log \frac{p(x,y)}{p(x)p(y)}
\label{eq:mutualinformation}
\ee
where $H(X) = -\sum_x p(x) \log p(x)$ is the Shannon entropy of $X$. The conventional definition of mutual information omits the $1/\sqrt{H(X) H(Y)}$ normalization factor, but we will find it convenient below to include it. $I(X; Y)$ is a good measure of the relationship between two variables because, unlike simpler correlation coefficients, it can handle arbitrary nonlinear relationships. That is, if $X$ is a deterministic function of $Y$ and vice versa, they are effectively the same variable. In this case, the mutual information between them, as we have defined it, will simply equal 1. On the other hand, if the variables are completely uncorrelated the mutual
information is zero.

Even if the network were performing exactly the same analysis as human physicists we would not necessarily expect the intermediate steps to be expressed in a human-readable format. However, if the variables identified by physicists were truly uniquely endowed with discriminative power, we would expect them to show up in some form in the latter layers of the network. Thus, we should find at least some neurons which have high mutual informations with these variables.

\subsection{Activation difference}
\label{subsec:activdiff}

We would also like to be able to independently examine the behavior of a neuron, irrespective of any knowledge of preexisting physics variables. The first tool we use to do this is the \emph{activation difference}\cite{robnik2008explaining , 2016arXiv160302518Z, 2017arXiv170204595Z}: a simple measurement of how much the output of a given neuron changes due to changes in the input variables. That is, given a neuron expressed as a function $z$ over a set of inputs
$\textbf{x}$, we define the activation difference to be
\be
AD_i(z) = \frac{1}{\sigma_z} \sum_{\textbf{x}} p(\textbf{x}) \sum_{x_i} p(x_i | \textbf{x}_{/i}) |z(\textbf{x}) - z(\textbf{x}_{/i}, x_i)|
\ee
where we sample input vectors $\textbf{x}$, replace a single element $x_i$ with another value drawn from the same data distribution, and record the average change in the output value $z(\textbf{x})$. For this purpose, we make the approximation that the distributions of the inputs are independent, ie. $p(x_i | \textbf{x}_{/i}) \approx p(x_i)$. We have normalized by the standard deviation of the neuron output over the dataset to account for differences in the activation range between neurons, although we will never actually try to directly compare the activation differences of different neurons.

The activation difference gives a clear, easily understandable measurement of which inputs are the most ``important'' to a given neuron; that is, which inputs have the strongest impact on the output. This technique can also be used to give an idea of which \emph{neurons} in an earlier hidden layer are most important in determining the output, simply by taking $\textbf{x}$ to be the neuron activations of this layer rather than the network inputs themselves. Since --- especially in
networks with non-trivial L1 regularization --- it can occur that only a small subset of neurons in a given layer contribute significantly to a given neuron activation in the next layer, we can often use this strategy to quickly reduce a neuron's activation pattern to a manageable number of terms with relatively good approximation.

\subsection{Activation maximization}

A more involved technique for visualizing network behavior is something we will refer to as \emph{activation maximization}. Much work has been done in the field of image processing on producing visual representations of the patterns being recognized by specific neurons or networks \cite{Erhan:2009, 2013arXiv1312.6034S, 2015arXiv151202017M, 2015arXiv150606579Y}; we will use a related technique here.

The basic idea is to assume that the behavior of a neuron can be qualitatively well described by the input patterns which provoke the most extreme activations. That is, if we consider the set of events for which the neuronal response is above a certain (high) threshold, we will presumably be able to deduce some sort of pattern in the inputs which we can reasonably say is being ``recognized'' by the neuron. Since in general we are only interested in the events which most strongly
activate the final neuron of the network, and this activation is usually related to extreme activations (either high or low) in previous layer neurons, we can often, to reasonable approximation, replace the entire neuron with simple, human-understandable variables which are high when the neurons are maximized and low otherwise. By doing this, we can often get a fairly good qualitative idea of what patterns the network has learned to
look for in the most signal-like events.

One might think that one could simply examine the events in the dataset the network flags as most likely to be signal to understand its behavior, and to some extent this is true. However, this approach has a few drawbacks. First, the statistics of this set of selected events can be too small to meaningfully infer the criteria the network used to select them, even in relatively large simulated datasets. Second, and more importantly, it can be very difficult to guess exactly what criteria the
network has utilized to pluck this specific privileged group from the undifferentiated hordes, especially if the input space is large and complicated. In order to do so, one must constantly compare with the background distribution for all possible combinations of inputs, looking for discrepancies. Even when discrepancies are found, they are not necessarily the patterns the network is selecting on, but may simply be mildly correlated with them in the data set.

Therefore, what we will do instead is perform gradient ascent on the space of physics inputs, looking for inputs which maximize the neuron output and stopping when we reach a given threshold. More precisely, starting with some initial input $\textbf{x} = \textbf{x}_0$, we compute the neuron activation $z(\textbf{x})$ and then take small steps in input space along the gradient $\partial{z}/\partial{\textbf{x}}$, synthesizing a series of artificial inputs until we find one which is above our
desired activation threshold. By repeating this process many times for different randomly chosen starting inputs, we can, in essence, generate what amounts to a single contour in a contour map of the network's classification function over the input space. Since we produce these maps without referencing the data at all (save in the sense that the network was originally trained on it), we can be sure that any patterns we find are actually caused by selection pressure from the network itself.
Any input variables which are relatively unimportant to the network will be distributed roughly according to their initial random distribution, and thus we can ignore them.

\subsubsection{Toy model: product network}

To get a feel for what these activation maximization plots look like and how they relate to the pattern the network is learning, let's take a look at a couple of simple toy networks. We start with a network which has been trained on a ``dataset'' of a million pairs of real numbers $(x, y)$ in the range $[-1, 1]$ and learned to return $-1$ when the product $xy \ge 0$ and $1$ when $xy < 0$. We use the architecture described in Section \ref{sec:nndesign}, although it is certainly more complex than necessary for this application.

\begin{figure}[!ht]
\begin{center}
\scalebox{0.54}{\includegraphics{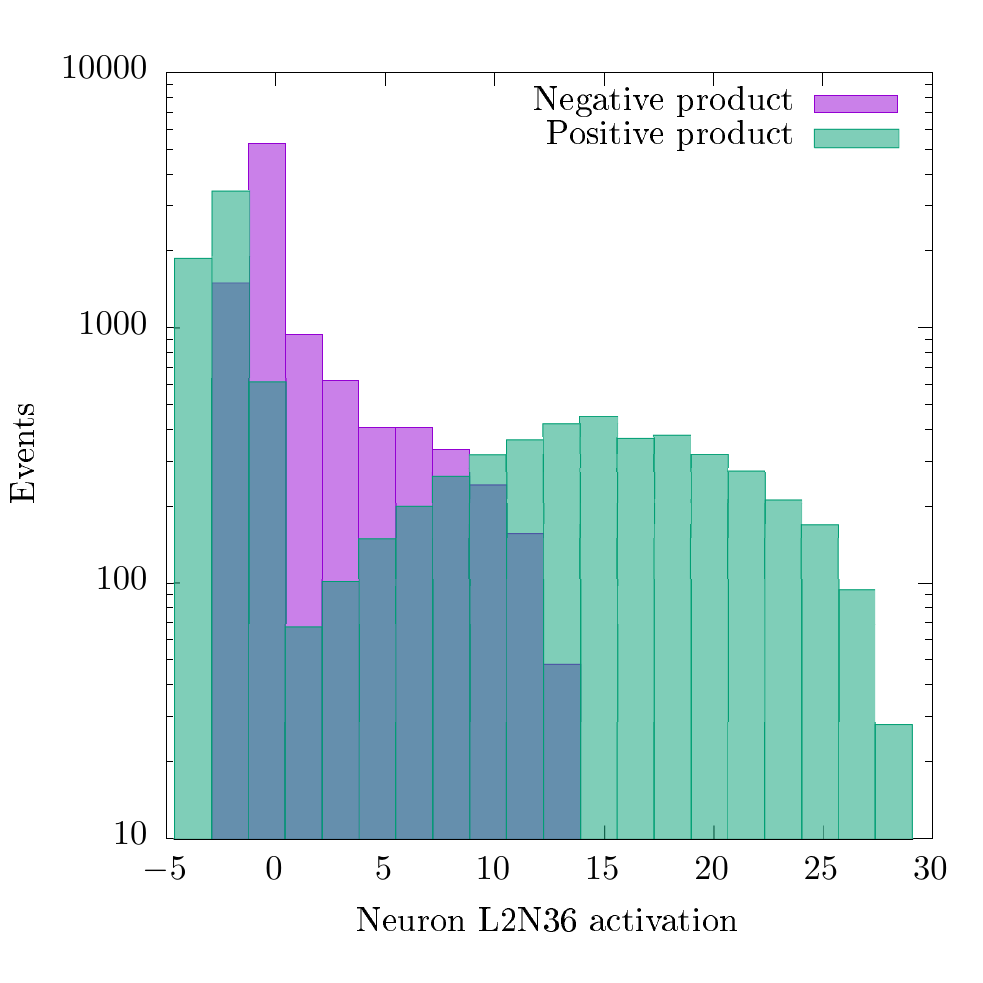}} \scalebox{0.54}{\includegraphics{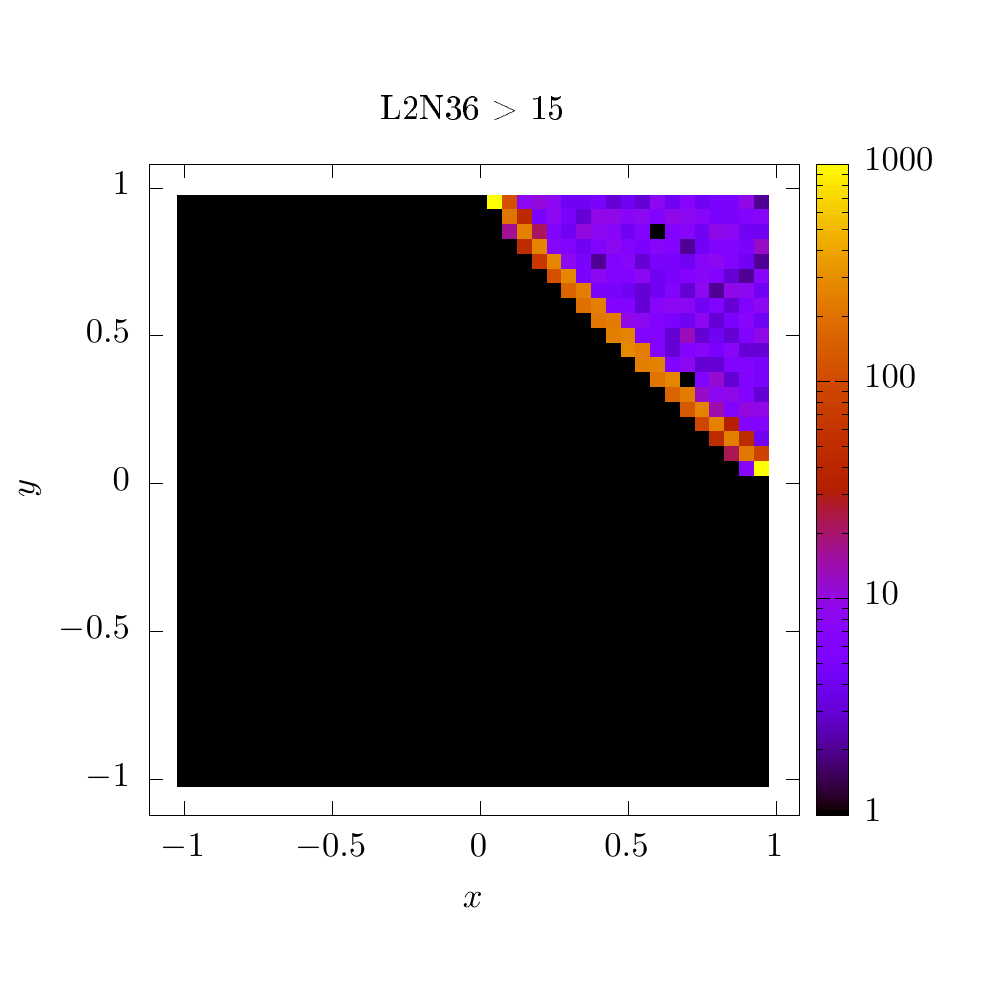}} \scalebox{0.54}{\includegraphics{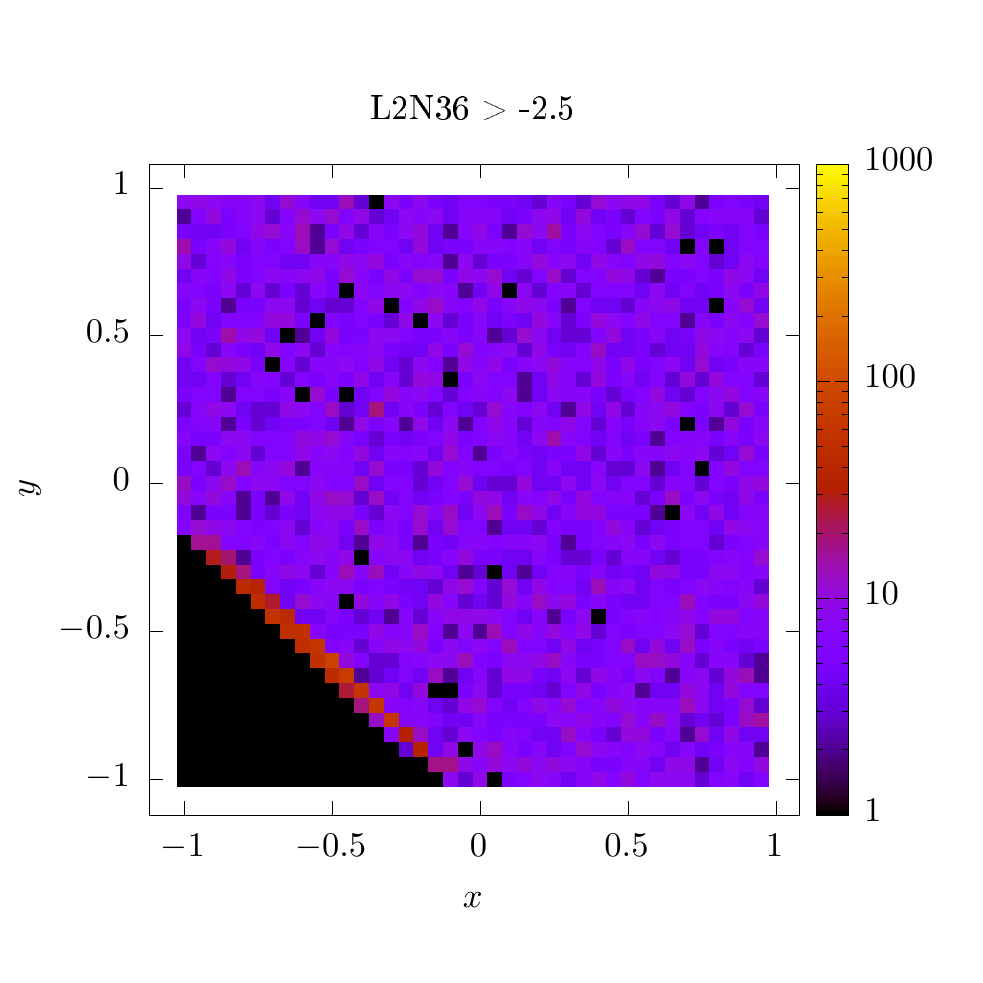}} \\
\caption{Analysis of neuron $L2N36$ of a network trained to distinguish the sign of the product of two input variables. Left: neuron activations on ``data.'' Right: activation maximization histograms.}
\label{fig:toydata_pos_L2N36}
\end{center}
\end{figure}

\begin{figure}[!ht]
\begin{center}
\scalebox{0.54}{\includegraphics{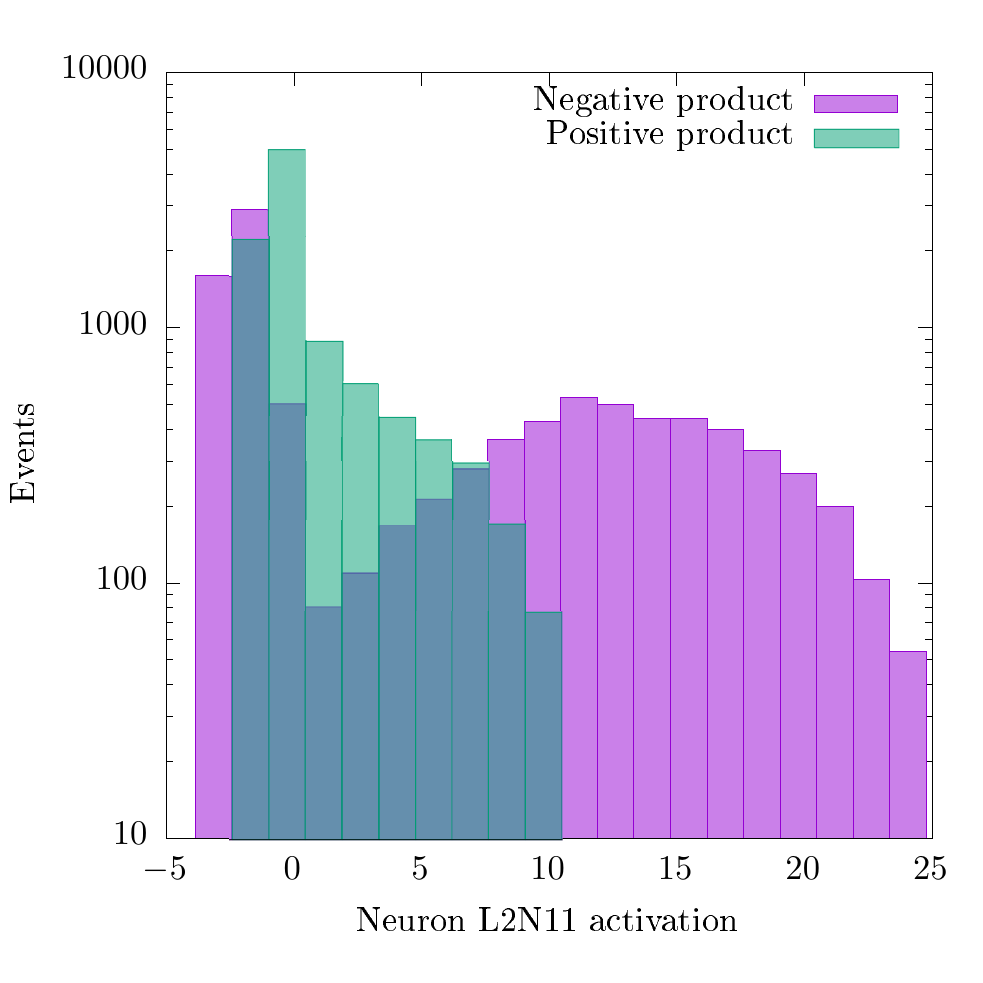}} \scalebox{0.54}{\includegraphics{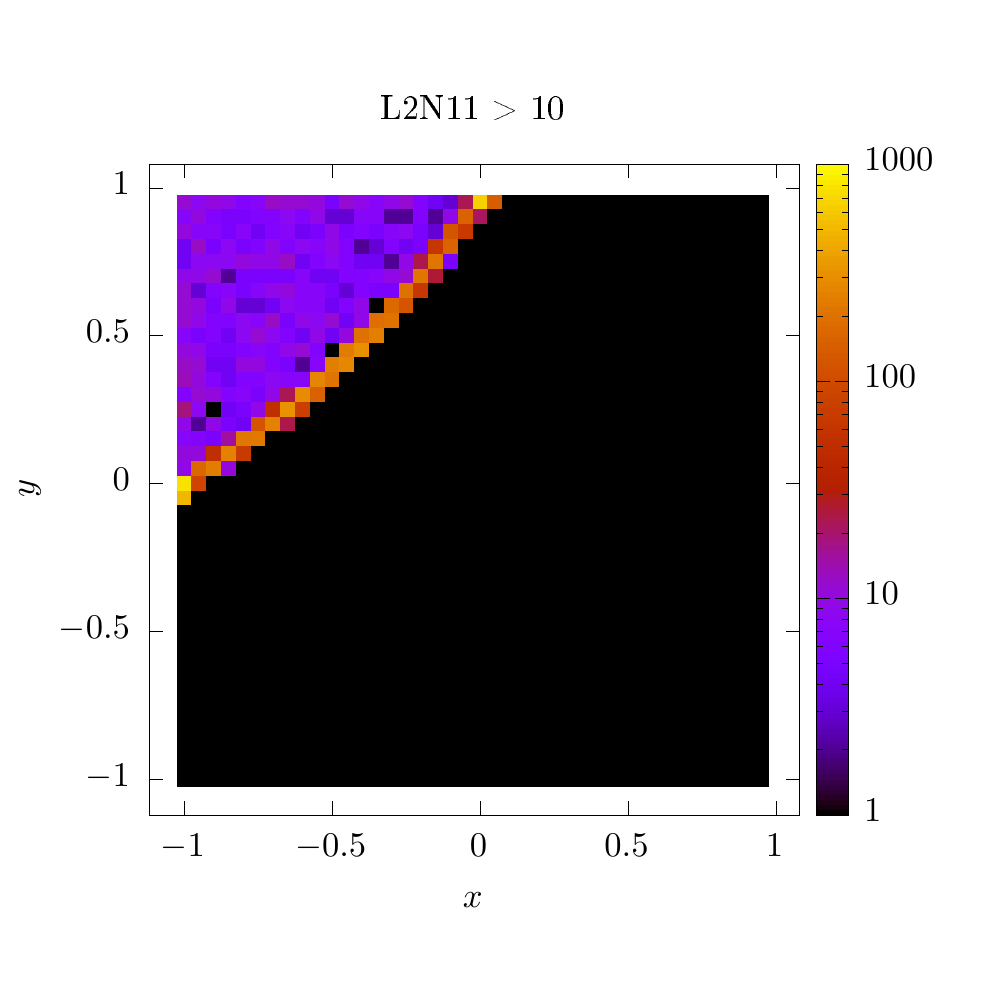}}\scalebox{0.54} {\includegraphics{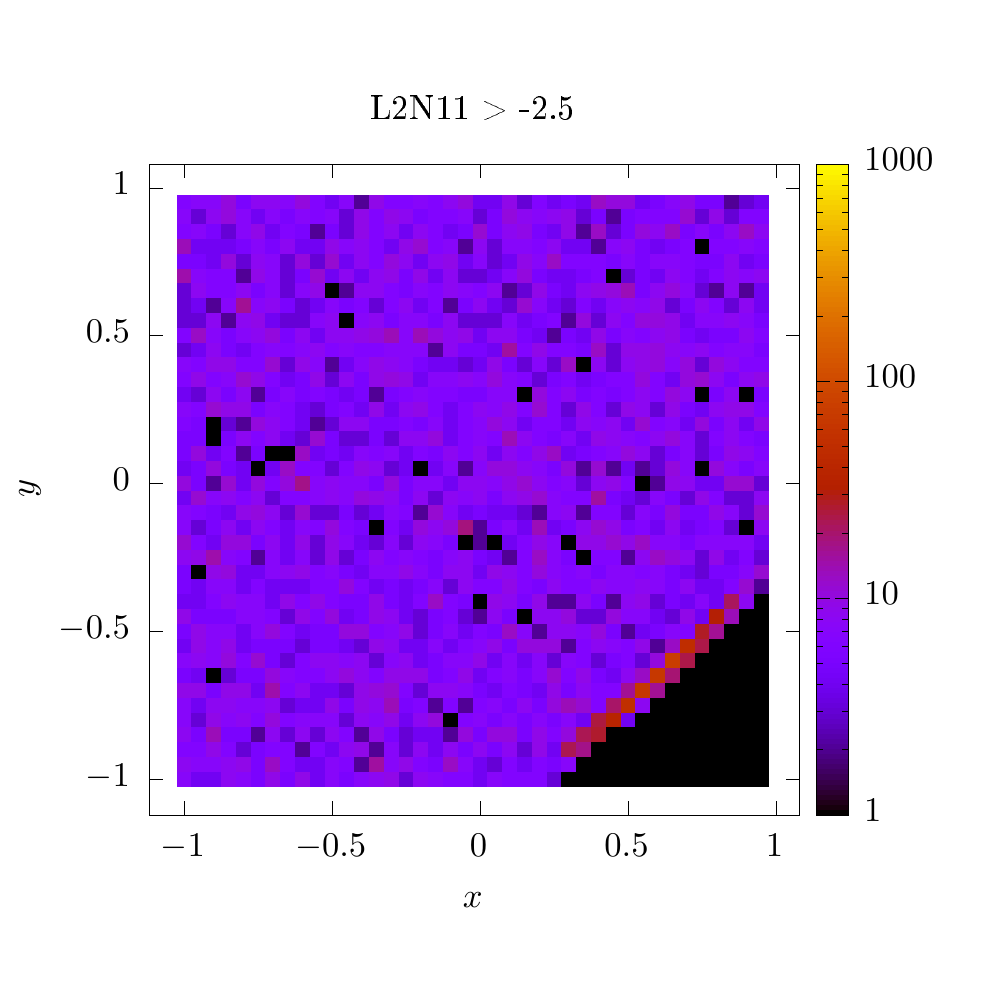}} \\
\caption{Analysis of neuron $L2N11$ of a network trained to distinguish the sign of the product of two input variables. Left: neuron activations on ``data.'' Right: activation maximization histograms.}
\label{fig:toydata_pos_L2N11}
\end{center}
\end{figure}

Activation maximization histograms for two early neurons in this network, $L2N36$ and $L2N11$, are shown in Figures \ref{fig:toydata_pos_L2N36} and \ref{fig:toydata_pos_L2N11}. We see a common pattern that will often come up later: in both cases the maximized inputs are clustered in a corner of the graph, with a relatively low density area marked off by a very high density boundary. The reason for this is simple: the points on the boundary have been pushed there from the empty regions of
the plot by the maximization procedure, while the points in the interior of the cluster were simply initially placed above the activation threshold. However, the important takeaway from these histograms is that the apparently sharp boundary we observe is \emph{not} actually a feature of the underlying activation function, but just a function of the threshold we have chosen; it is a single contour on the contour map of the neuron's activation function. The activation function itself is likely smoothly
increasing as it crosses the boundary --- for example, these test neurons appear to be learning something akin to $x+y$ for $L2N36$ and $y-x$ for $L2N11$.

\begin{figure}[!htbp]
\begin{center}
\scalebox{0.55}{\includegraphics{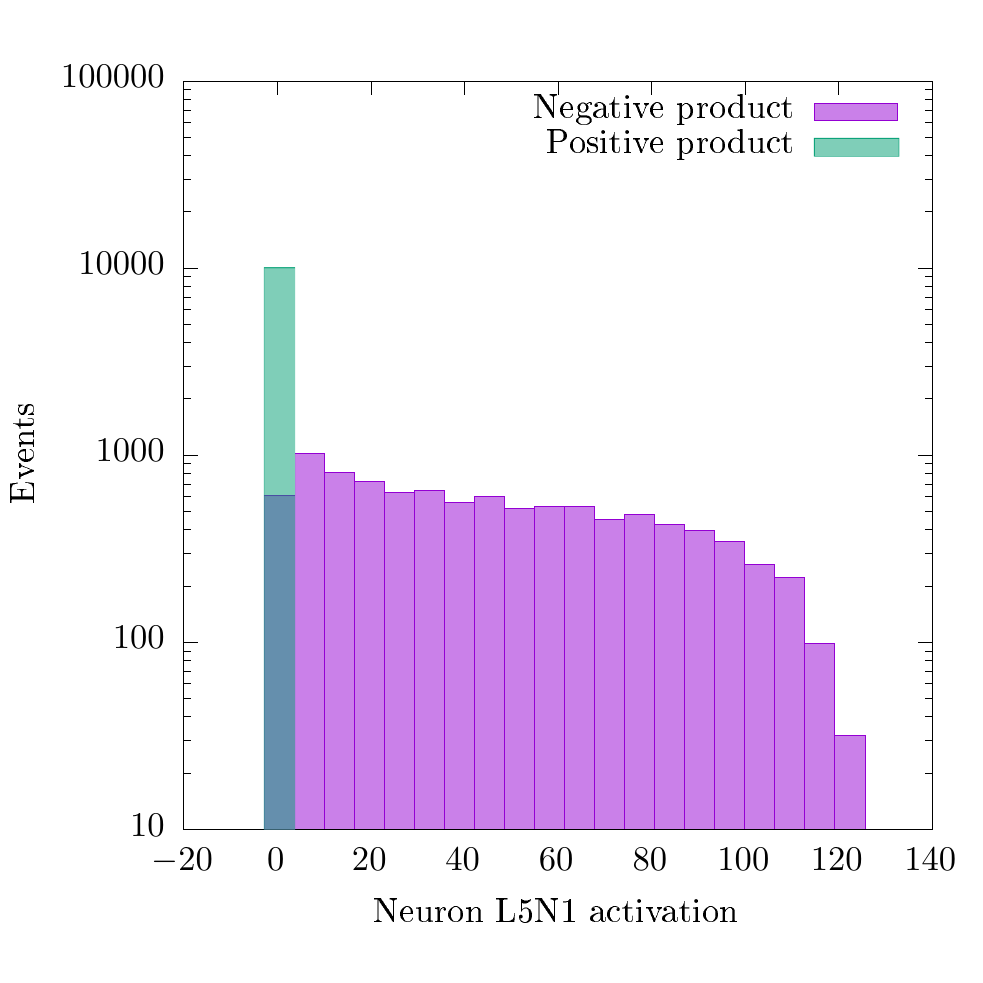}}\scalebox{0.55}{\includegraphics{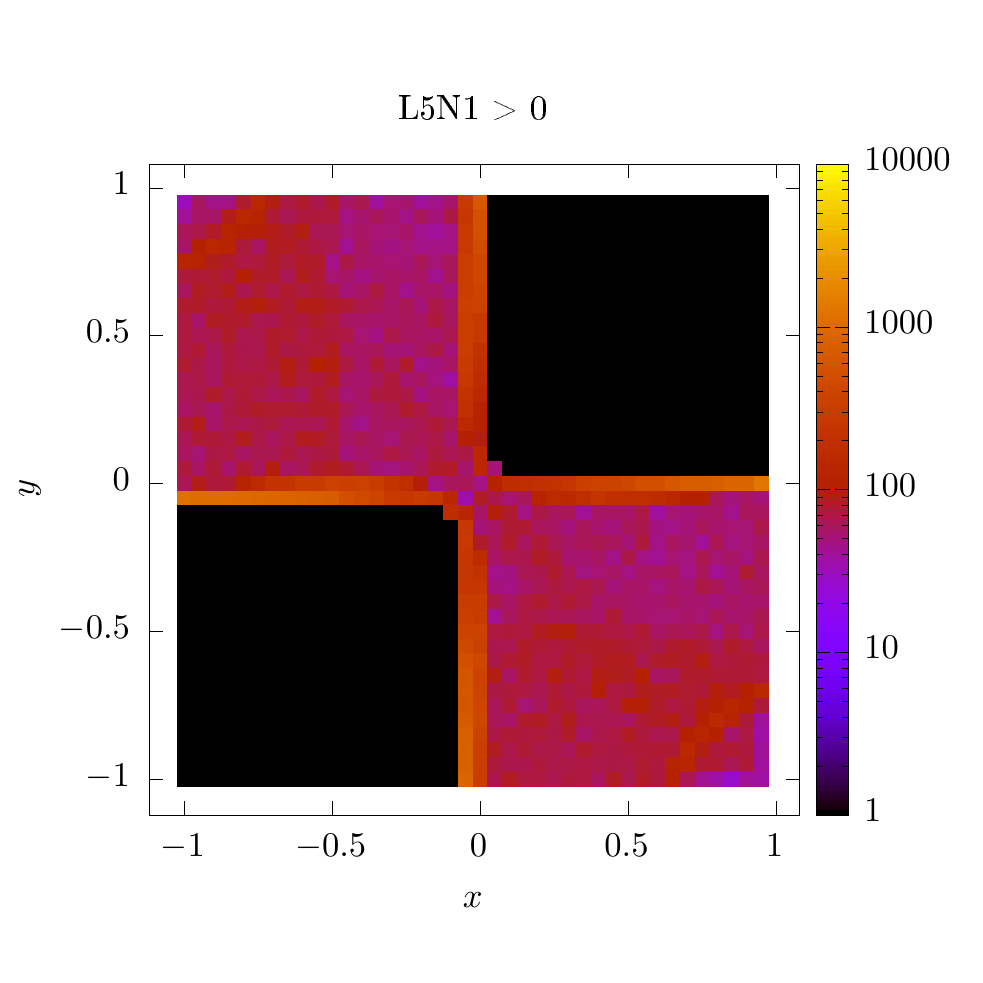}}\scalebox{0.55}{\includegraphics{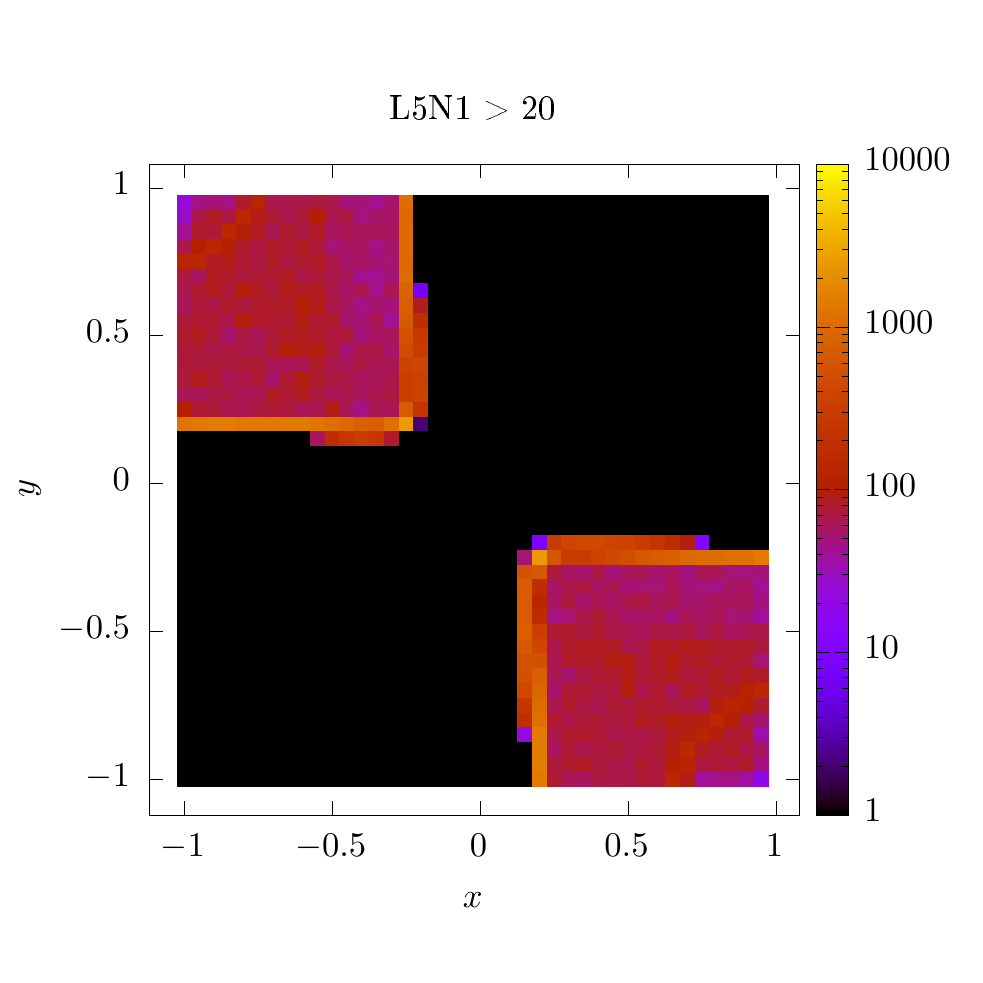}}
\caption{Analysis of neuron $L5N1$ of a network trained to distinguish the sign of the product of two input variables. Left: neuron activations on ``data.'' Right: activation maximization histograms.}
\label{fig:toydata_pos_L5N1}
\end{center}
\end{figure}

If we go down a few layers, we see that the network is already able to more or less solve the problem by the 5th hidden layer (Figure \ref{fig:toydata_pos_L5N1}). We also see that it has come up with a certainty measurement which more confidently predicts the output of points in the outer corners, although these points are not explicitly privileged by the data.

\subsubsection{Toy model: Cartography}
\label{sec:hawaii}

\begin{figure}[!htbp]
\begin{center}
\begin{tabular}{|c|c|}
\hline
\includegraphics[width=0.4\textwidth]{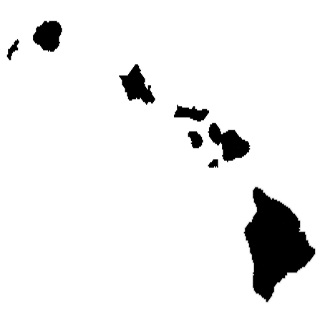} \\
\hline
\end{tabular}
\caption{The map of Hawaii that serves as the underlying truth distribution the network is trained on. The network is then trained to, given a pair of real numbers between 0 and 1, return -1 if the corresponding pixel is black and 1 if it is white.}
\label{fig:toydata_hawaii_image}
\end{center}
\end{figure}

\begin{figure}[!htbp]
\begin{center}
\scalebox{0.54}{\includegraphics{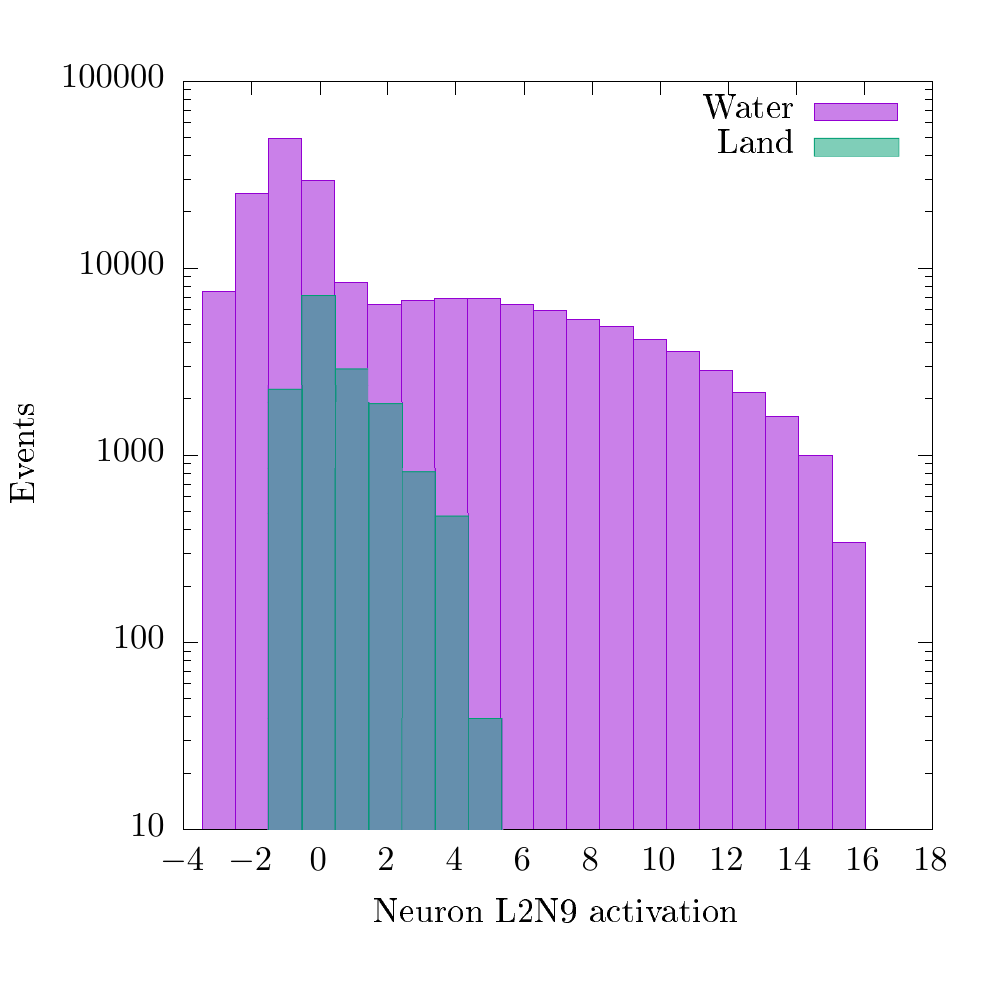}} \scalebox{0.54}{\includegraphics{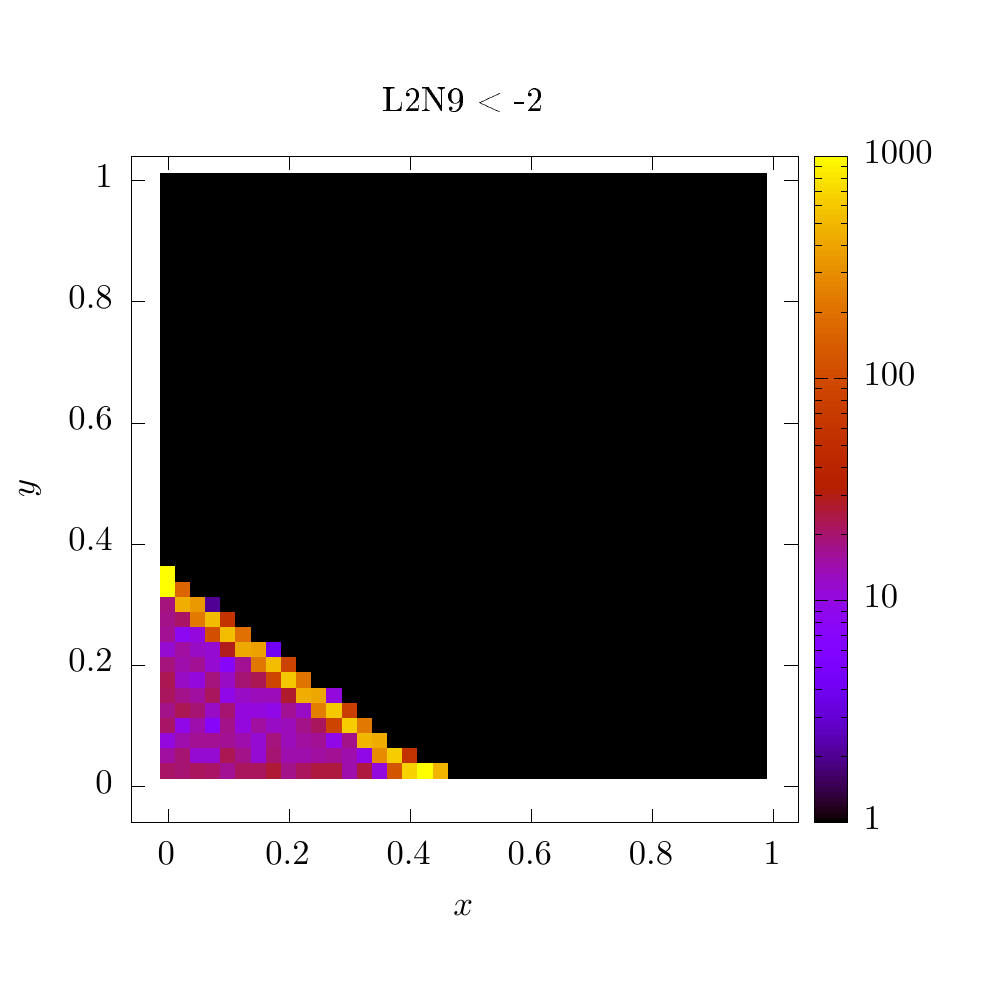}} \scalebox{0.54}{\includegraphics{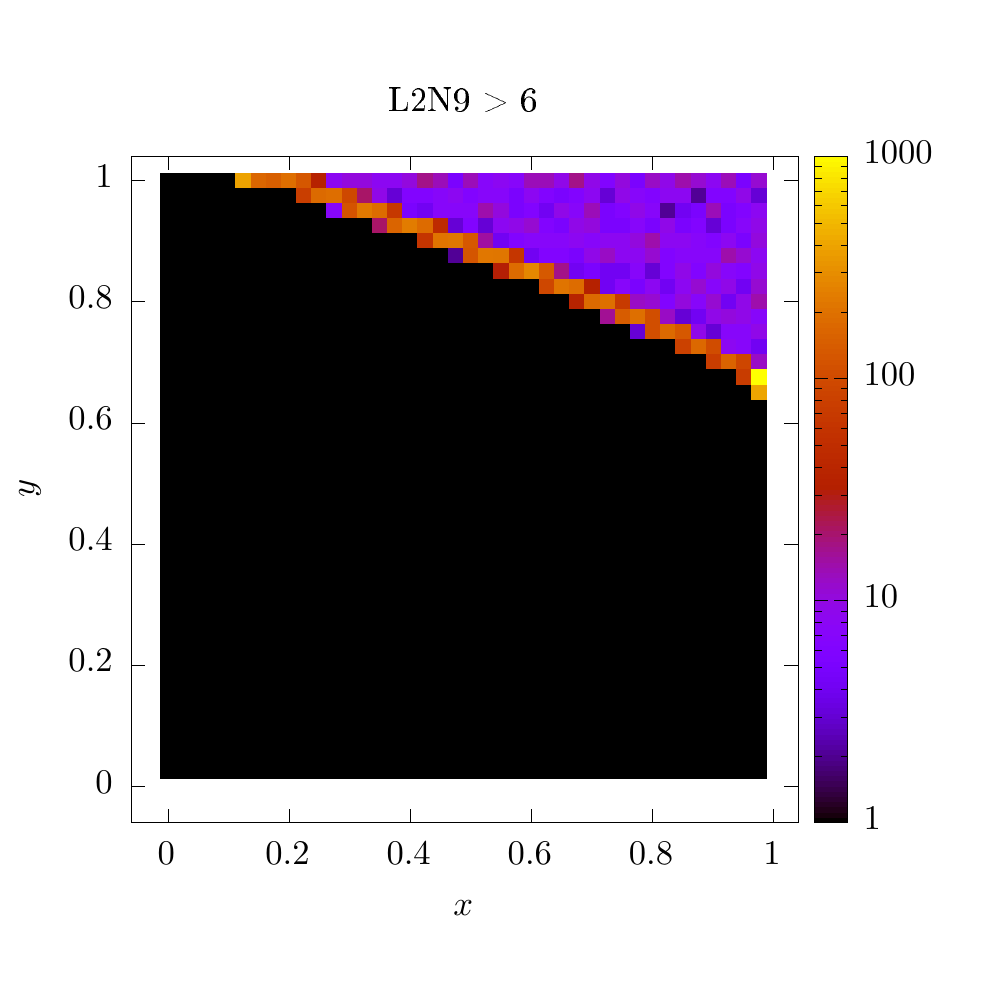}} \\
\scalebox{0.54}{\includegraphics{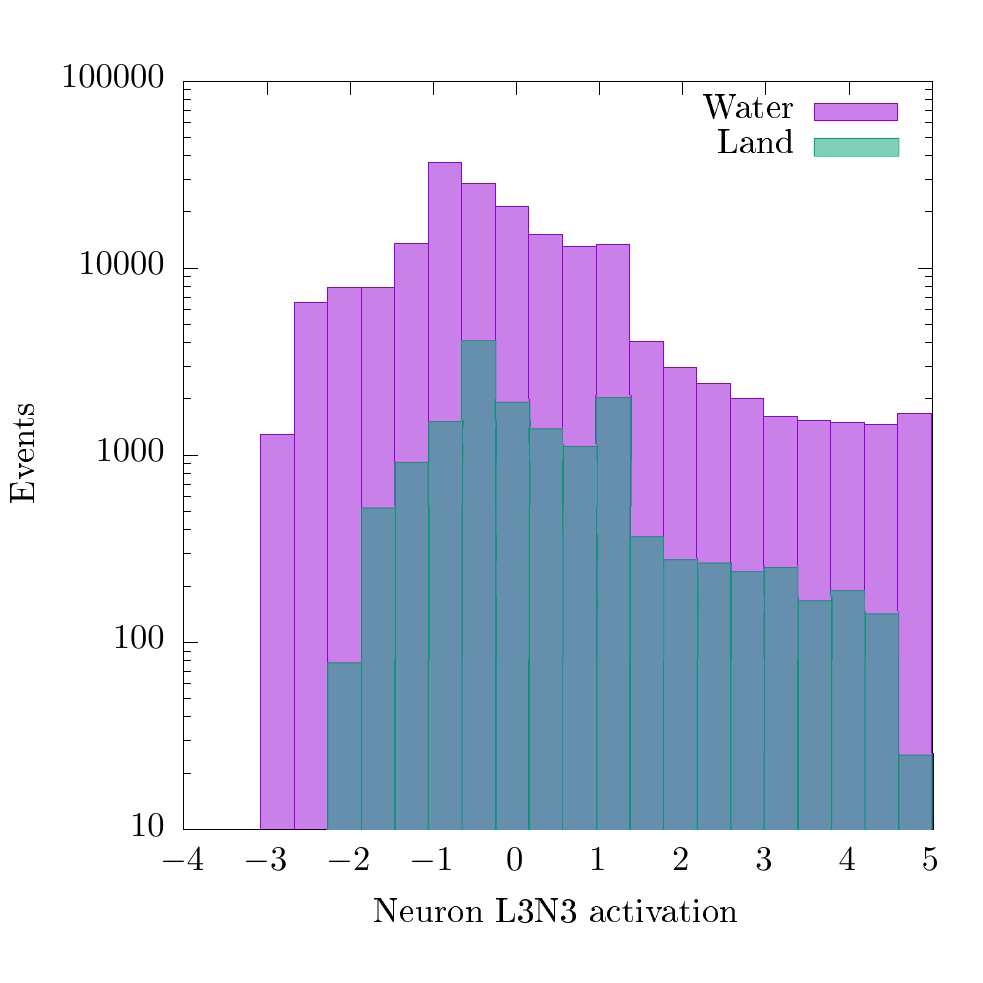}} \scalebox{0.54}{\includegraphics{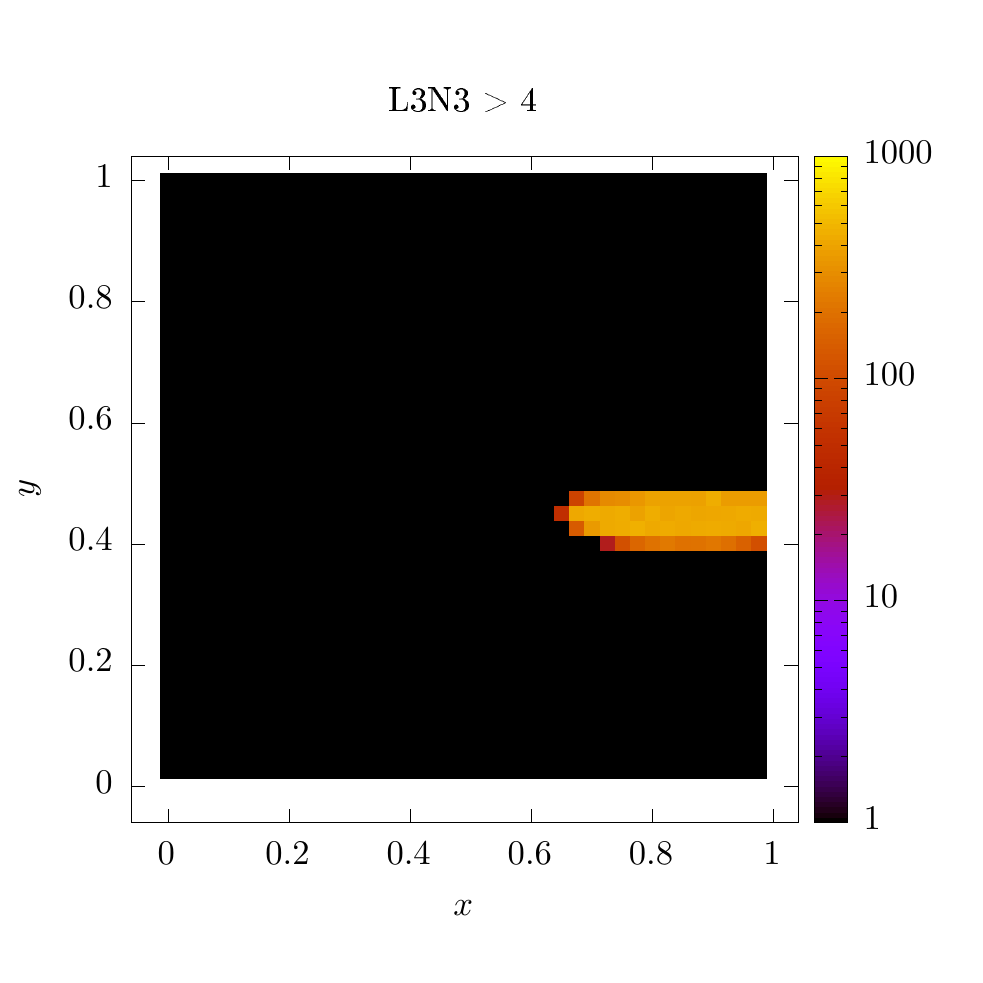}} \scalebox{0.54}{\includegraphics{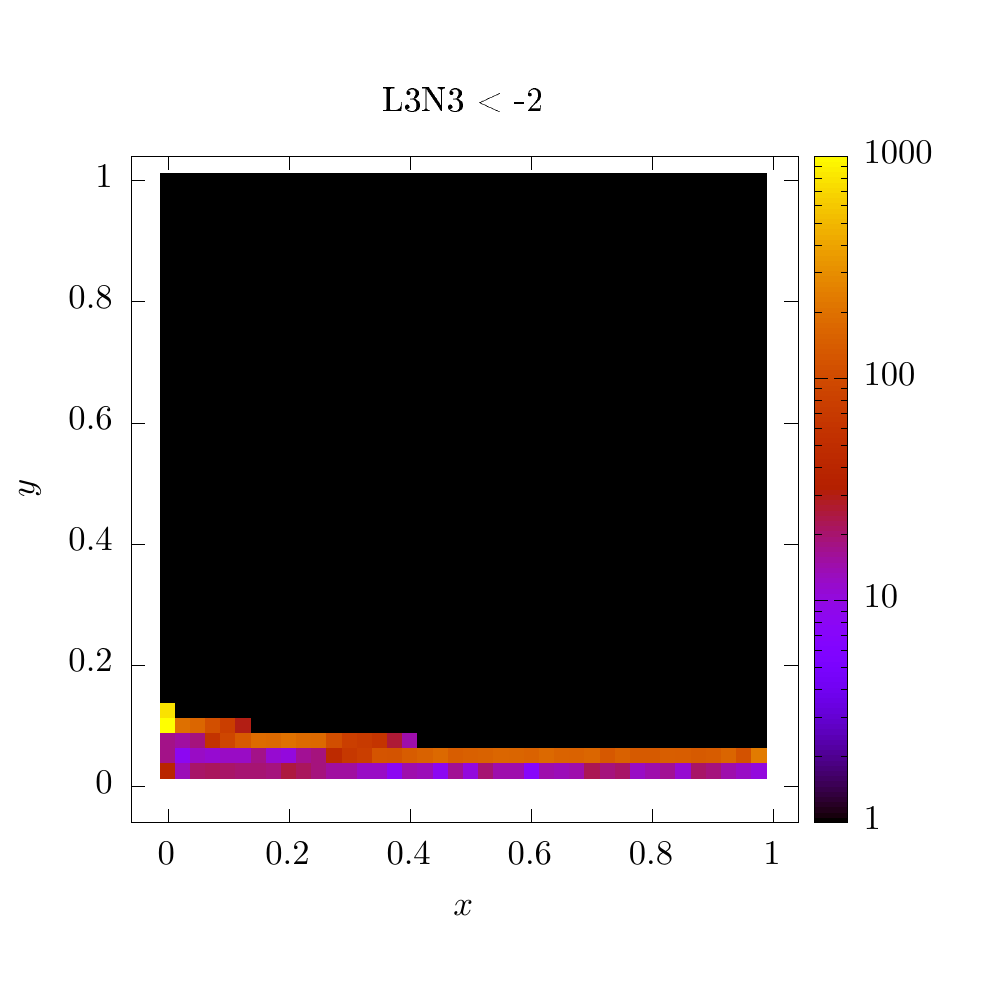}} \\
\scalebox{0.54}{\includegraphics{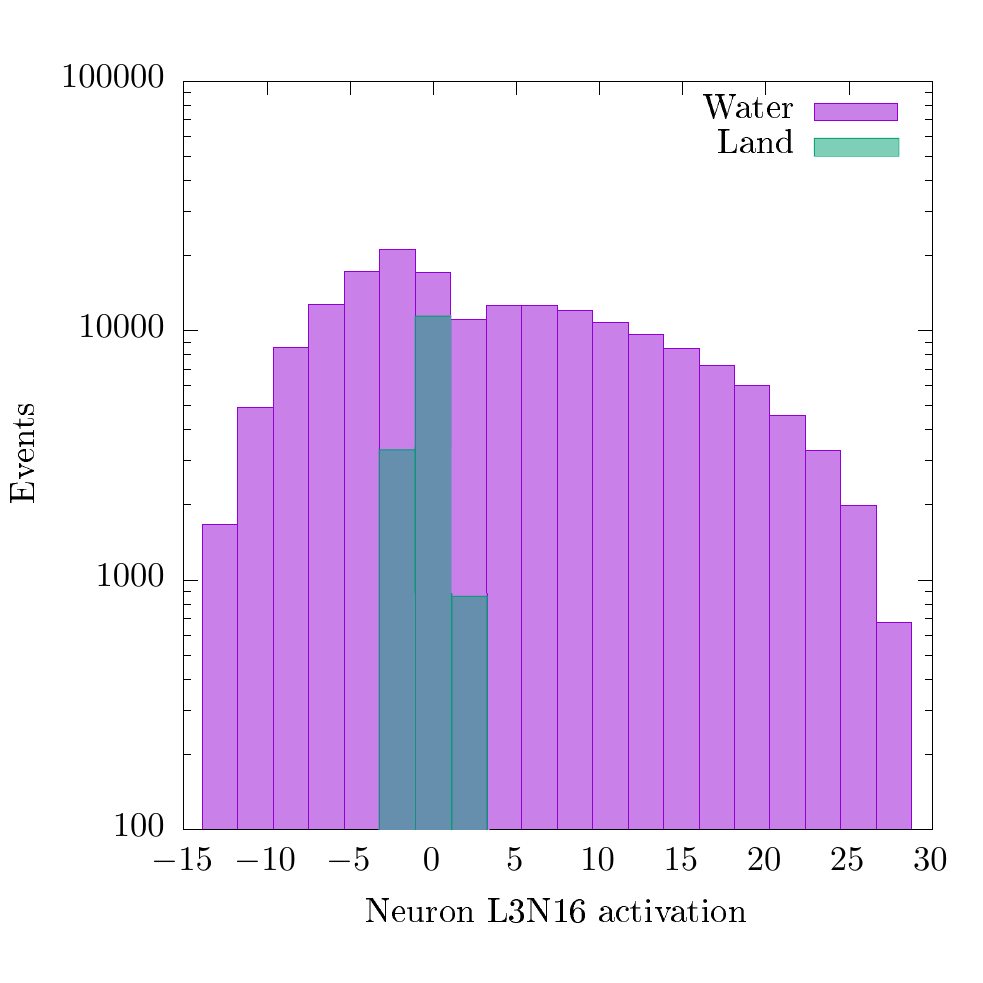}} \scalebox{0.54}{\includegraphics{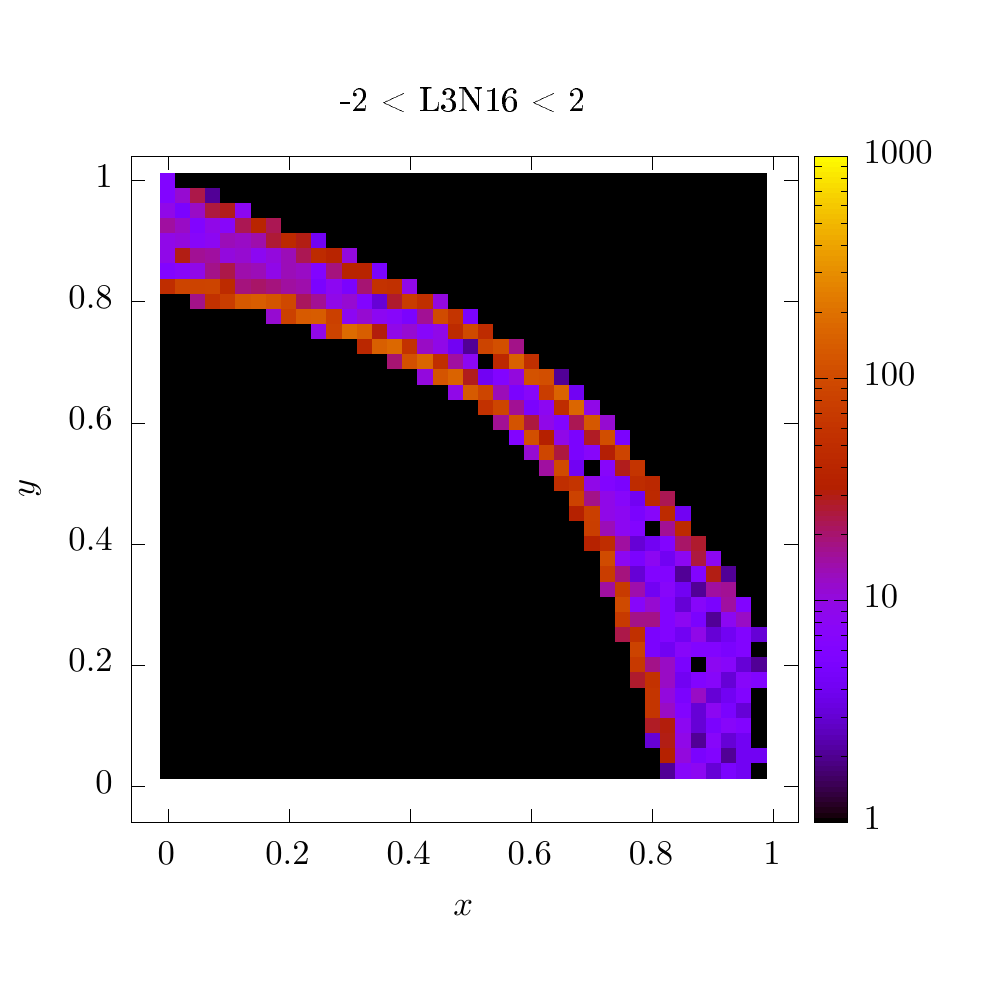}} \\
\caption{Activation maximization histograms of various neurons trained to recognize the islands of Hawaii.}
\label{fig:toydata_hawaii_1}
\end{center}
\end{figure}

\begin{figure}[!htbp]
\begin{center}
\scalebox{0.54}{\includegraphics{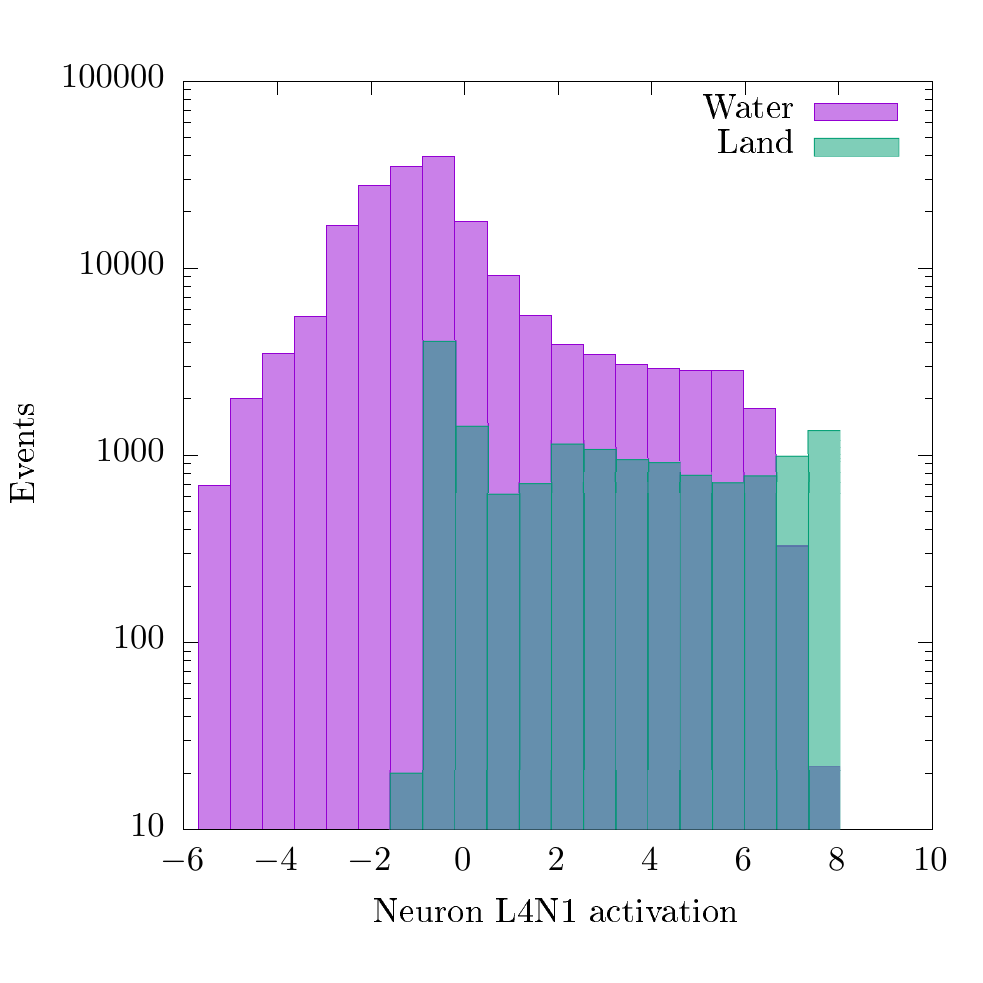}} \scalebox{0.54}{\includegraphics{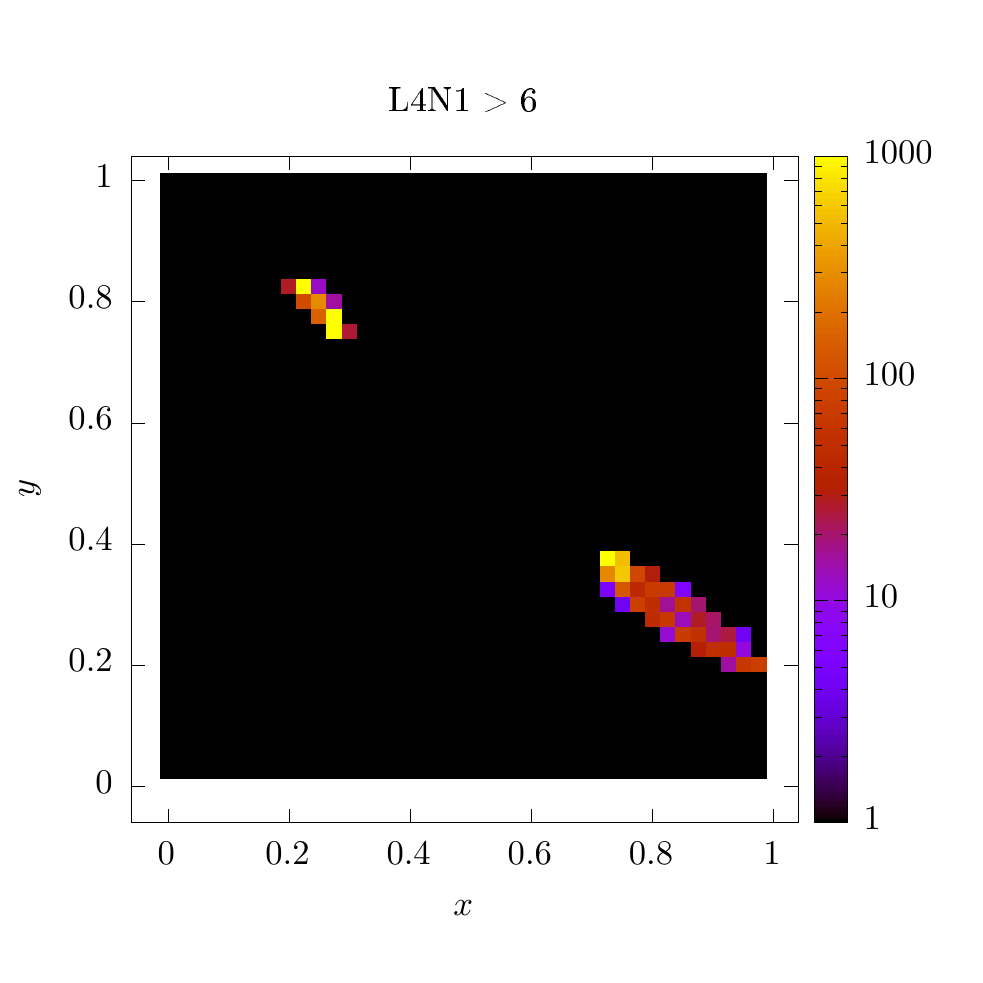}} \scalebox{0.54}{\includegraphics{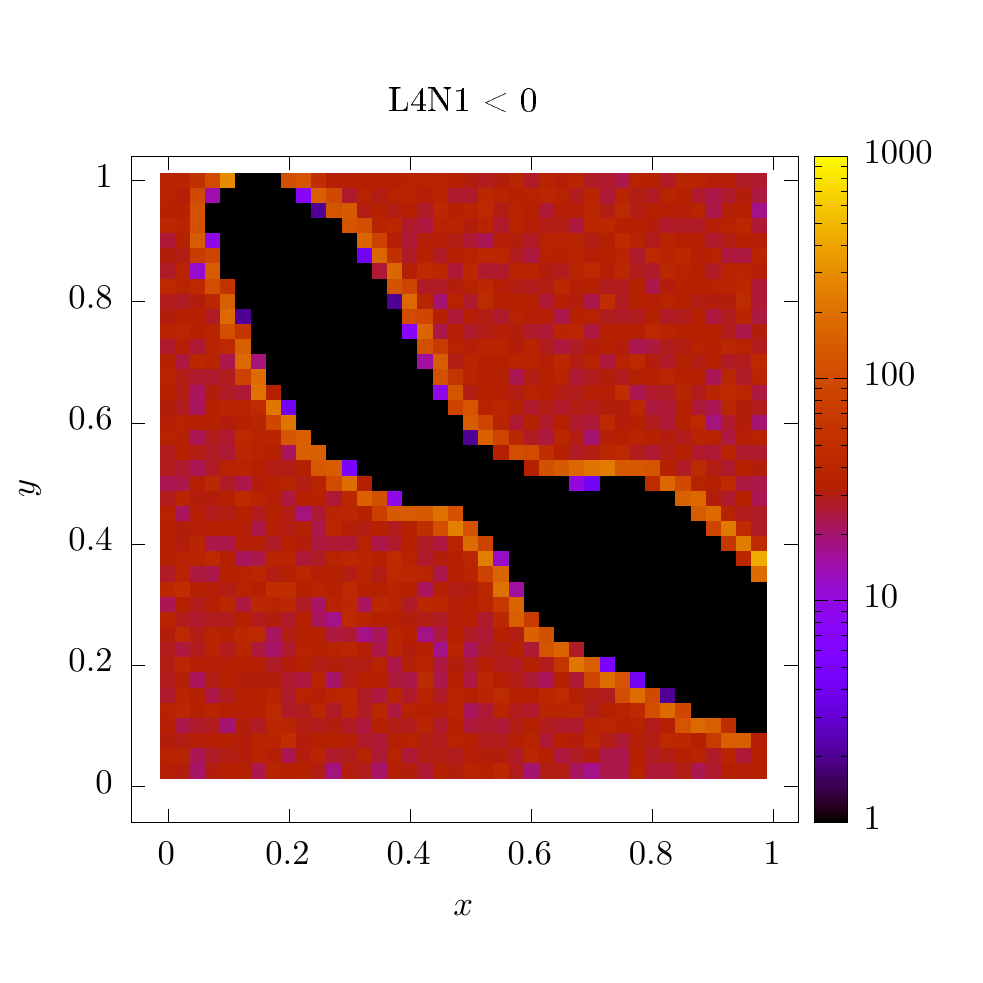}}\\
\scalebox{0.54}{\includegraphics{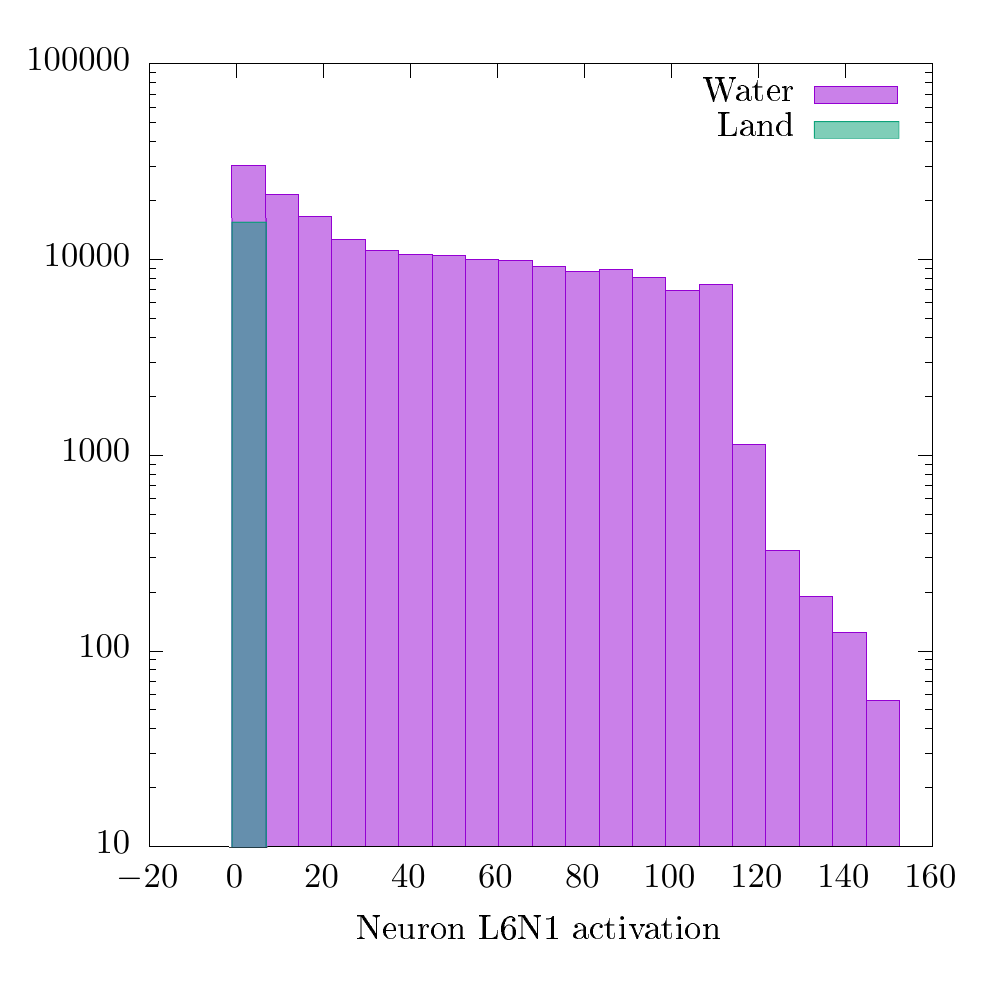}} \scalebox{0.54}{\includegraphics{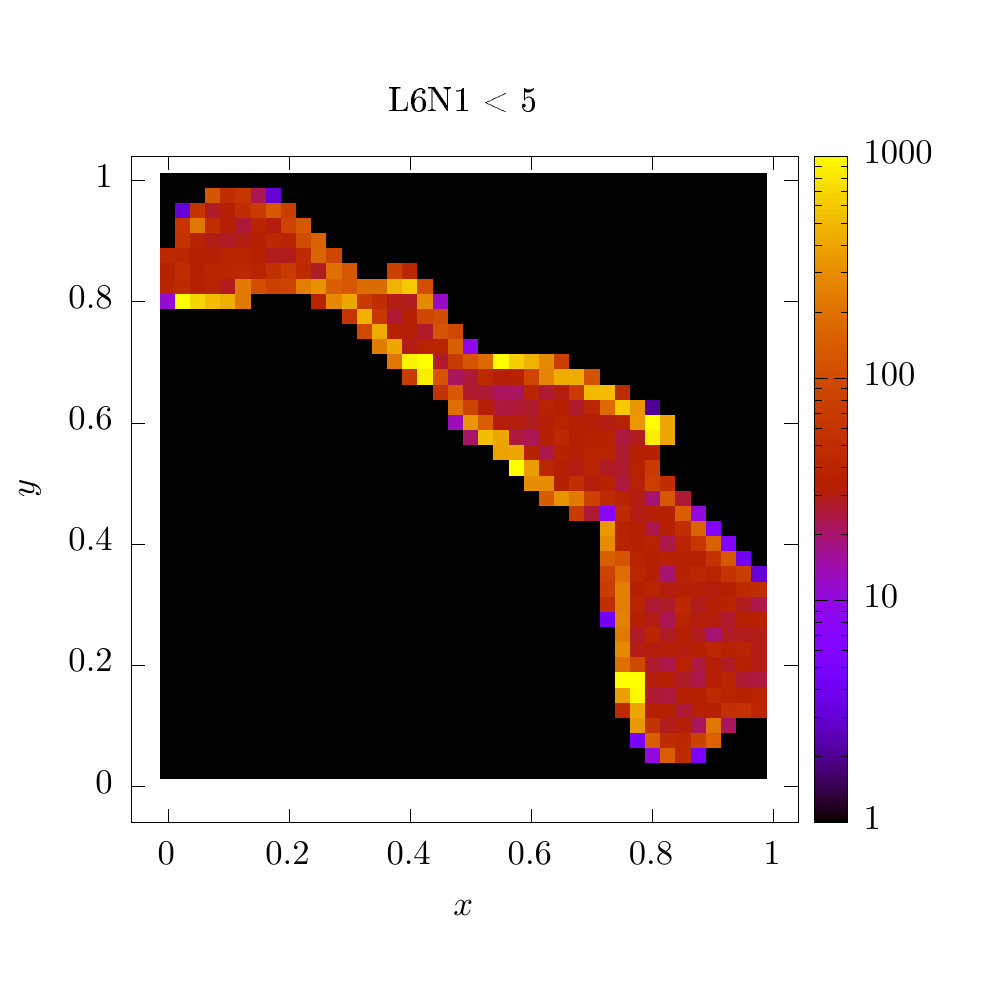}} \scalebox{0.54}{\includegraphics{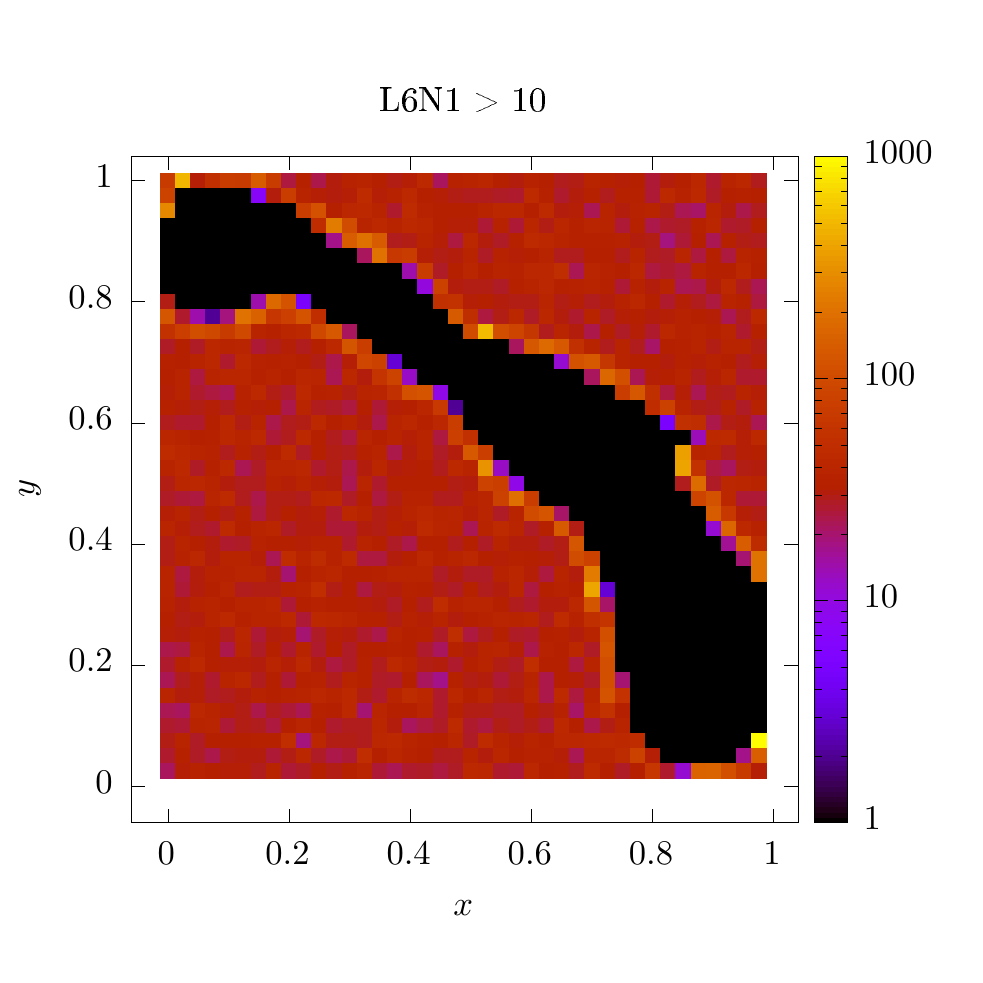}} \\
\scalebox{0.54}{\includegraphics{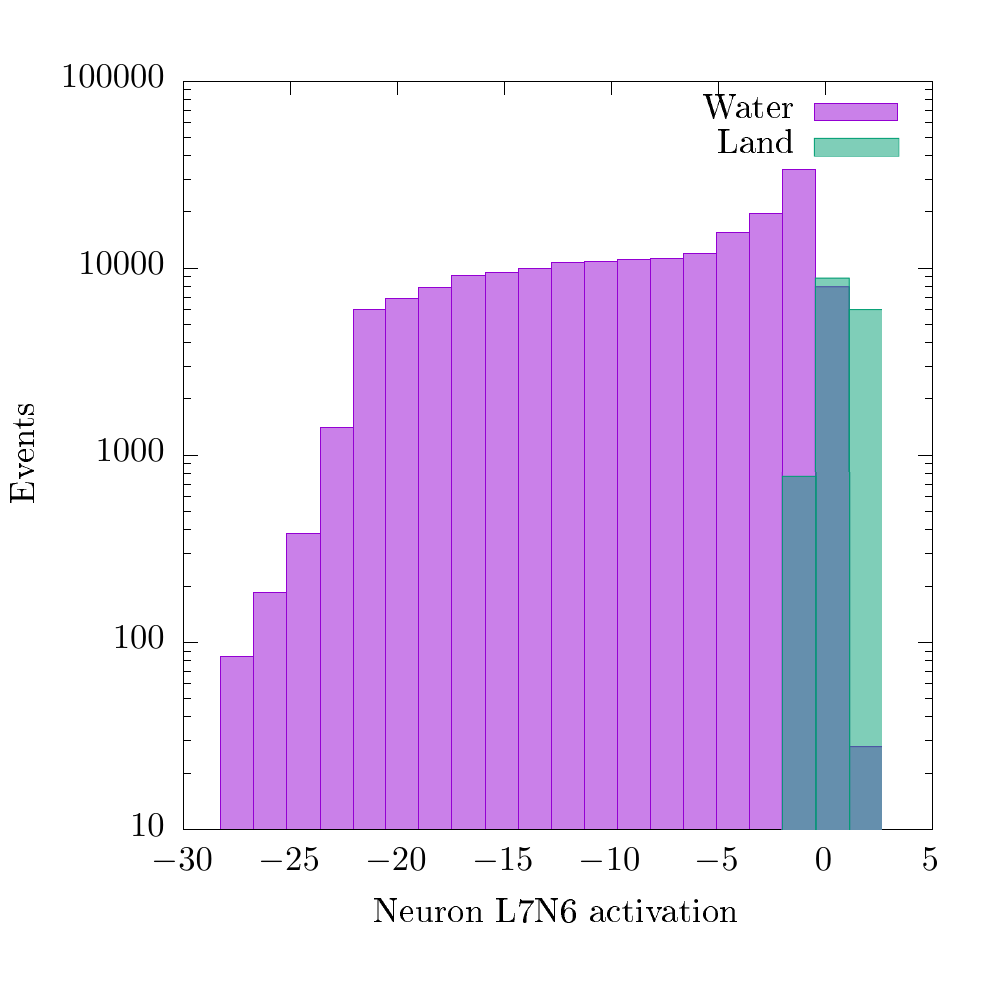}} \scalebox{0.54}{\includegraphics{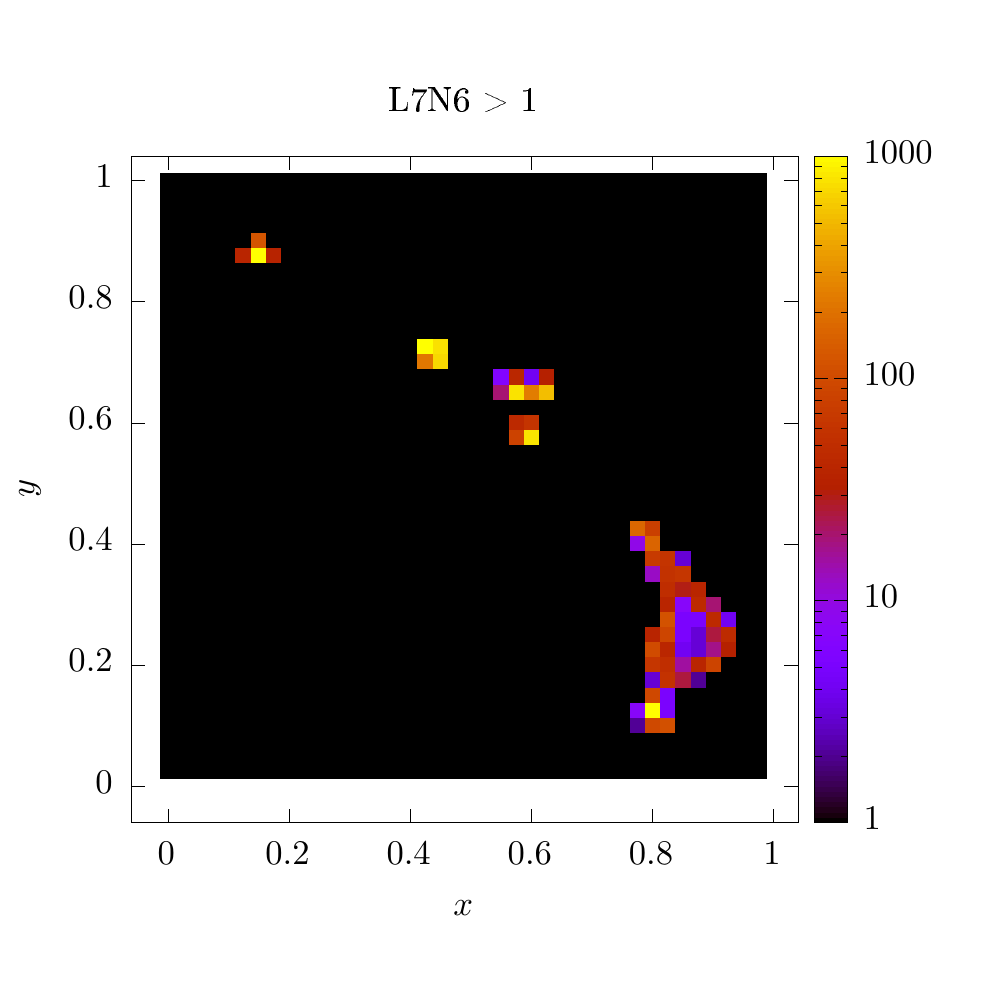}} \scalebox{0.54}{\includegraphics{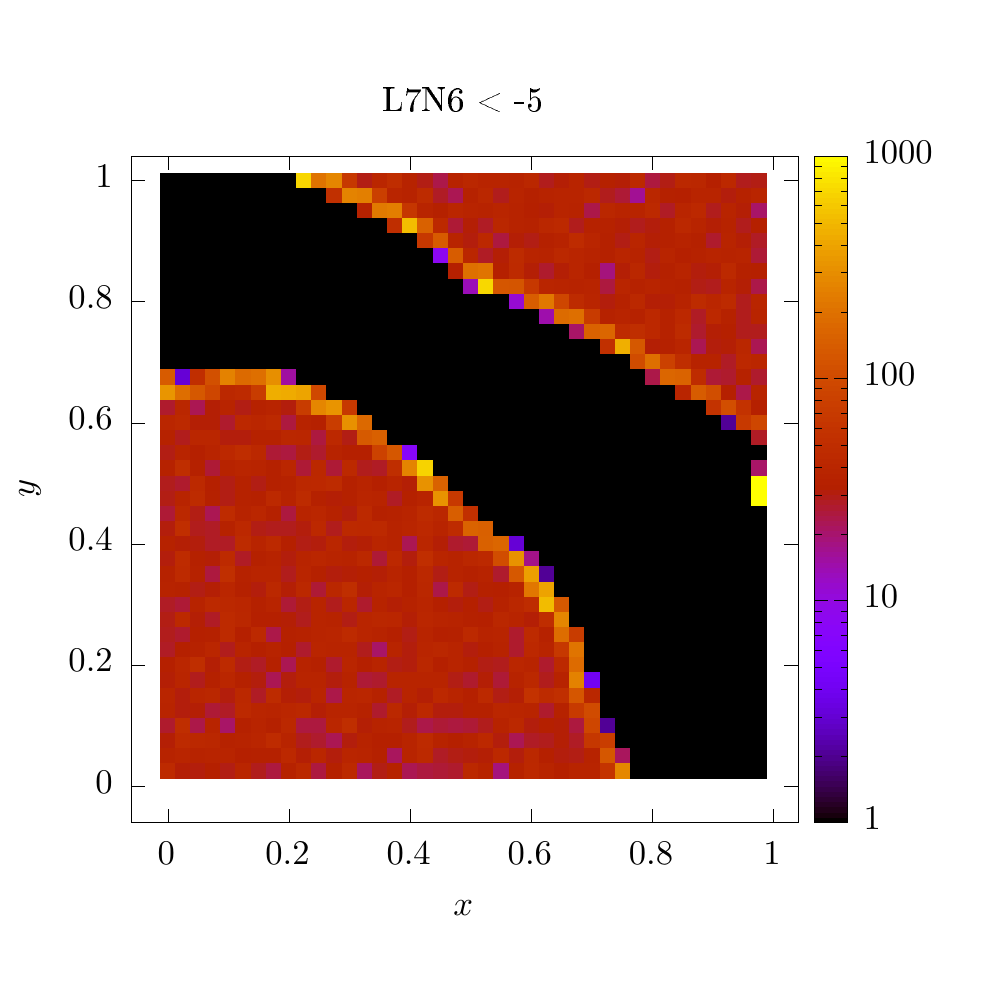}}\\
\scalebox{0.54}{\includegraphics{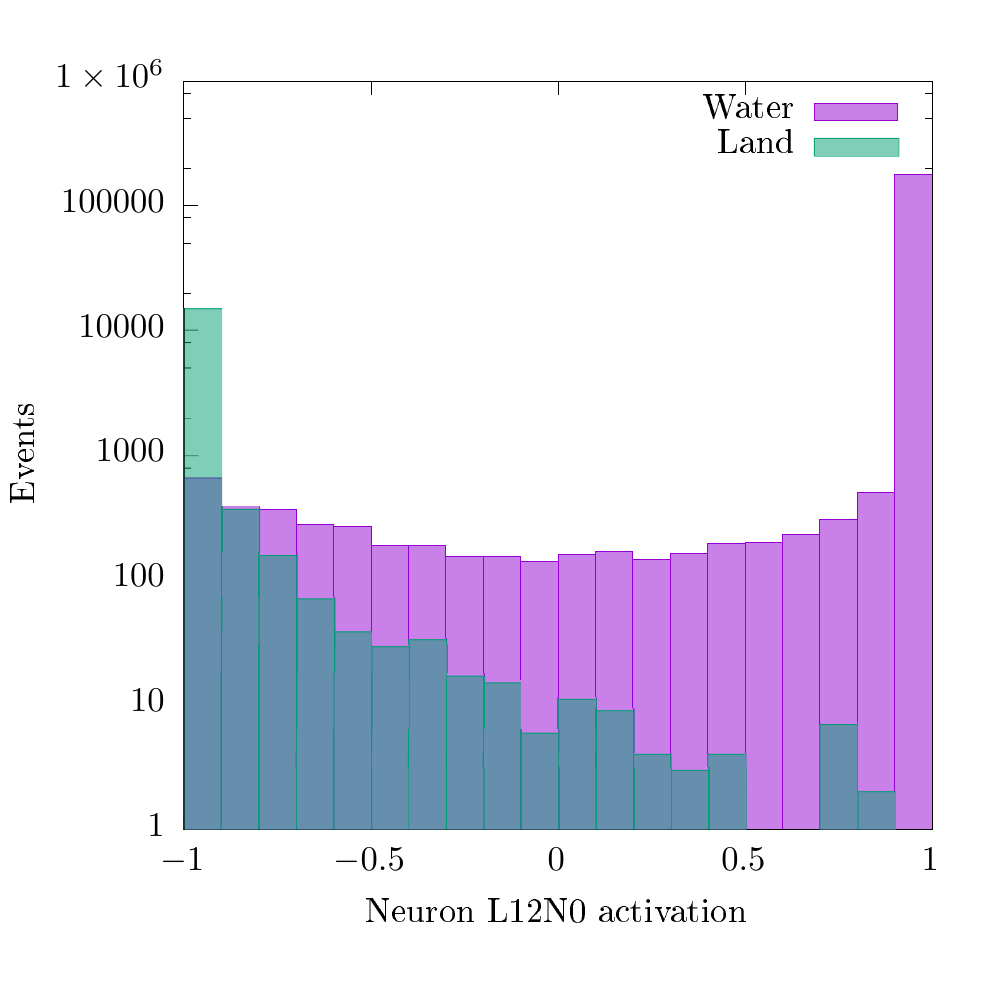}} \scalebox{0.54}{\includegraphics{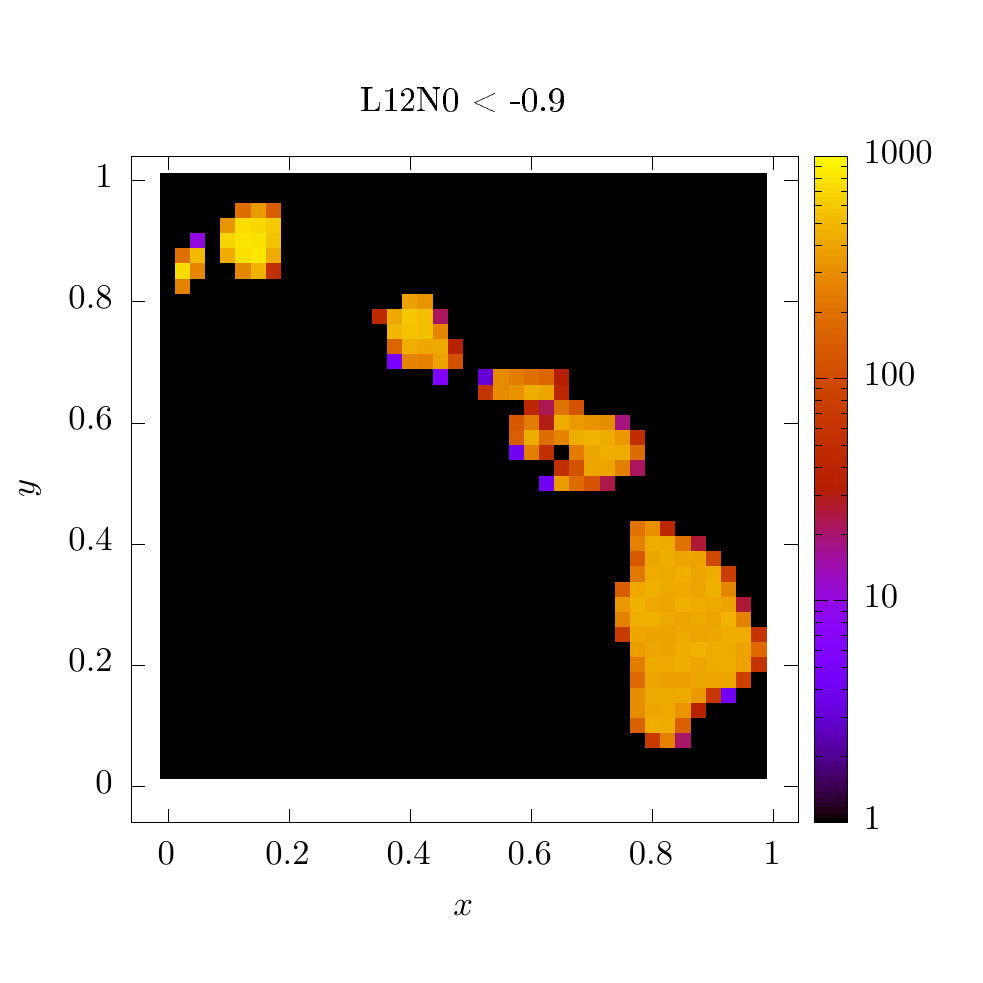}} \\
\caption{Activation maximization histograms of various neurons trained to recognize the islands of Hawaii.}
\label{fig:toydata_hawaii_2}
\end{center}
\end{figure}

Now let's try a slightly more difficult problem. Specifically, given a latitude-longitude pair, we would like our network to tell us whether the coordinates refer to a location on the landmass of Hawaii, as depicted on the map in Figure \ref{fig:toydata_hawaii_image}. This problem is actually fairly complicated because it requires the network to remember a detailed image without any underlying pattern to guide it. Nonetheless, the network is able to succeed in its task; we can see how by examining the activation maximization histograms in Figures \ref{fig:toydata_hawaii_1} and \ref{fig:toydata_hawaii_2}.

These images merely show a few representative neurons from a few layers, but they are enough to get a fairly good general picture of how the network operates: namely, each neuron encodes a small slice or scoop of the input space which is predominantly either signal or background, and these slices are then gradually combined together in later layers to make more and more complicated shapes.

\clearpage

\subsection{The caveman variable}
\label{sec:caveman}

As a sanity check, we would like to demonstrate that it is actually possible to extract meaningful approximations of the network behavior from its activation maximization histograms. To this end, we consider what we will refer to as ``caveman'' variables,  crude approximations of the network activations constructed out of a set of human-readable terms drawn from the activation maximization patterns. These variables are not meant to provide a replacement or equivalent formulation for the network itself, but merely to act
as a demonstration that the relatively simple patterns extracted by observing only the highest-activation events can actually serve as useful proxies for the more complicated behavior of the full network.

The basic idea behind these variables is that the points in an activation maximization histogram tend to group themselves in clusters or ``islands'' much like the islands of Hawaii in \ref{sec:hawaii}. We can then try to very roughly model these islands with functions that have support only within their domain. Summing up a number of such functions can serve as a very crude approximation of the network (or at least the highest activation region) that can be directly comprehended by a human observer. A more complete description of the algorithm for constructing these variables is in Appendix \ref{app:caveman}.

\section{The simplest case: uncompressed stops}
\label{sec:uncompressed}

The problem of differentiating stop decay from background events is relatively tractable in the region $m_{\widetilde t} \gg m_{\widetilde \chi}$, since the additional energy released by the decay of the very massive stops can result in events with much higher missing momentum than any background events. A well-known method for exploiting this observation is to use the \emph{stransverse mass} or $m_{\rm T2}$, used for instance in the CMS search \cite{Sirunyan:2017leh}. We begin with the familiar {\em transverse mass} constructed from the momentum of a lepton $\ell$ and neutrino $\nu$,
\begin{equation}
m_T^2 = 2 p_T(\ell) p_T(\nu) \left[1 - \cos\left(\phi(\ell) - \phi(\nu)\right)\right],
\end{equation}
which is invariant under boosts in the $z$ direction and is bounded above by the $W$ mass if both the lepton and neutrino originate from the same parent $W$. Given a measured missing transverse momentum and the hypothesis that it originates from neutrinos from two $W$ decays in an event, we can scan over all possible decompositions into two separate momenta to form the \emph{lepton-based stransverse mass} ($m_{\text{T2}}^{\ell \ell}$) defined as \cite{Lester:1999tx, Burns:2008va}
\begin{equation}
m_{\text{T2}}^{\ell \ell} = \min_{\vec{p}_{\text{T1}}^{\text{miss}} + \vec{p}_{\text{T2}}^{\text{miss}} = \vec{p}_{\text{T}}^{\text{miss}}} \left( \max \left[ m_\text{T}(\vec{p}_\text{T}^{\ell_1}, \vec{p}_\text{T1}^\text{miss}), m_\text{T}(\vec{p}_\text{T}^{\ell_2}, \vec{p}_\text{T2}^\text{miss})\right] \right).
\label{eq:mT2ll}
\end{equation}
The minimization is performed over all possible partitions of the missing energy. If there is no extra missing energy in the event, this variable will have an upper end-point at the $W$ mass. However, if there are additional invisible particles such as ${\widetilde \chi}$ as in the case of stop decays, we can have $m_{\text{T2}}^{\ell\ell} > m_W$; thus, such observations would be a strong sign of new physics.

\begin{figure}[!h]
\begin{center}
    \scalebox{0.55}{\includegraphics{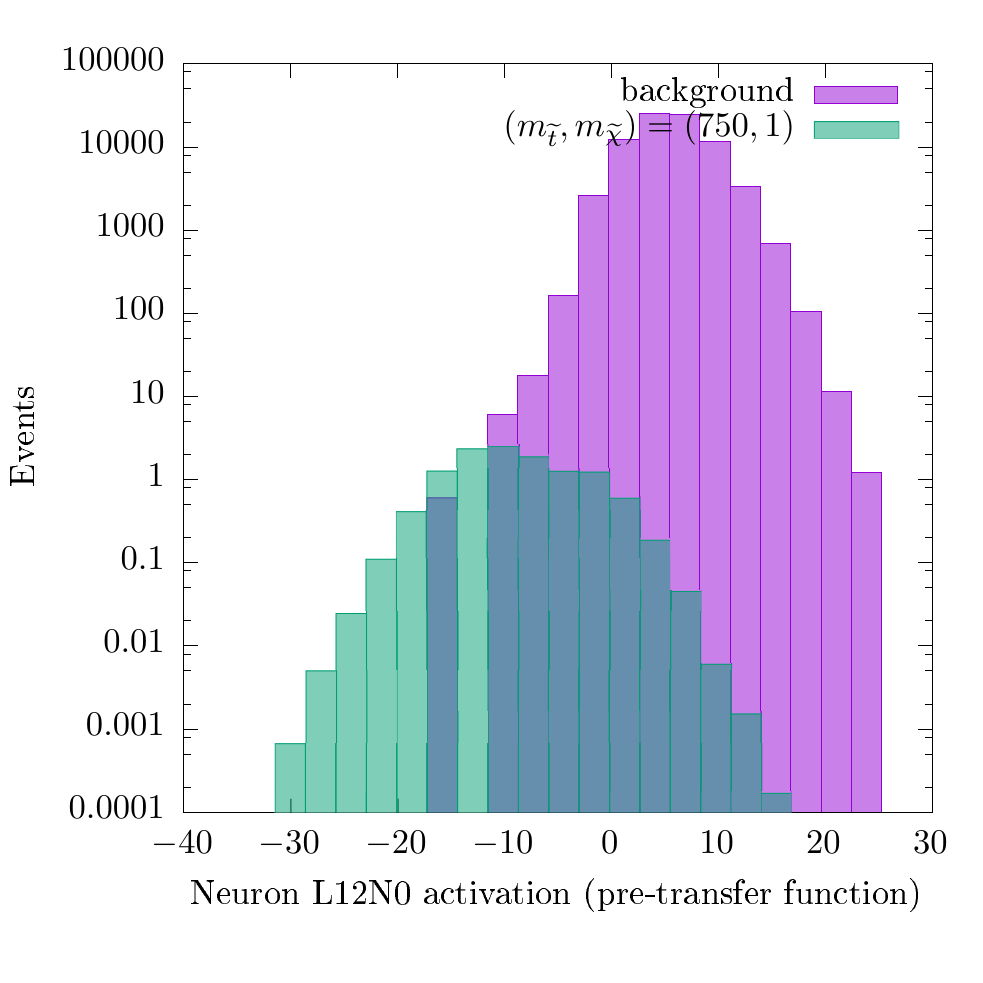}}
\end{center}
\caption{Activation histogram for the output neuron of a network trained to distinguish $(m_{\widetilde t}, m_{\widetilde \chi})$ = $(750, 1)$ stop decays. Although the network was trained with a $\tanh$ transfer function on the final neuron, the neuron output is plotted here before applying this function for more comprehensible viewing.}
\label{fig:nhist_deep750-1TN_L12N0}
\end{figure}

What happens when we train a neural network to solve the problem? For concreteness, we focus on a benchmark point where $m_{\widetilde t} = 750$ GeV and $m_{\widetilde \chi} = 1$ GeV, well within the uncompressed region. (The challenge for this parameter point is that the signal cross section is quite small.) Will the neural network settle upon the same discrimination strategy that is familiar from human-designed studies? To find out, we trained a neural network with the architecture
given in section \ref{sec:nndesign} on a dataset constructed as described in \S\ref{sec:simulation} with approximately 2 million signal and 8 million background events. A histogram of the activation of the output neuron for signal and background distributions is shown in Figure \ref{fig:nhist_deep750-1TN_L12N0}; it appears that the network is capable of distinguishing a subset of the data for which signal matches or exceeds background.

To quantify the efficacy of the network at distinguishing a background distribution from a signal + background distribution, we turn to the ``approximate median significance'' (AMS) variable used in the Higgs Machine Learning Challenge \cite{Adam-Bourdarios:2015pye, Cowan:2010js}. This variable is designed to give a value approximately equal to the number of sigma of the putative discovery; it is given by
\begin{equation}
AMS = \sqrt{2\left((s+b+b_r) \text{log}\left(1+\frac{s}{b+b_r}\right)-s\right)}
\end{equation}
where $s$ and $b$ are the expected number of true and false positives, respectively, and $b_r = 10$ is a constant regularization term.

By this metric, extrapolating our simulation to a total luminosity of 35.9 $\text{fb}^{-1}$, we find that our network has an AMS of 1.72 on this dataset for events with a pre-transfer output of less than $-10.4$. On the other hand, a simple cut on $m_{\text{T2}}$ is optimized at 156 GeV with an AMS of 1.56, so we conclude that our network is performing approximately as well as or slightly better than this variable alone. 


\begin{figure}[!h]
\begin{center}
    \scalebox{0.5}{\includegraphics{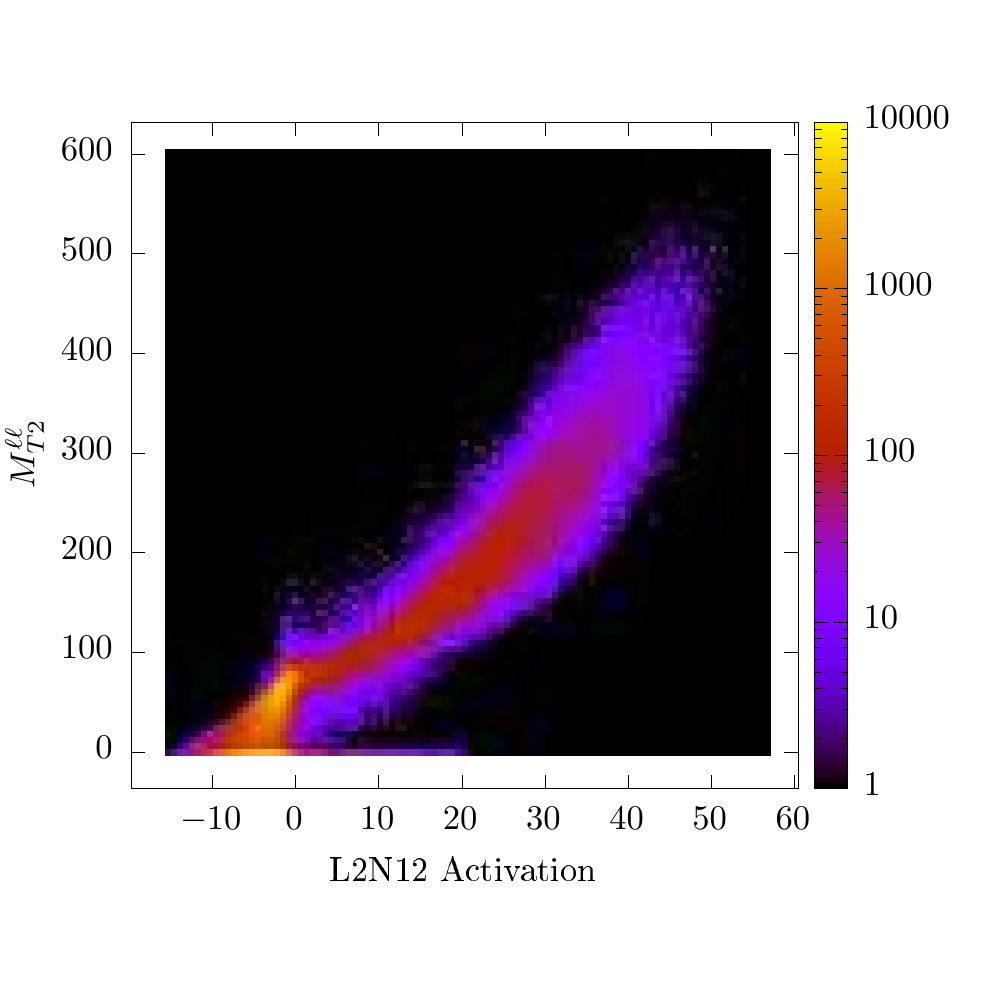}} \scalebox{0.5}{\includegraphics{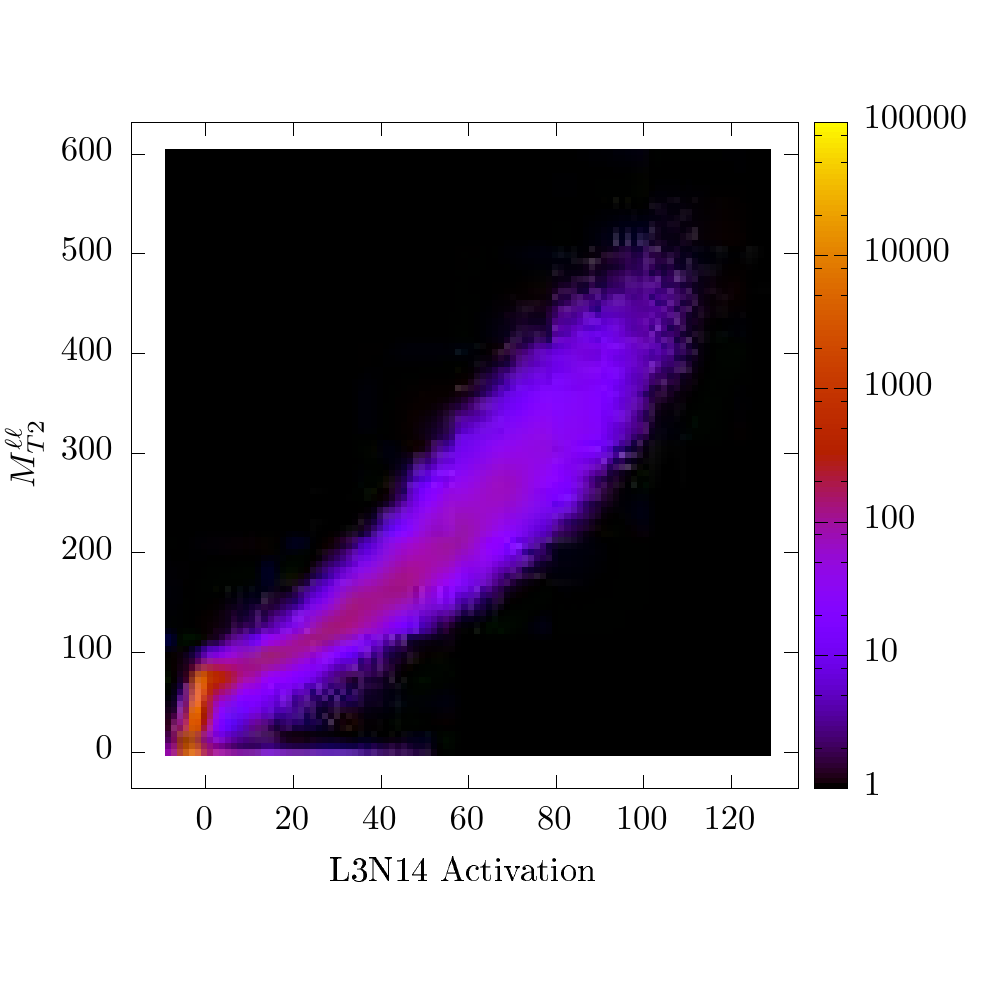}} \scalebox{0.5}{\includegraphics{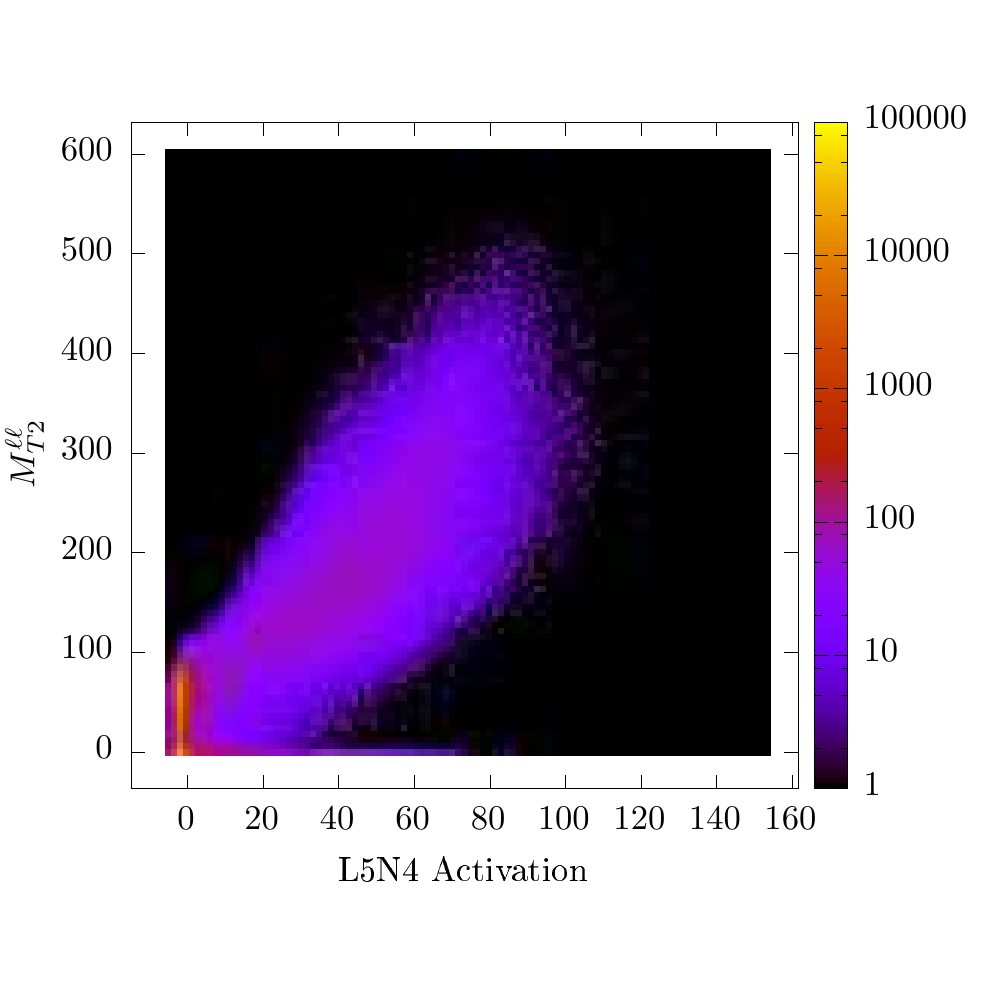}} \\
    \scalebox{0.5}{\includegraphics{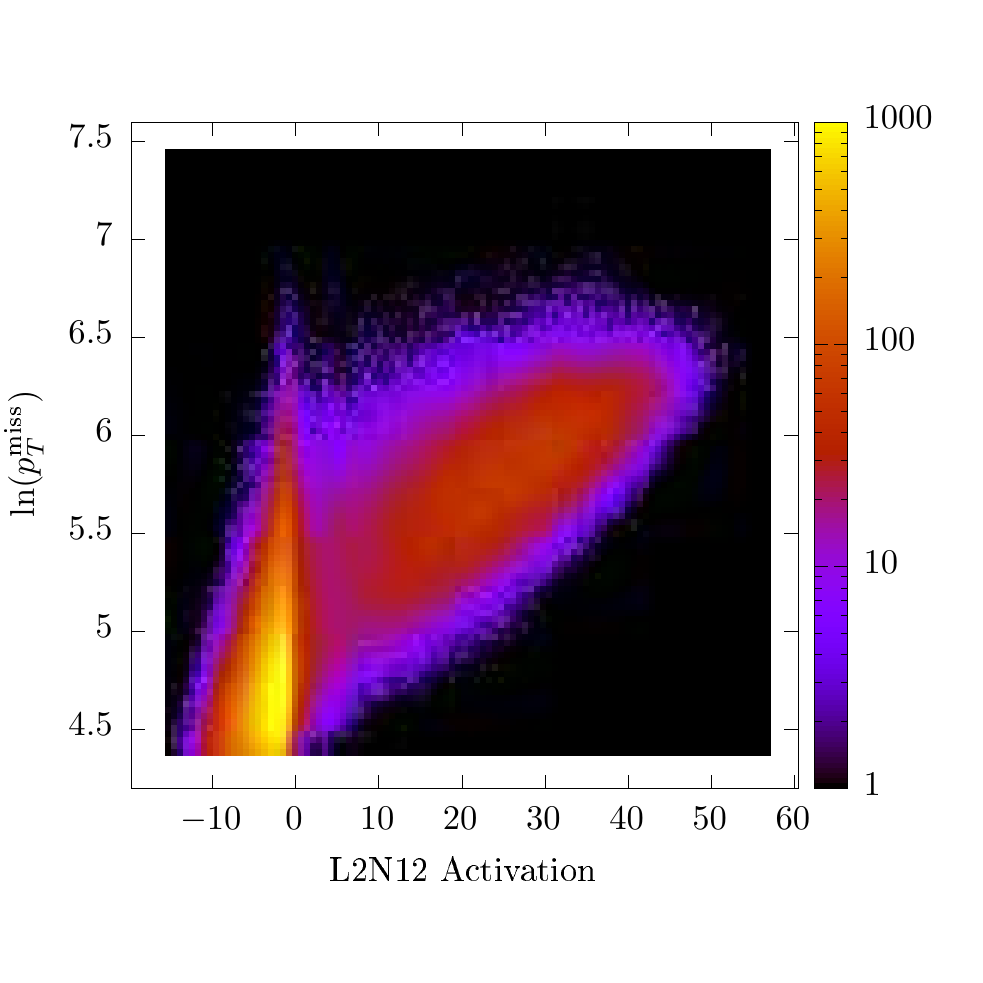}} \scalebox{0.5}{\includegraphics{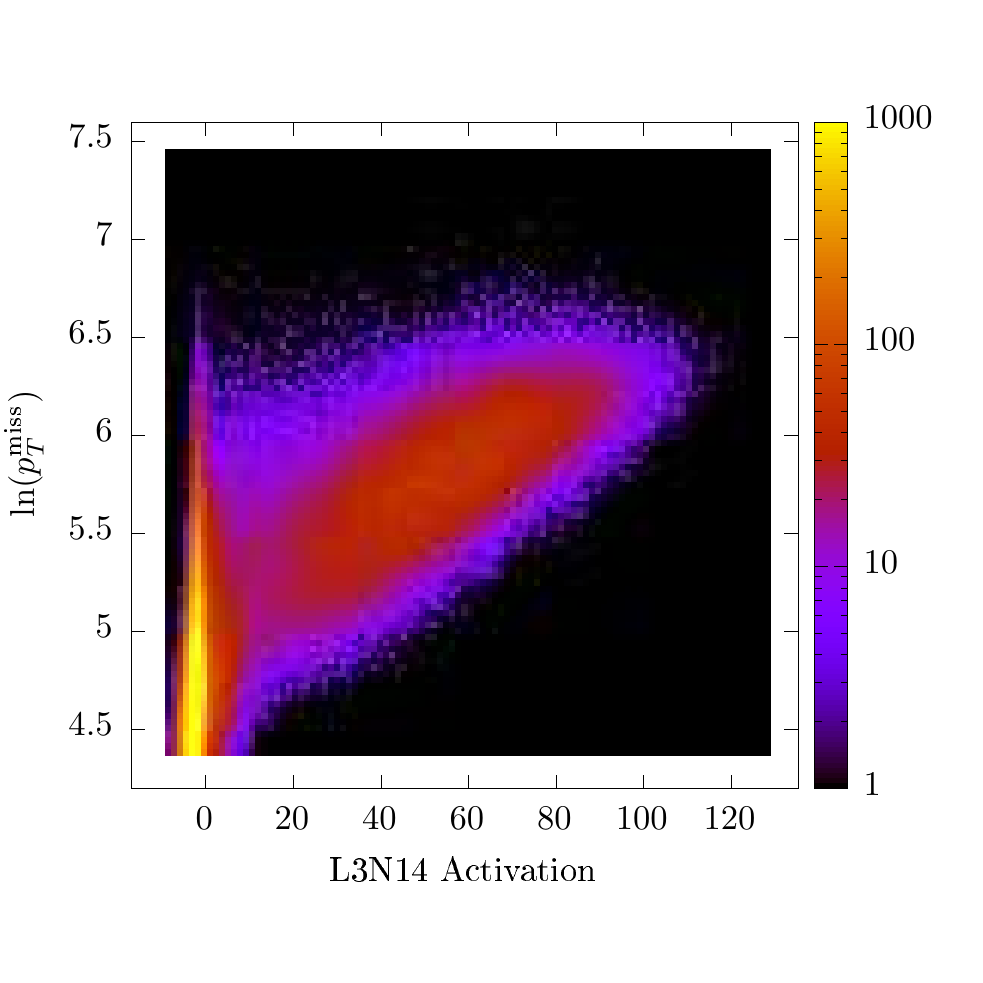}} \scalebox{0.5}{\includegraphics{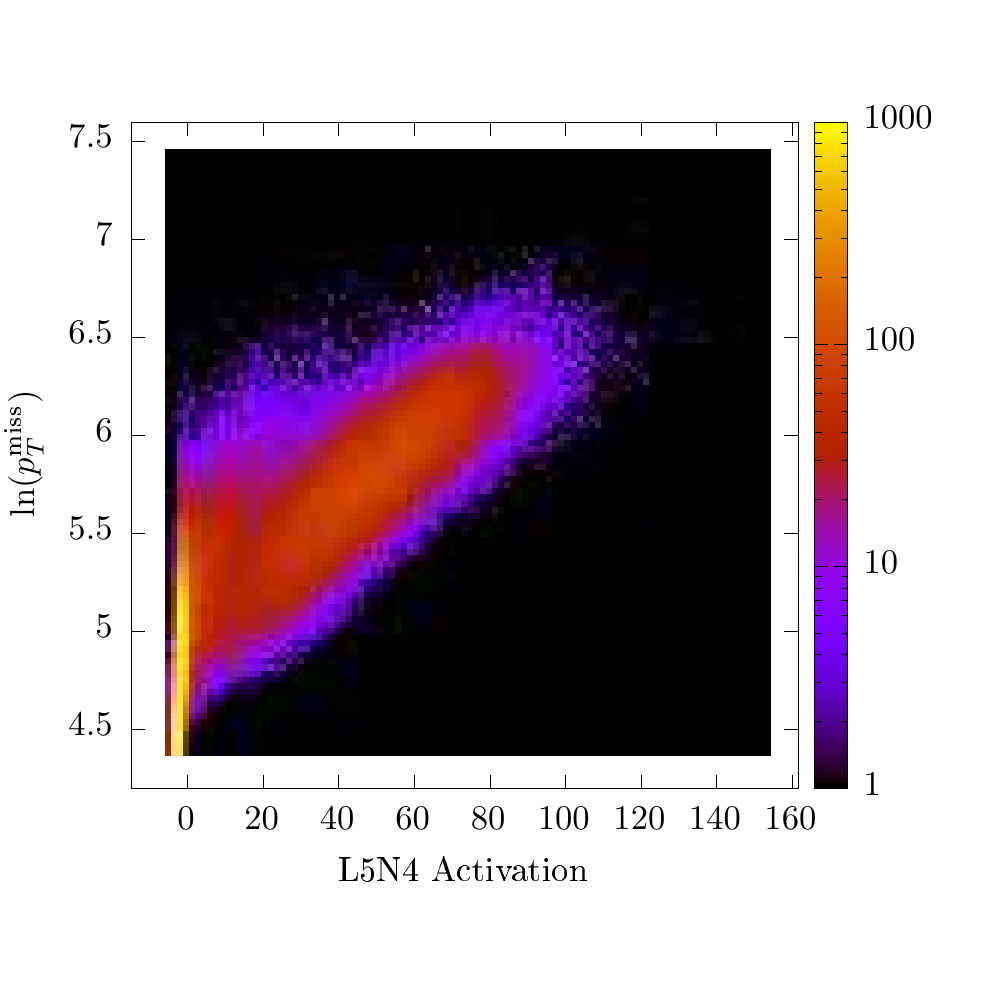}} \\
    \scalebox{0.5}{\includegraphics{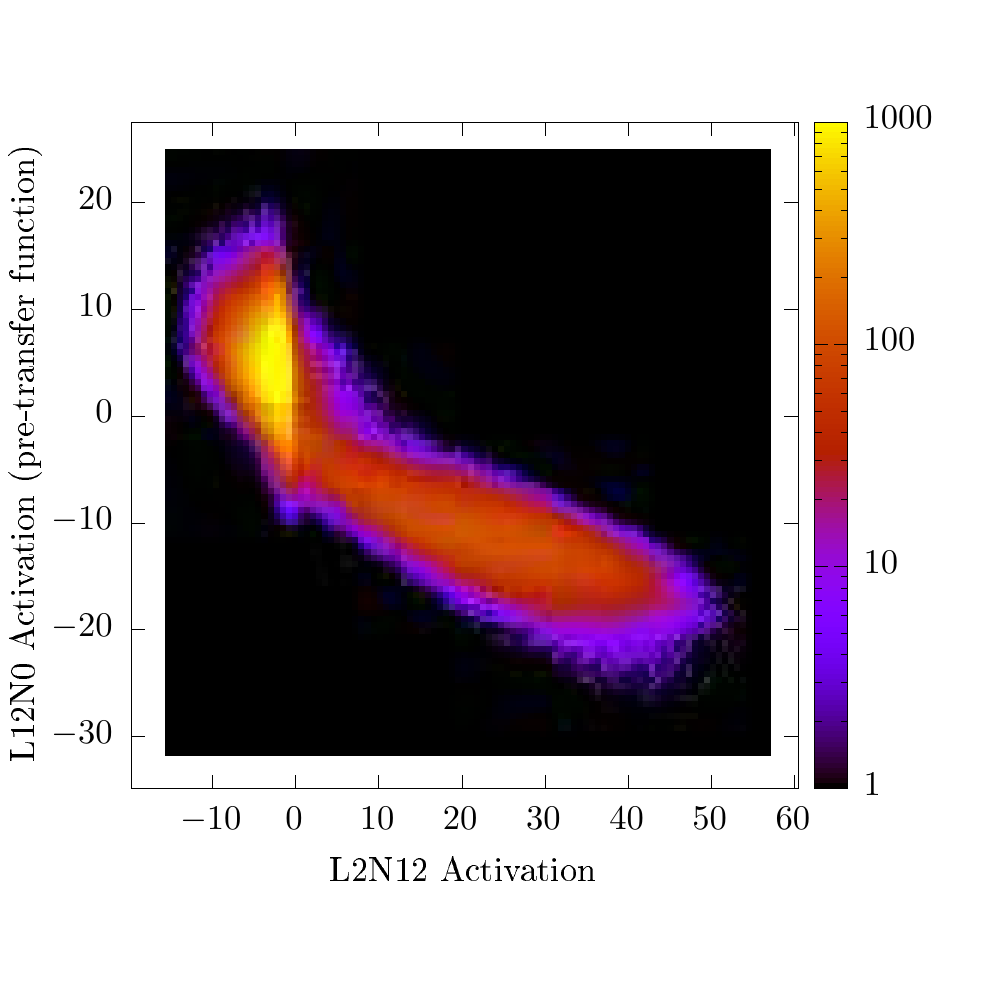}} \scalebox{0.5}{\includegraphics{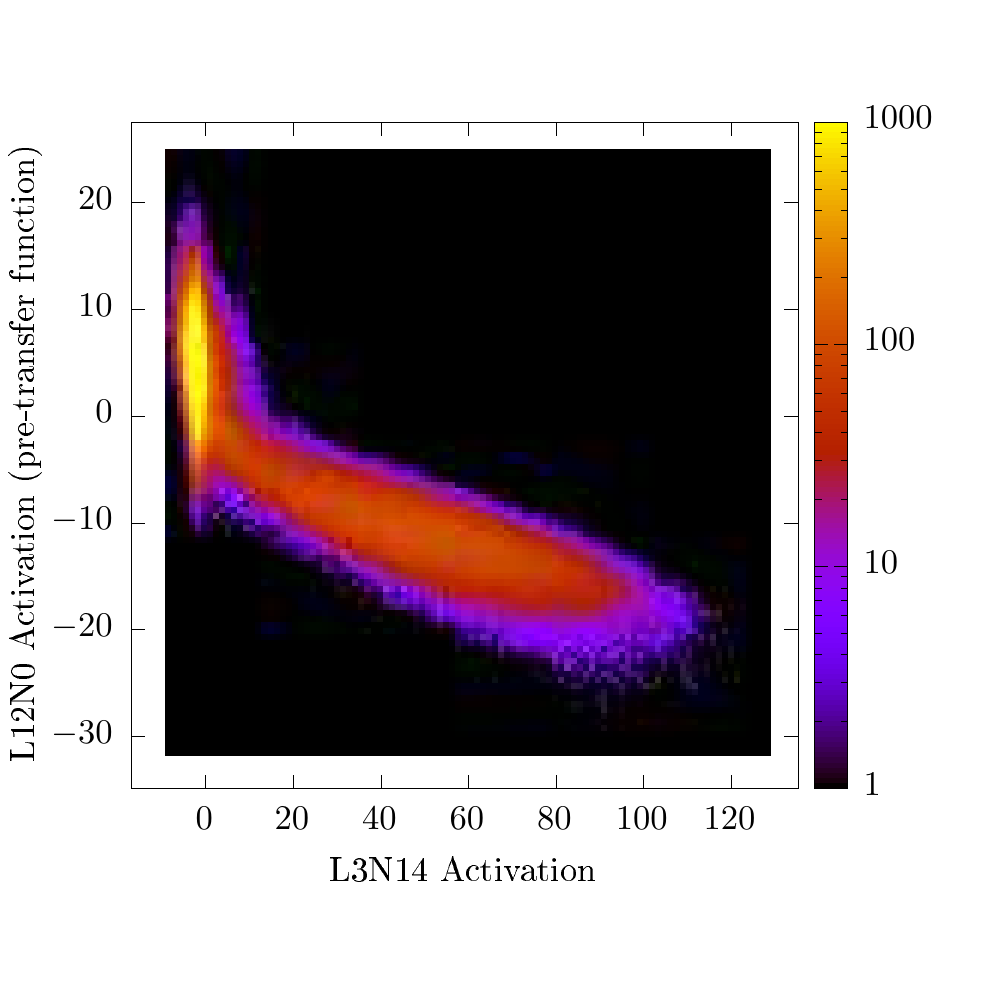}} \scalebox{0.5}{\includegraphics{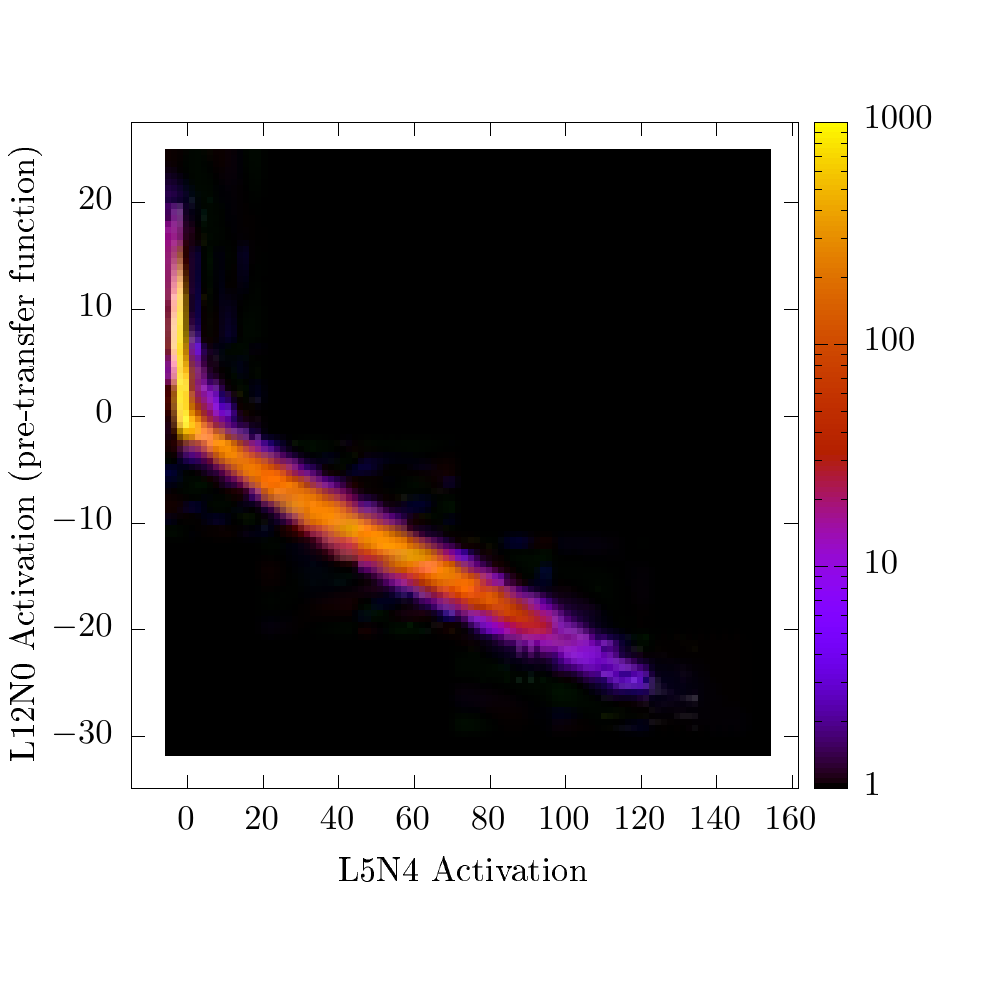}} \\
    \scalebox{0.5}{\includegraphics{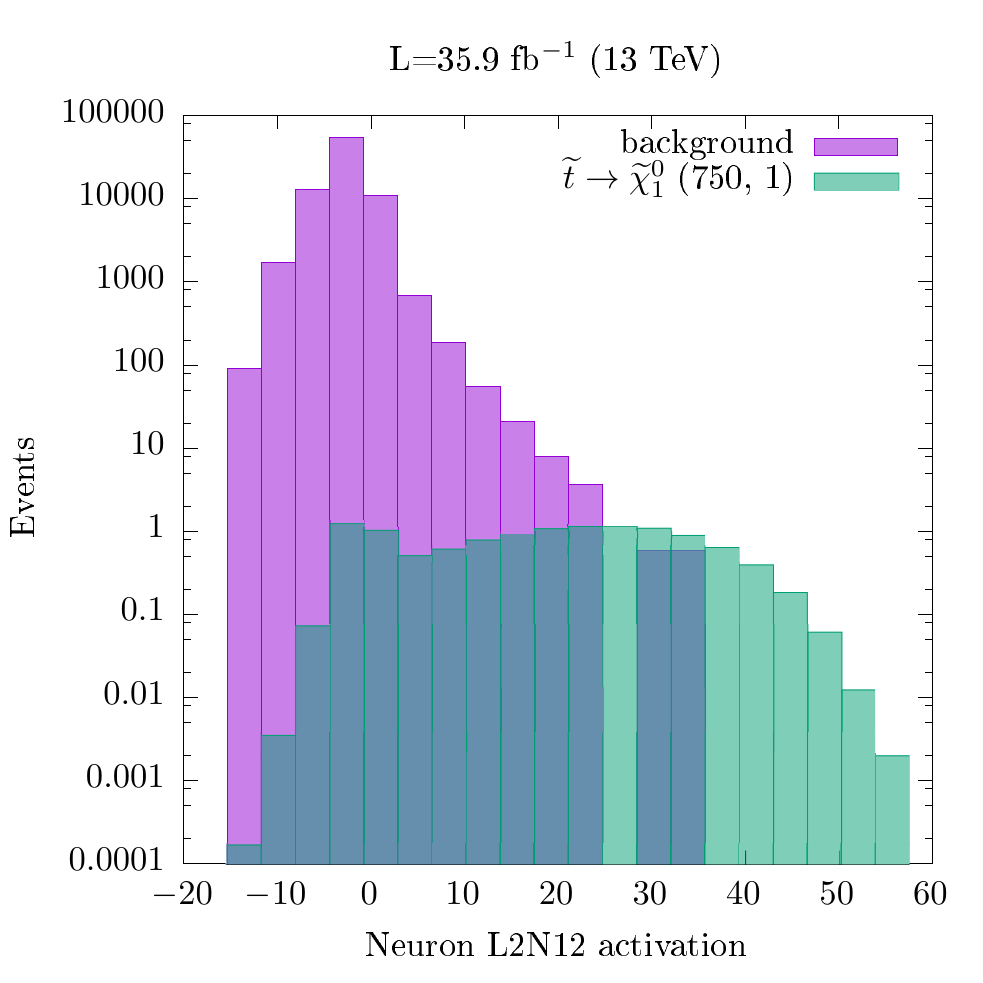}} \scalebox{0.5}{\includegraphics{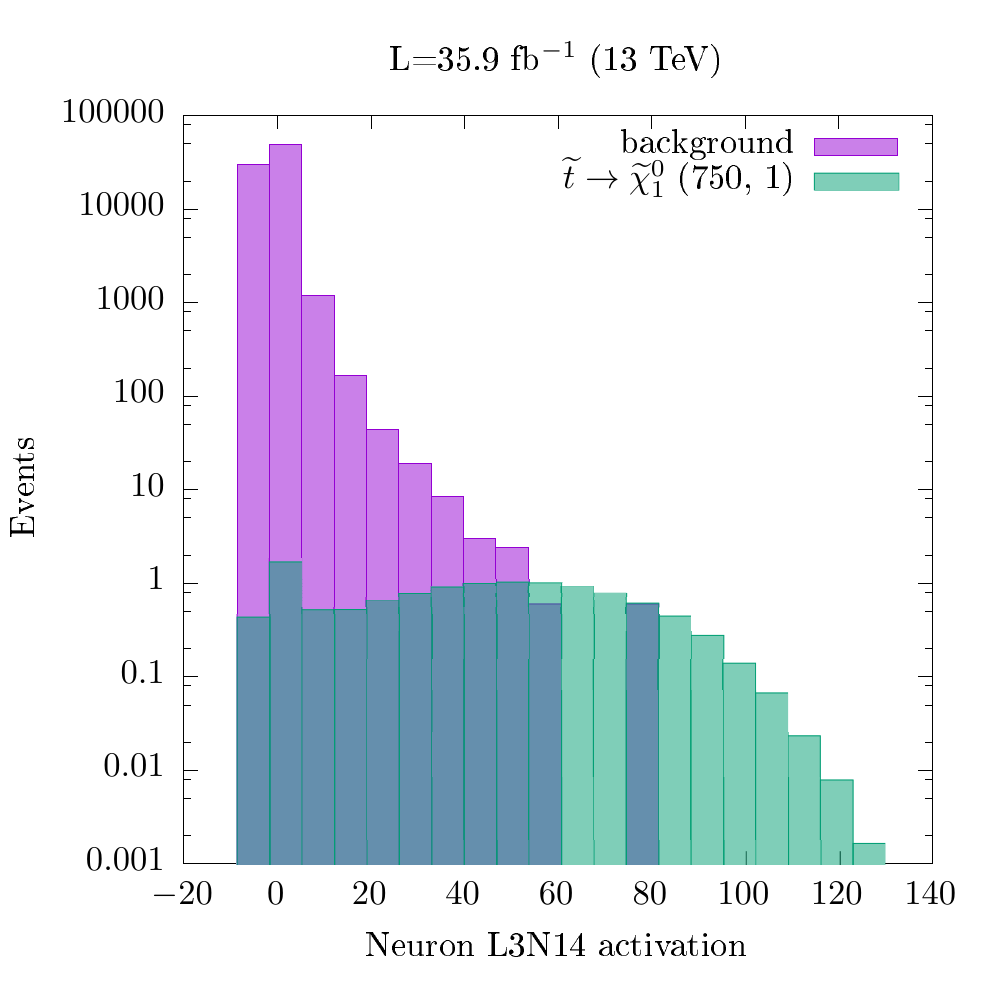}} \scalebox{0.5}{\includegraphics{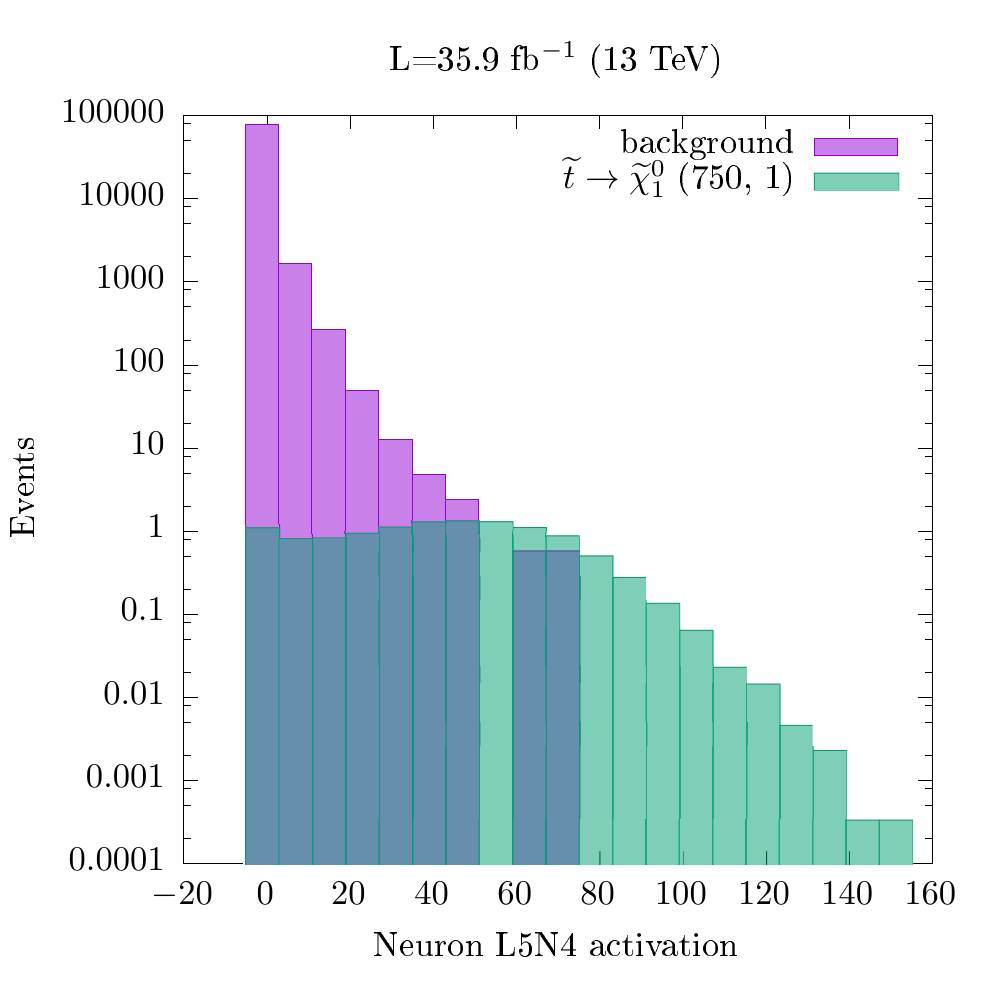}} \\
\end{center}
\caption{Top row: comparison histograms between $m_{\text{T2}}$ and the highest activation-difference neurons in the 2nd, 3th, and 5th hidden layers. The normalized mutual informations, eq.~\eqref{eq:mutualinformation}, associated with these plots are, respectively: 0.31, 0.31, and 0.20. Second row: histograms of the same neurons against $p_T^{\text{miss}}$, for comparison. Normalized mutual informations: 0.13, 0.14, 0.22. Third row: the same neurons against the final neuron of the network,
$L12N0$. Mutual informations: 0.20, 0.22, 0.36. Bottom row: activation histograms showing how well these neurons discriminate signal from background.}
\label{fig:mt2comp}
\end{figure}

Is our network actually computing $m_{\text{T2}}$ or something like it? We compare several neuron activations to the $m_{\text{T2}}$ variable in Figure \ref{fig:mt2comp} and compute the corresponding mutual information. It seems that even as early as the second hidden layer the network has learned to recognize something similar to $m_\text{T2}$; in fact, by the fifth layer the network seems to have progressed to calculating something different (and perhaps more effective) than
$m_\text{T2}$. Moreover, the neurons appear to have positive activations only when $m_\text{T2}$ is above the $W$ mass. This is probably to be expected, since the variable $m_\text{T2}$ more or less lacks discriminative power below the $W$ mass, and thus it makes sense for the non-linearity of the neuron (i.e.~zero) to lie at this critical point. Neurons with leaky ReLU transfer functions, such as the ones we are discussing, are in some sense a strange hybrid of a binary and a continuous variable.
In this case, we can say that the neuron only appears to ``turn on'' when $m_\text{T2}$ does.

\subsection{$L2N12$}

Let's try to open the black box of neuron $L2N12$, which seems to be well-correlated with $m_{\text{T2}}$. The five most important inputs, ranked by activation difference (as defined in \S\ref{subsec:activdiff}), are given in Table \ref{tab:750-1_L2N12_actDiff}. We see that, as one would expect of a variable correlated with $m_{\text{T2}}^{\ell\ell}$, the most important variables are the missing energy and the kinematics of the two leptons (for reference, the sixth most important variable is $\phi^{j_1}$, whose activation difference clocks
in at a distant 0.02).

\begin{table}[!h]
\begin{center}
\begin{tabular}{|c|c|c|c|c|c|}
    \hline
    Variable & $\phi^{\ell_2}$ & $p_T^{\text{miss}}$ & $\phi^{\ell_1}$ & $p_T^{\ell_2}$ & $p_T^{\ell_1}$ \\
    \hline
    Activation Difference & 0.41 & 0.35 & 0.27 & 0.17 & 0.11 \\
    \hline
\end{tabular}
\end{center}
\caption{Activation differences of the 5 most important input variables for neuron $L2N12$ of a network trained on $(m_{\widetilde t}, m_{\widetilde \chi}) = (750, 1)$ data.}
\label{tab:750-1_L2N12_actDiff}
\end{table}

\begin{figure}[!htbp]
\begin{center}
\scalebox{0.55}{\includegraphics{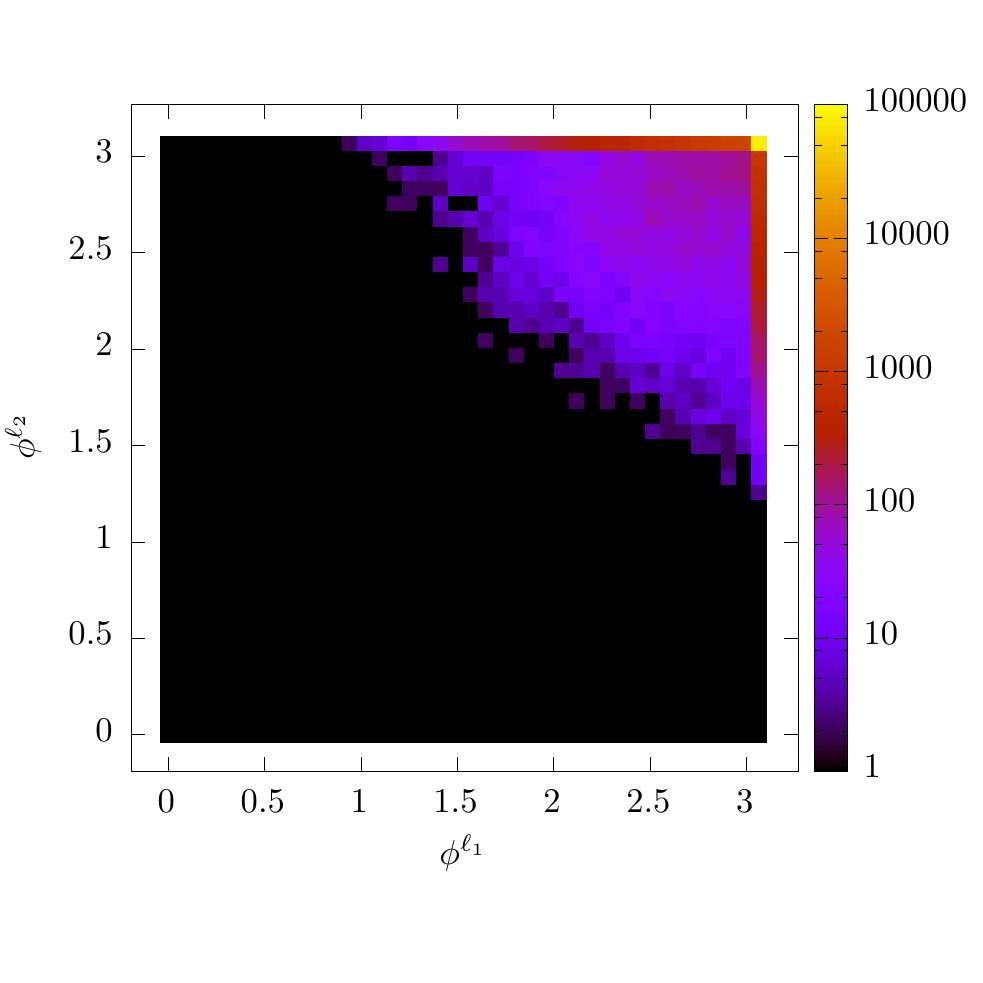}}\scalebox{0.55}{\includegraphics{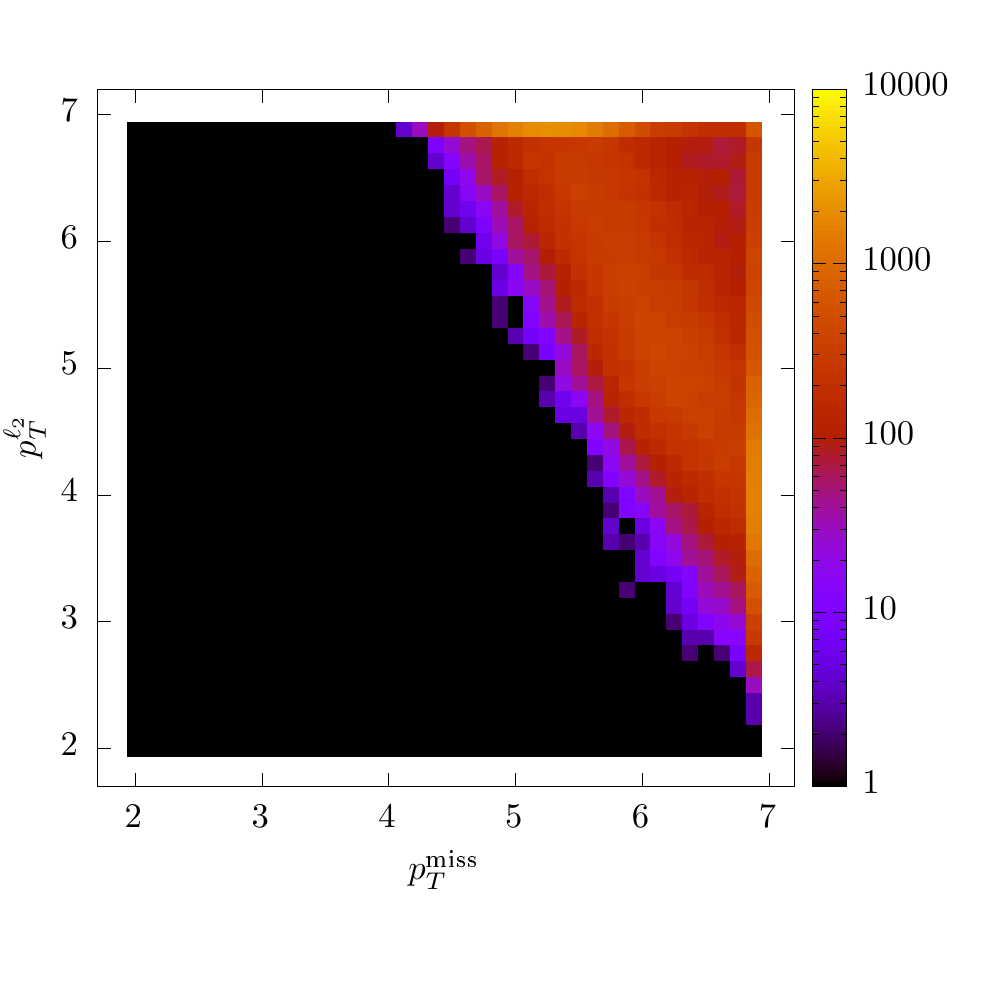}}\scalebox{0.55}{\includegraphics{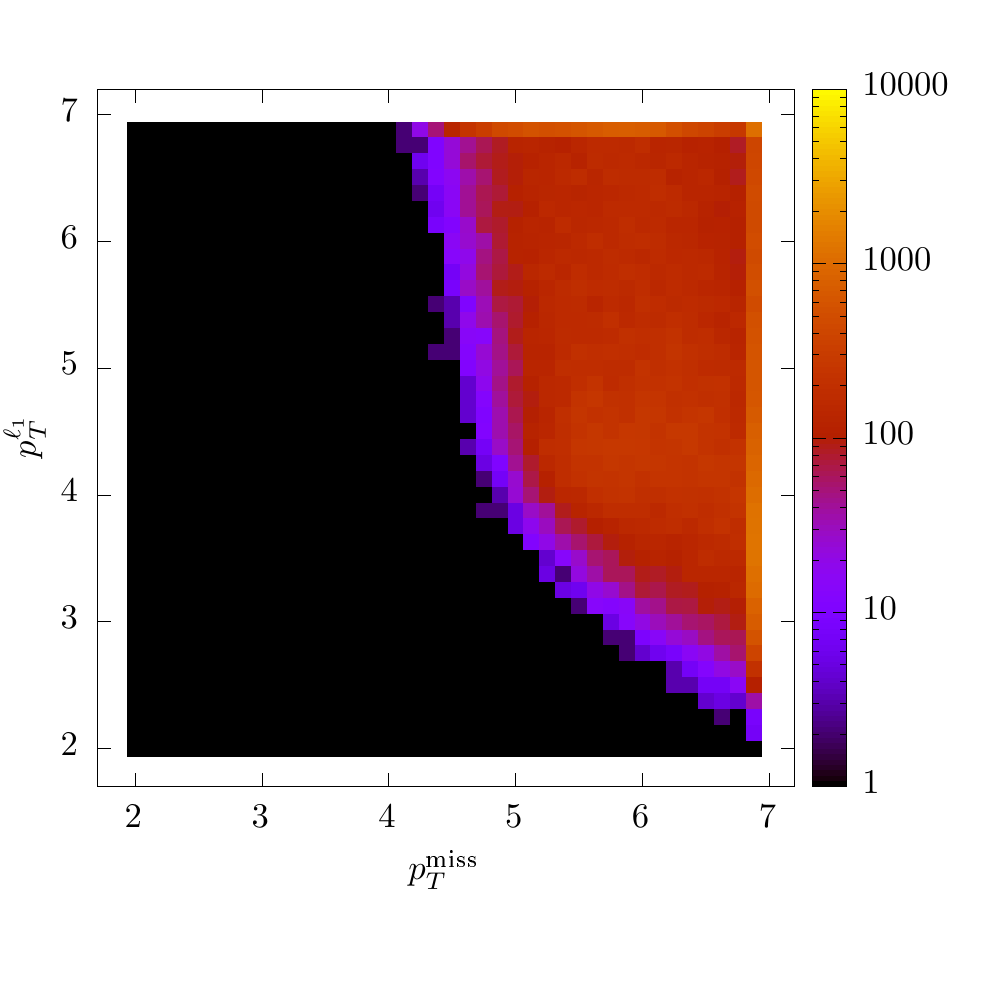}}
\end{center}
\caption{Activation maximization histograms for neuron $L2N12$ trained on $(m_{\widetilde t}, m_{\widetilde \chi}) = (750, 1)$ data, using an activation threshold of 50.}
\label{fig:750-1_L2N12_actMax}
\end{figure}

By looking at the activation maximization histograms of the most important inputs (Figure \ref{fig:750-1_L2N12_actMax}), we can see that this neuron is actually quite simple: it is looking for events with high missing energy and high-$p_T$ leptons where the leptons are rotated away from the missing energy in the azimuthal direction. This is evident from the position of the filled bins in each histogram: they are clustered in the upper right of each graph, indicating that all
high-activation events must have large values of each of these variables. Since the clusters border the edges of the histograms, which represent the edges of the input space allowed by the maximization procedure, we interpret them as a general desire of the neuron for higher input values, rather than a preference for any specific value.

\begin{figure}[!htbp]
\begin{center}
\scalebox{0.55}{\includegraphics{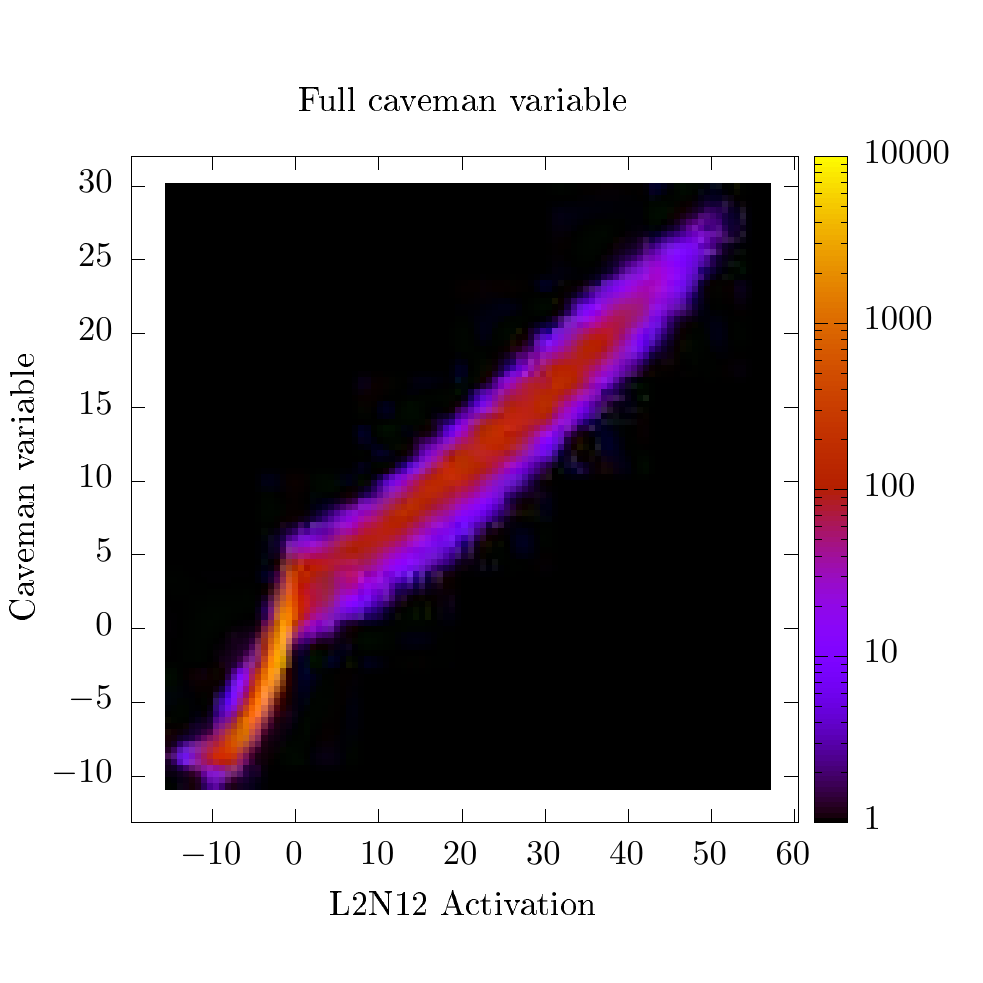}}\scalebox{0.55}{\includegraphics{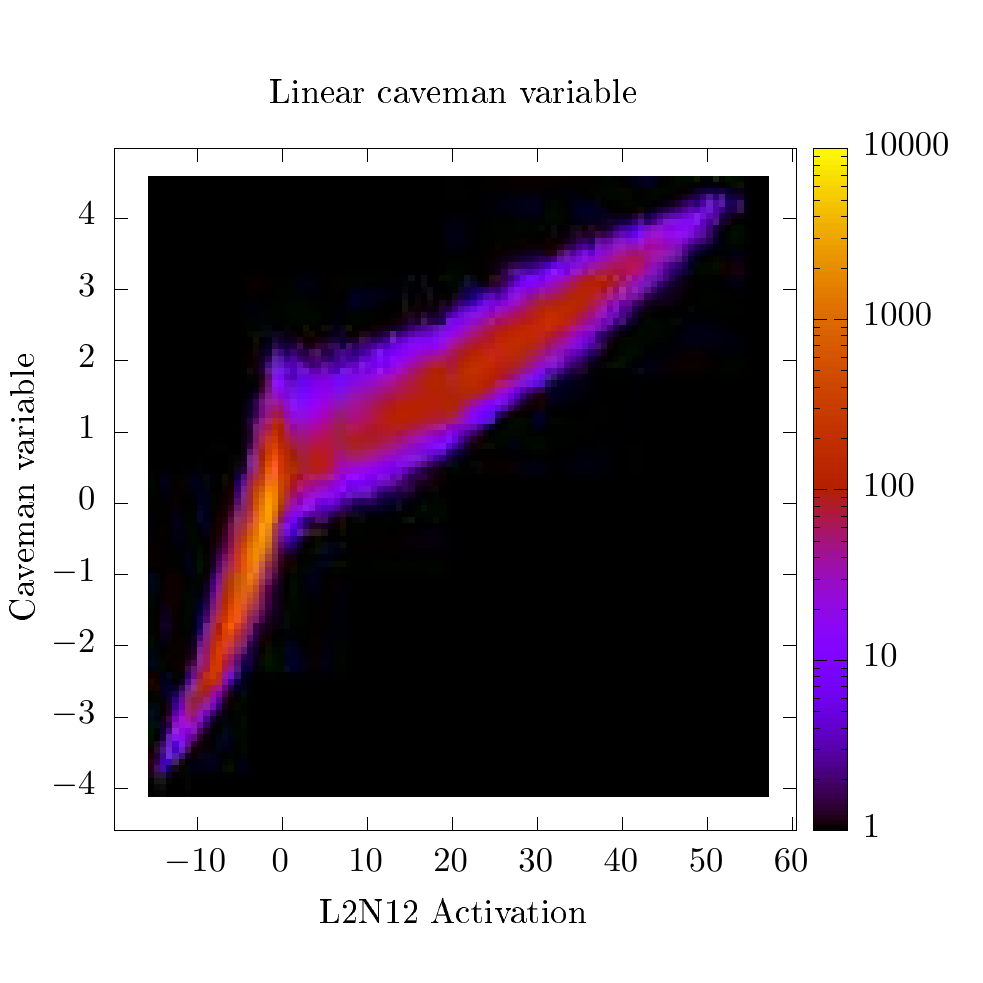}}
\scalebox{0.55}{\includegraphics{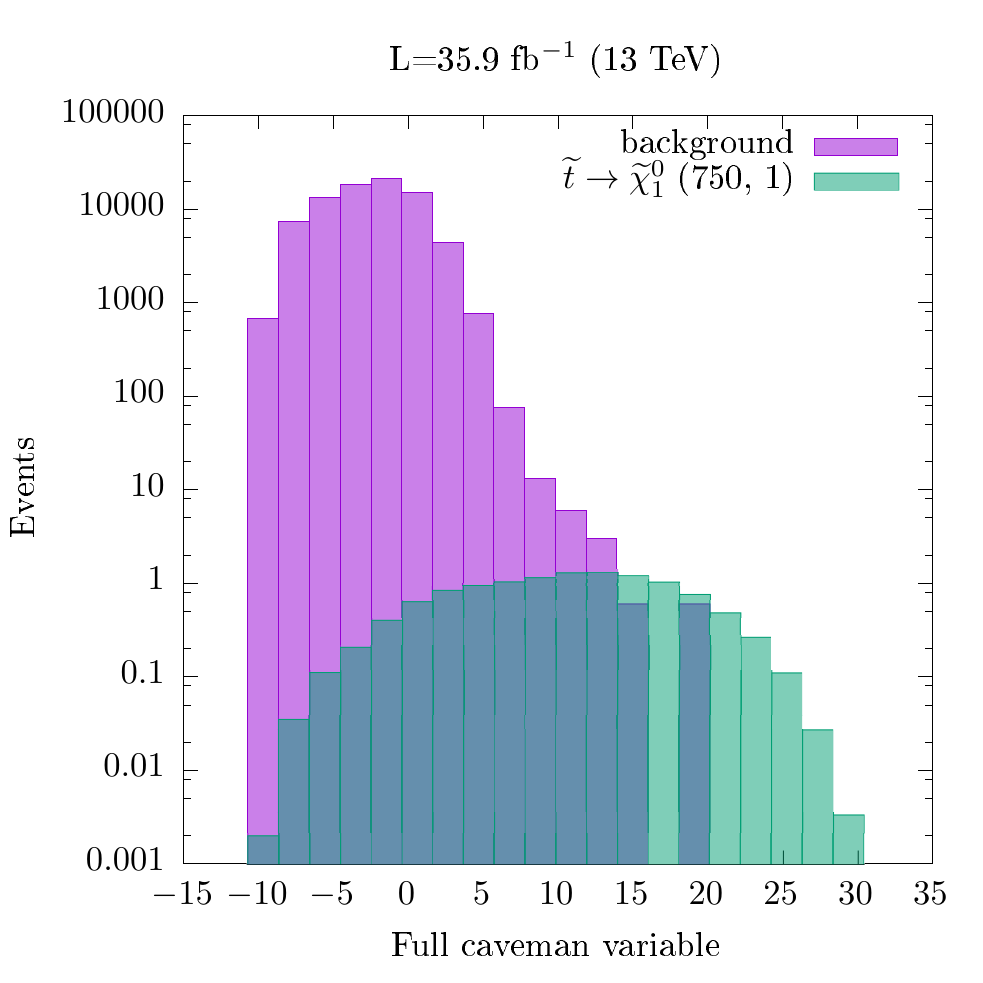}}\scalebox{0.55}{\includegraphics{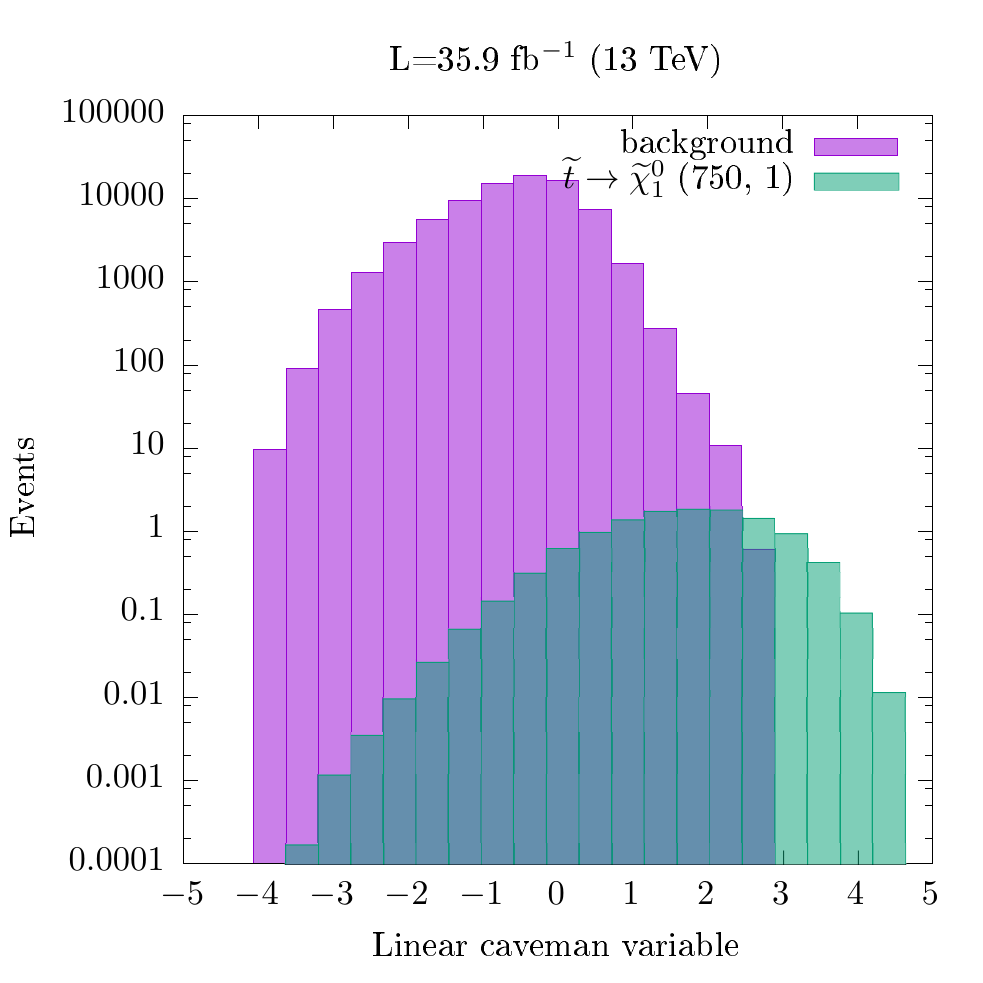}}
\end{center}
\caption{Top: comparison of the full caveman variable created via the procedure in Section \ref{sec:caveman} (left) and a simple linear combination of inputs (right) with the neuron $L2N12$. Bottom: discrimination power of these caveman variables.}
\label{fig:750-1_L2N12_cave}
\end{figure}

As corroboration of this interpretation, we can construct a caveman variable following the procedure explained in Section \ref{sec:caveman} and Appendix \ref{app:caveman}. Even restricting the variable to only the 5 inputs in Table \ref{tab:750-1_L2N12_actDiff}, we find that it reasonably reproduces the behavior of the neuron (Figure \ref{fig:750-1_L2N12_cave}), indicating that the neuron is well-described by its highest activations. Indeed, we see that even a caveman variable which is a simple linear combination of these inputs (i.e.~omitting the 2- and 3-variable correlations) is a fairly good description of the neuron, as suggested by the simplicity of the activation histograms.

\clearpage
\subsection{$L5N4$}

\begin{figure}[!htbp]
\begin{center}
\scalebox{0.65}{\includegraphics{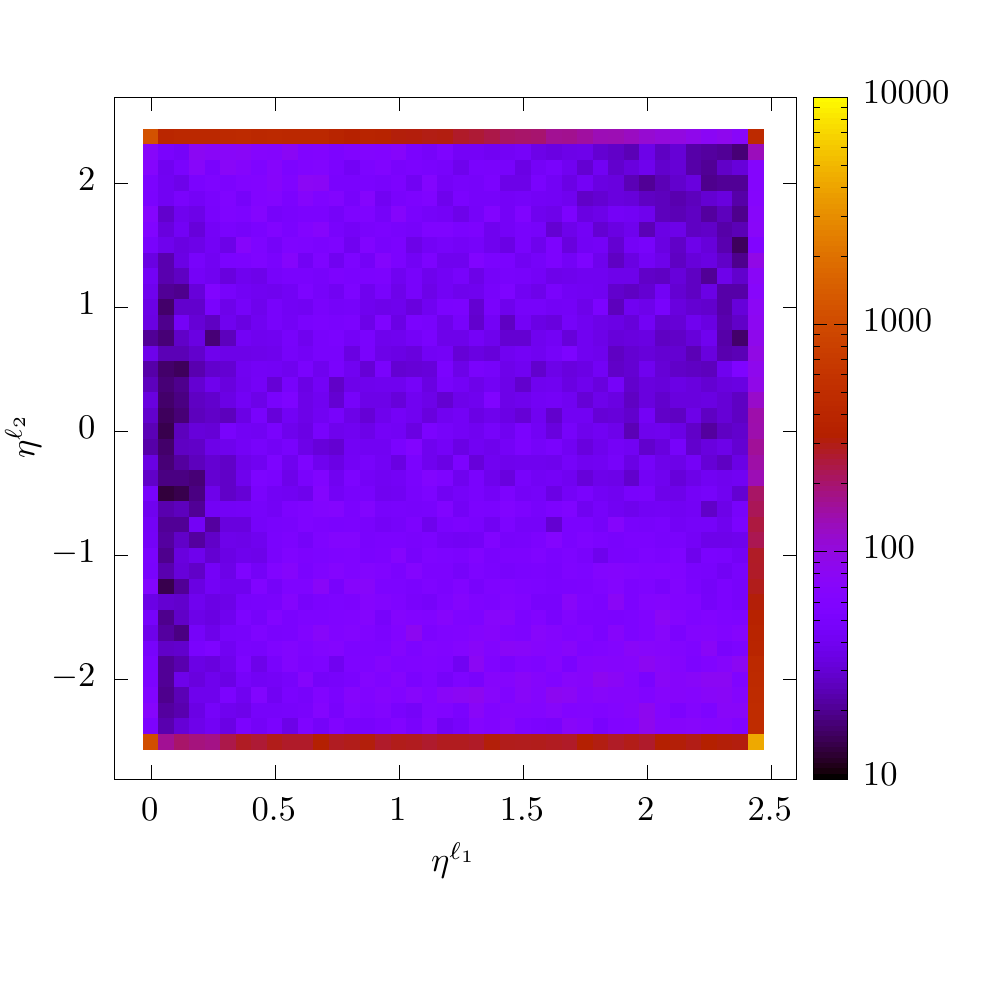}}
\end{center}
\caption{Activation maximization histogram for neuron $L5N4$ trained on $(m_{\widetilde t}, m_{\widetilde \chi}) = (750, 1)$ data. The network prefers both leptons to be at high absolute pseudorapidity, and seems to have a mild preference for events where their $\eta$s have opposite signs, i.e.~for events with high mass rather than just a large overall boost along the beamline.}
\label{fig:750-1_L5N4_eta}
\end{figure}

If we move to examining the dominant neuron in the fifth layer, $L5N4$, we find that it is qualitatively very similar to $L2N12$. At least, it is still dominated by the same five variables with the same basic behavior. However, the relative importance of the inputs has shifted slightly, with $p_T^{\text{miss}}$ clearly taking the top spot (Table \ref{tab:750-1_L5N4_actDiff}). In addition, the subdominant variables have become relatively more important: the 6th most important variable is now
$\eta^{\ell_2}$ at 0.06. This means that the neuron has become more sensitive to, for example, the $\eta$ values of the leptons as well as $p_T$ and $\phi$ (see Figure \ref{fig:750-1_L5N4_eta}).

\begin{table}
\begin{center}
\begin{tabular}{|c|c|c|c|c|c|}
    \hline
    Variable & $p_T^{\text{miss}}$ & $\phi^{\ell_2}$ & $\phi^{\ell_1}$ & $p_T^{\ell_2}$ & $p_T^{\ell_1}$ \\
    \hline
    Activation Difference & 0.57 & 0.18 & 0.12 & 0.10 & 0.10 \\
    \hline
\end{tabular}
\end{center}
\caption{Activation differences of the 5 most important input variables for neuron $L5N4$ of a network trained on $(m_{\widetilde t}, m_{\widetilde \chi}) = (750, 1)$ data.}
\label{tab:750-1_L5N4_actDiff}
\end{table}

\section{Understanding the lepton and missing $p_T$ anti-correlations in $\phi$}
\label{sec:lepmetphi}
\afterpage{\clearpage}

In the previous section, we saw that neurons like $L2N12$ computed a quantity similar to $m_{\text{T2}}^{\ell\ell}$, a variable that is quite sophisticated, involving a minimization over all possible partitions of missing momentum into two particles. However, it is also apparent from Figure \ref{fig:750-1_L2N12_actMax} and from the success of the caveman variable construction shown in Figure \ref{fig:750-1_L2N12_cave} that it may be possible to capture similar information in a
relatively straightforward way. In particular, it appears that selecting events with large $p_T^{\text{miss}}$ and with both leptons pointing away from the missing momentum in azimuthal angle $\phi$ captures much of the same information. It is worth taking a closer look at this phenomenon, as we will see it re-appear when we study 3-body stop decays in \S\ref{sec:threebody}.

\begin{figure}[!h]
\begin{center}
    \includegraphics[width=1.0\textwidth]{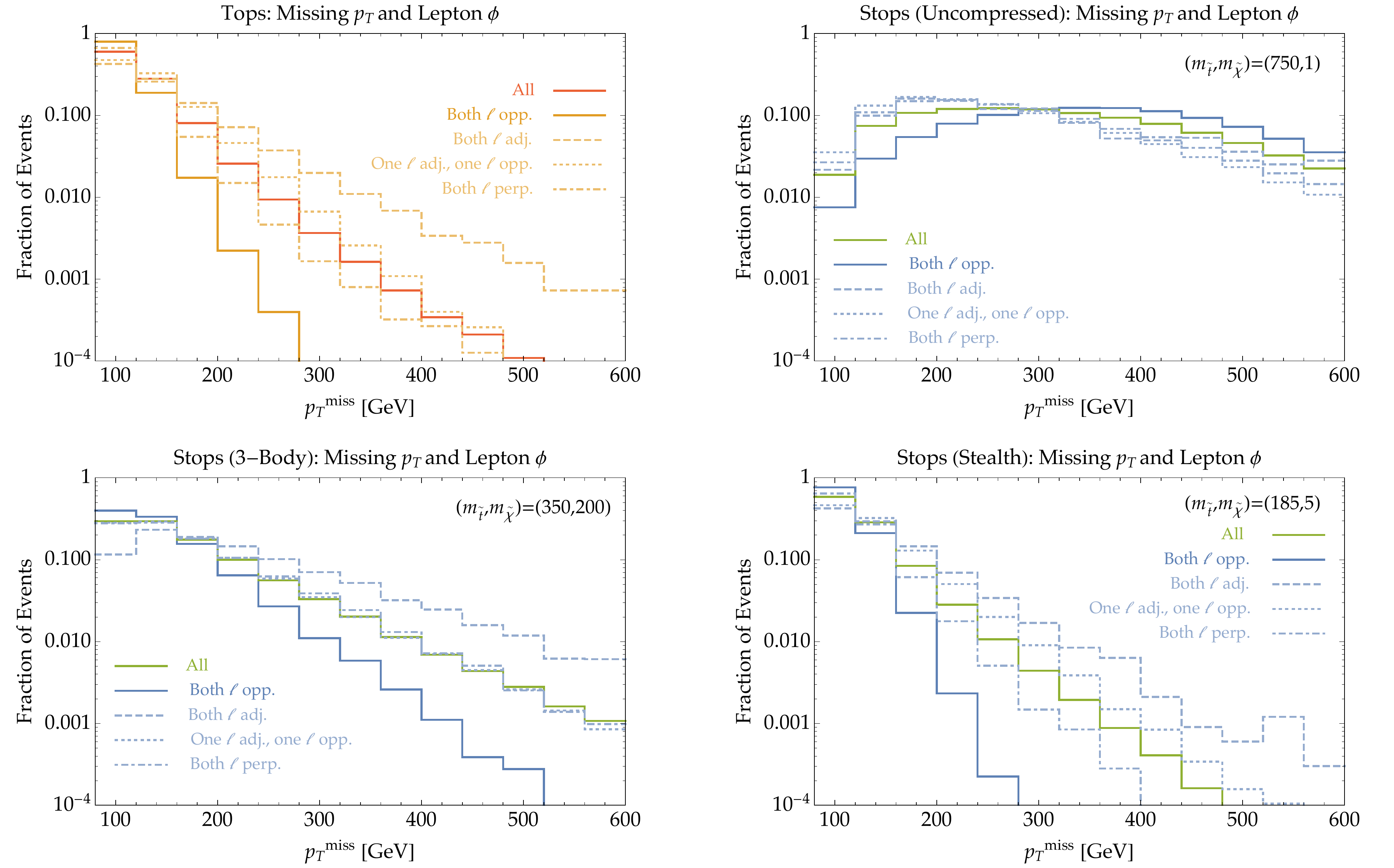}
\end{center}
\caption{Effect of cuts on angles $\Delta \phi(\ell, p_T^{\text{miss}})$ between the leptons and missing momentum on the $p_T^{\text{miss}}$ distribution for $t\overline{t}$ events (left) and for ${\widetilde t}{\widetilde t}^*$ events (other three panels) in simulated data. Clockwise from top right, the three stop parameter points are $(m_{\widetilde{t}}, m_{\widetilde{\chi}}) = (750,1)$ GeV; $(185,5)$ GeV; and $(350,200)$ GeV. Here ``$\ell$ opp.'' means $\cos(\Delta \phi) < -\frac{1}{2}$, a lepton {\em opposite} to missing $p_T$; ``$\ell$ adj.'' means $\cos(\Delta \phi) > \frac{1}{2}$, a lepton {\em adjacent} to missing $p_T$; and ``$\ell$ perp.'' means $-\frac{1}{2} < \cos(\Delta \phi) < \frac{1}{2}$, a lepton {\em perpendicular} to missing $p_T$. We see that the missing momentum in the background is much more sensitive to angular cuts than in the non-stealthy signal, and in particular requiring both leptons opposite the missing momentum strongly suppresses the missing $p_T$. The stealth signal is kinematically extremely similar to the background.}
\label{fig:METLepPhi}
\end{figure}

In Figure \ref{fig:METLepPhi}, we show the missing $p_T$ distribution on signal and background events after making various selections on the azimuthal angle between the leptons and missing $p_T$. We include three different signal points: one uncompressed, one with 3-body decays, and one stealthy point. We loosely group leptons by angle relative to missing $p_T$, categorizing them as ``opposite'' to missing momentum when $\cos(\Delta \phi(\ell, p_T^{\text{miss}})) < -\frac{1}{2}$; ``adjacent'' to missing momentum when $\cos(\Delta \phi(\ell, p_T^{\text{miss}})) > \frac{1}{2}$; and ``perpendicular'' to missing momentum when $-\frac{1}{2} < \cos(\Delta \phi(\ell, p_T^{\text{miss}})) < \frac{1}{2}$. We see from the figure that in the background, 3-body signal, and stealth signal points, asking for both leptons opposite to missing momentum significantly reduces the typical $p_T^{\text{miss}}$, while asking for both leptons adjacent to missing momentum significantly increases $p_T^{\text{miss}}$. Other selections, like both leptons perpendicular to the missing momentum or one lepton adjacent and the other opposite, have relatively little effect on the $p_T^{\text{miss}}$ distribution. Effects are muted in the $(750,1)$ uncompressed signal point studied above, which has very large $p_T^{\text{miss}}$ regardless of the angular cut imposed.

In the top background events and in the kinematically very similar stealth stop signal point, requiring two leptons opposite missing $p_T$ {\em strongly} suppresses the missing momentum, while the effect is more moderate for the 3-body signal point. In fact, this selection is related to the variable $m_{T2}^{\ell\ell}$ defined in eq.~\eqref{eq:mT2ll}: in events with both leptons opposite missing $p_T$, no matter how we partition the missing momentum into two hypothetical neutrinos, at least one of them will be energetic and pointing away from the leptons so that $1 - \cos(\Delta \phi)$ is not too small, producing a significant transverse mass.

From activation maximization images like Fig.~\ref{fig:750-1_L2N12_actMax}, we have extracted an interesting physics insight from the neural network: much of the power of the technically demanding $m_{T2}^{\ell\ell}$ cut can be obtained by considering events at large $p_T^{\text{miss}}$ with the constraint that leptons are azimuthally well-separated from missing momentum. This insight will again shed light on the neural network's behavior in the study of the 3-body point below.

\section{A more challenging case: three body decays}
\label{sec:threebody}


We now turn to the region where the decay is a three-body process, ${\widetilde t} \to W^+ b {\widetilde \chi}$, specifically the mass pair $(m_{\widetilde t}, m_{\widetilde \chi})$ = $(350, 200)$ GeV. Notice that this parameter point has a compressed phase space, as $m_{\widetilde t} - m_{\widetilde \chi} - m_W - m_b \approx 65~{\rm GeV} \ll m_{\widetilde t}$. Each of the daughter particles has relatively little momentum in the rest frame of the decaying stop, and so we expect that
kinematic variables like $p_T^{\text{miss}}$ are suppressed in the majority of the signal events (relative to the uncompressed two-body decay scenario studied above). Nonetheless, the neural network can separate signal events from background events well enough to produce at least a small subsample dominated by signal.

\begin{figure}[!htbp]
\centering
\scalebox{0.55}{\includegraphics{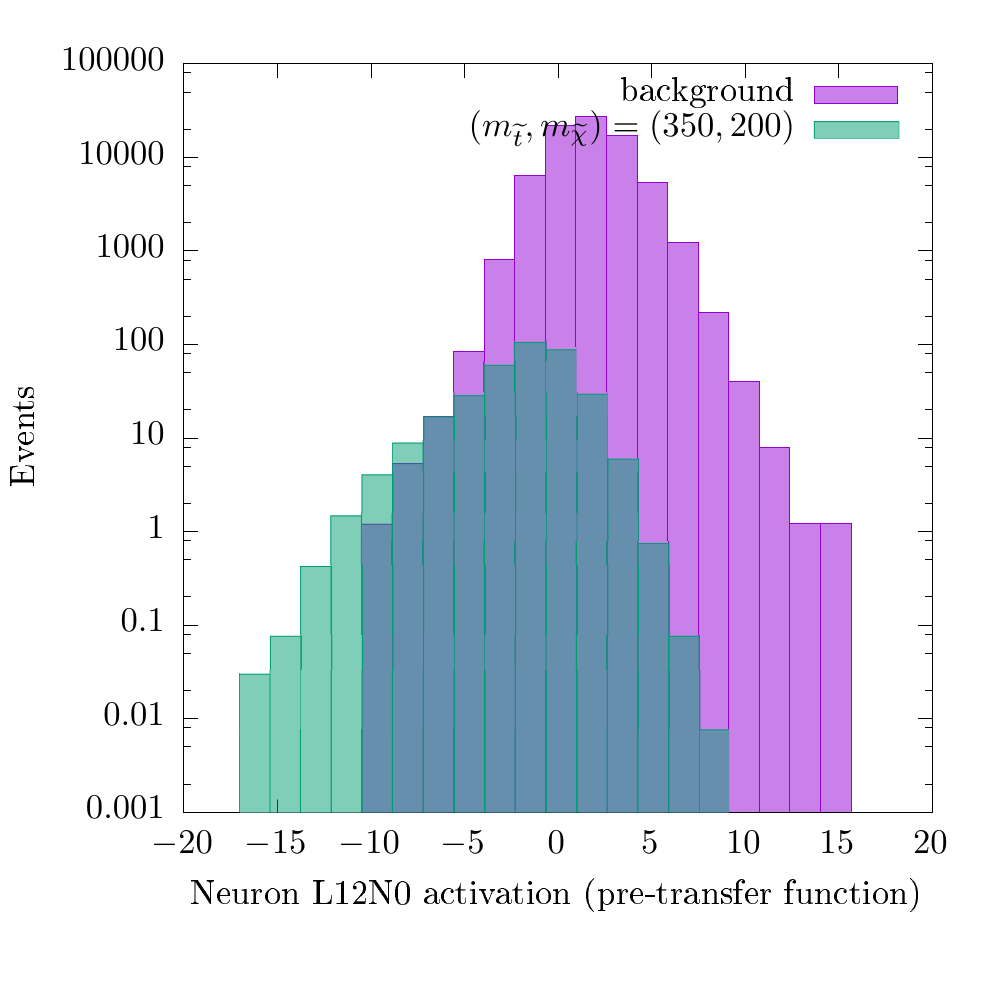}}\scalebox{0.55}{\includegraphics{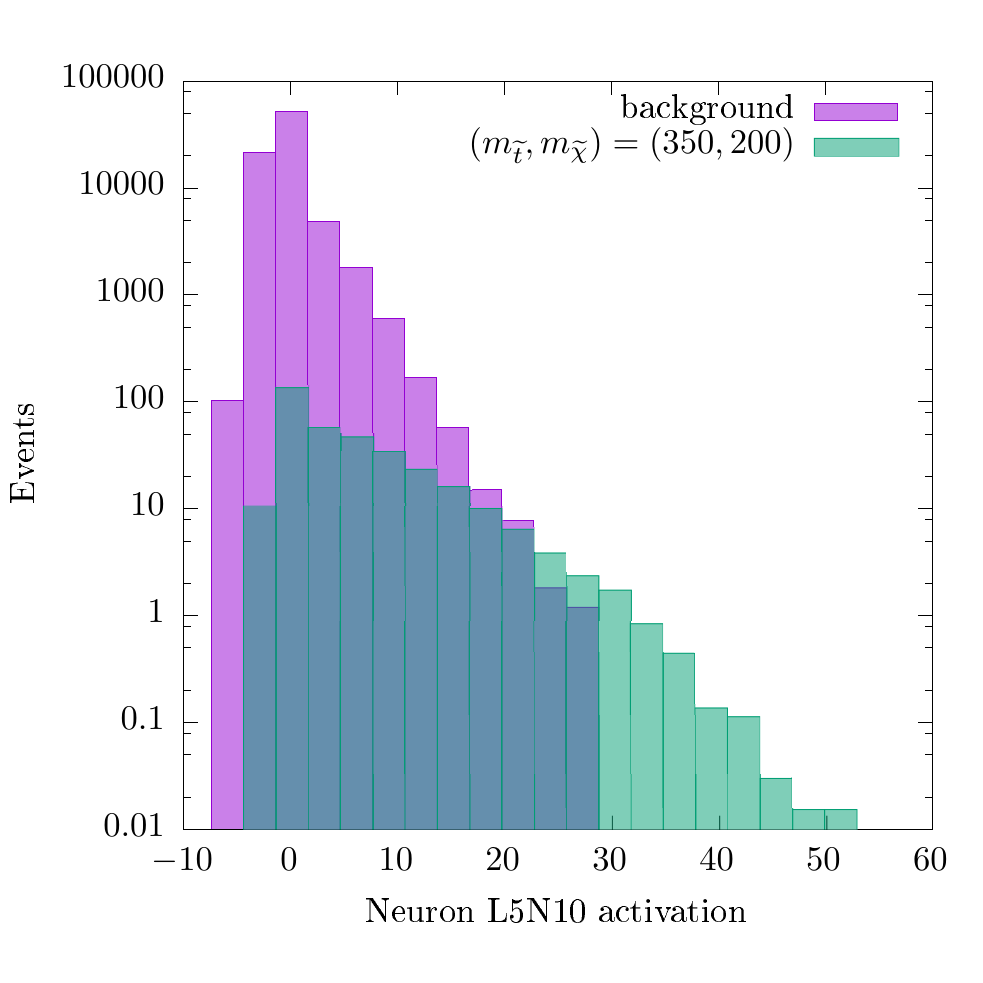}}
\caption{Neuron activations for the final neuron (left) and a neuron in the fifth hidden layer (right). As in the $(750, 1)$ example, the output of the final neuron is plotted before applying its transfer function. We see that the ability of the network to select a signal-rich sample is quite solidified as early as layer 5.}
\label{fig:nhistL5N10}
\end{figure}

As a representative example, we consider the activation patterns of one of the most important neurons in the fifth hidden layer, $L5N10$. The discrimination power of this neuron is shown in Figure \ref{fig:nhistL5N10}; we can see that the performance of even this relatively early neuron is quite good and comparable to that of the entire network. Since the results are qualitatively similar but early neurons tend to be a bit simpler than later ones, we will choose to primarily study the output of
$L5N10$ in lieu of the entire network, though we will also show plots for the final neuron $L12N0$.

\begin{table}[!htbp]
\begin{center}
\begin{tabular}{|c|c|c|c|c|c|}
    \hline
    Variable & $p_T^{\text{miss}}$ & $\phi^{\ell_1}$ & $\phi^{\ell_2}$ & $p_T^{j_1}$ & $p_T^{j_2}$ \\
    \hline
    Activation Difference & 0.50 & 0.30 & 0.27 & 0.24 & 0.23 \\
    \hline
    & $\eta^{j_2}$ & $p_T^{\ell_1}$ & $p_T^{\ell_2}$ & $\eta^{j_1}$ & $\eta^{\ell_2}$\\
    \hline
    & 0.22 & 0.20 & 0.19 & 0.17 & 0.16 \\
    \hline
    & $\phi^{j_1}$ & $\phi^{j_2}$ & $\eta^{\ell_1}$ & $b^{j_2}$ & $\phi_s^{j_2}$ \\
    \hline
    & 0.15 & 0.14 & 0.14 & 0.12 & 0.12 \\
    \hline
    & $\phi_s^{\ell_1}$ & $\phi_s^{\ell_2}$ & $\phi_s^{j_2}$ & $b^{j_1}$ & \\
    \hline
    & 0.11 & 0.08 & 0.07 & 0.06 & \\
    \hline
\end{tabular}
\end{center}
\caption{Activation differences of the input variables for neuron $L5N10$ of a network trained on $(m_{\widetilde t}, m_{\widetilde \chi}) = (350, 200)$ data. $b^{j_i}$ refers to the $b$-tag flag of the $i$th jet; $\phi_s^{i}$ refers to the sign of the $i$th $\phi$ variable (i.e.~whether it is measured clockwise or counterclockwise relative to the $\phi$ of the missing energy).}
\label{tab:350-200_L5N10_actDiff}
\end{table}

The input activation differences for this neuron are shown in Table \ref{tab:350-200_L5N10_actDiff}. We see that, unlike in the (750, 1) case, most or all of the inputs are used, not only the kinematics of the leptons.

\begin{figure}[!htbp]
\begin{center}
\scalebox{0.54}{\includegraphics{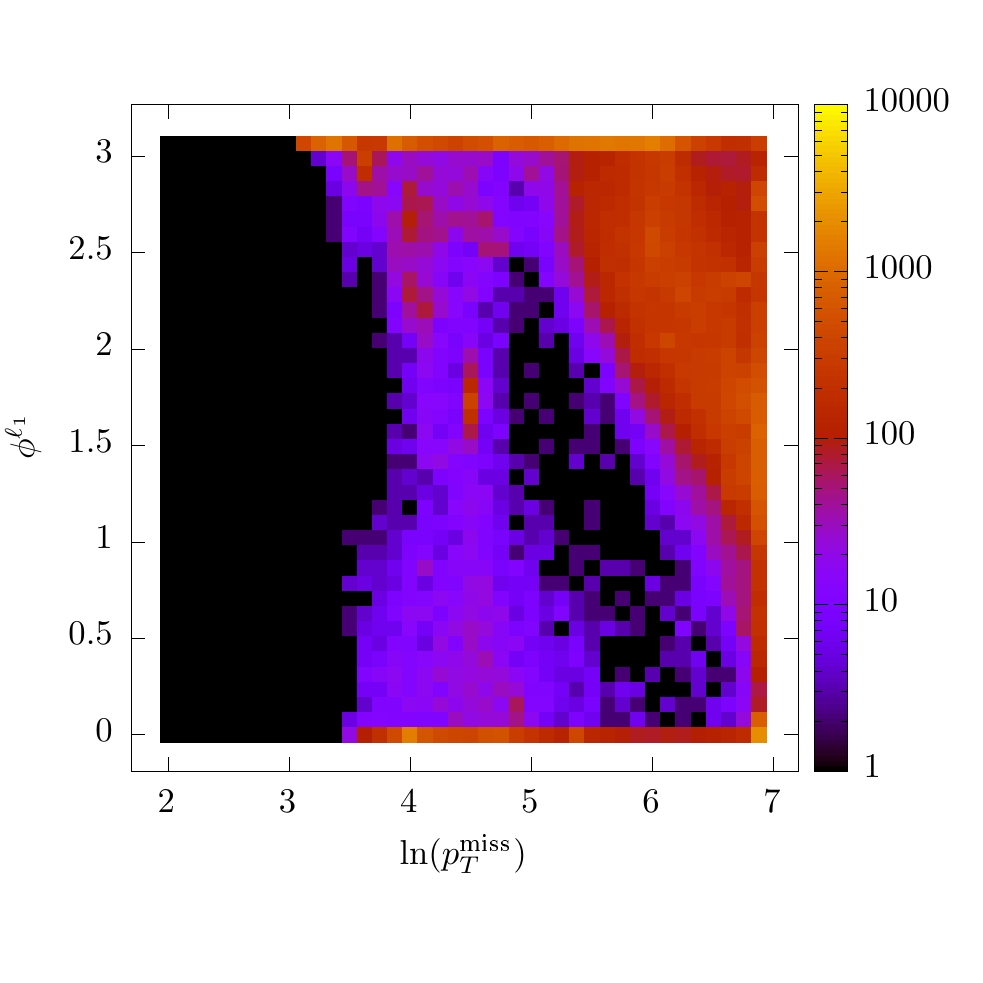}} \scalebox{0.54}{\includegraphics{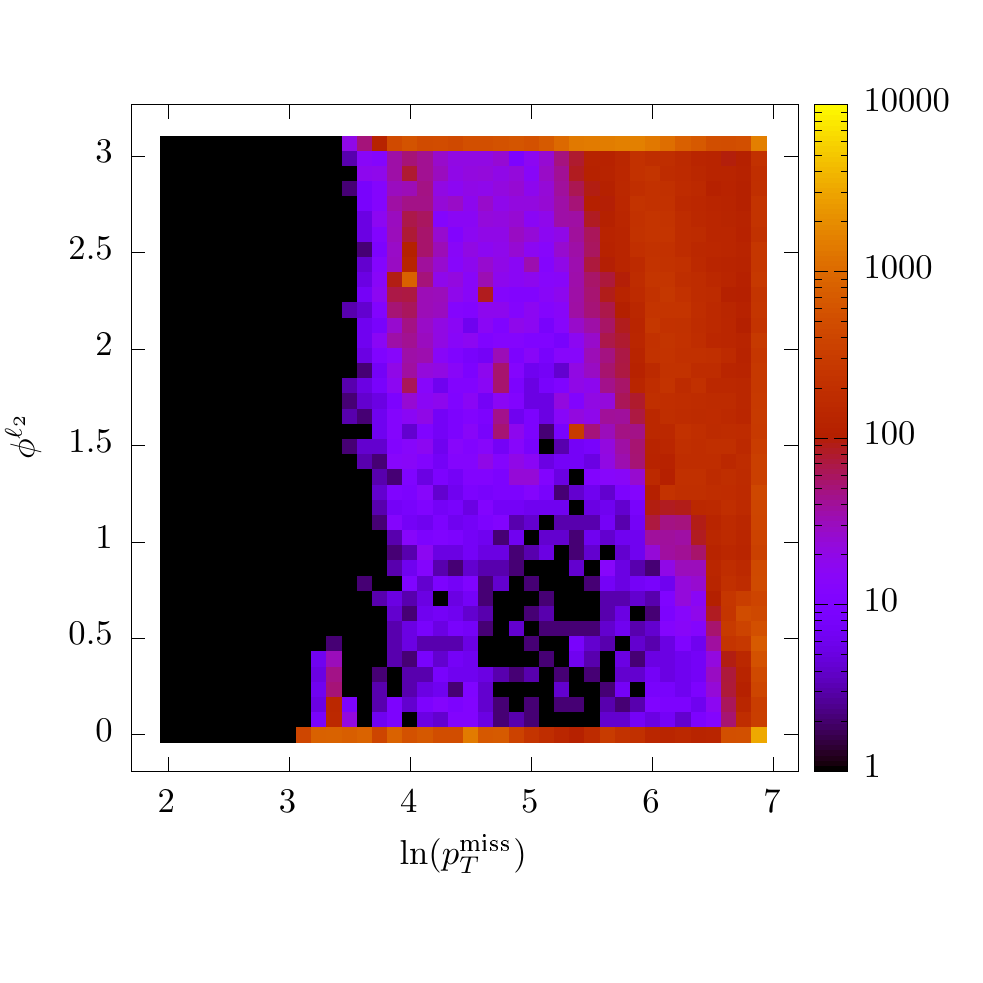}} \scalebox{0.54}{\includegraphics{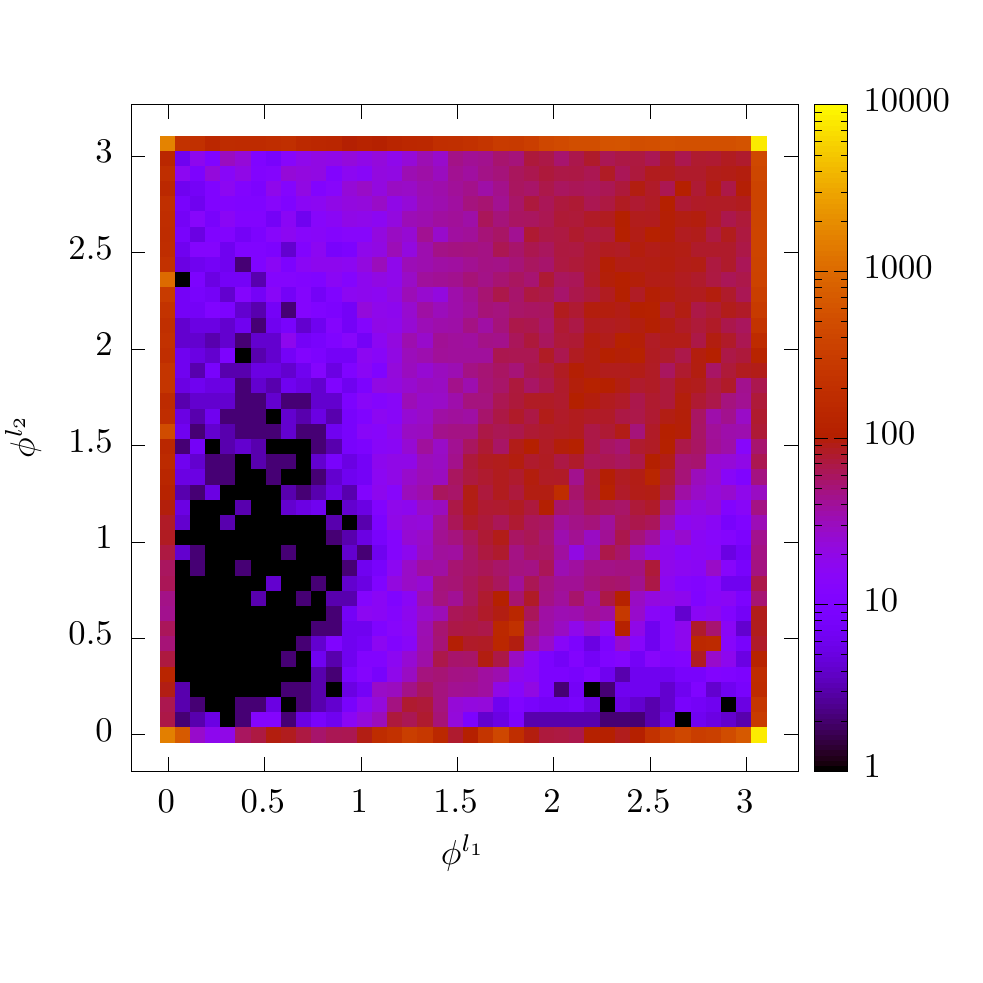}}
\end{center}
\caption{(350, 200) $L5N10$ activation maximization histograms for combinations of the variables $p_T^{\text{miss}}$, $\phi^{\ell_1}$, and $\phi^{\ell_2}$.}
\label{fig:dispL5N10_0-11-13}
\end{figure}

\subsection{Lepton position} \label{sec:simplepattern}

To begin, we can examine the correlations between one of the most important input triplets: $p_T^{\text{miss}}$, $\phi^{\ell_1}$, and $\phi^{\ell_2}$. A quick glance at the activation maximization histograms (Figure \ref{fig:dispL5N10_0-11-13}) reveals that the qualitative pattern is similar to the $(750, 1)$ point: the network likes events with high $p_T^{\text{miss}}$ and leptons rotated away from the missing energy. As discussed in \S\ref{sec:lepmetphi}, such events are suppressed in the top background, tending to have a (forbidden) large $m_{\text{T2}}^{\ell\ell}$. However, the network's preference for high lepton $\phi$s is significantly less pronounced in the 3-body case, particularly at high $p_T^{\text{miss}}$. In fact, from the $\phi^{\ell_1}-\phi^{\ell_2}$ plot in Figure \ref{fig:dispL5N10_0-11-13}, we see that the $\phi$ variables tend to move roughly together as they drift away from the upper right corner (at least for the interior points), and so by comparing to the other two panels we conclude that the accepted region of signal-like events systematically shifts to lower $\phi$ as $p_T^{\text{miss}}$ increases.
\begin{figure}[!htbp]
\begin{center}
\scalebox{0.54}{\includegraphics{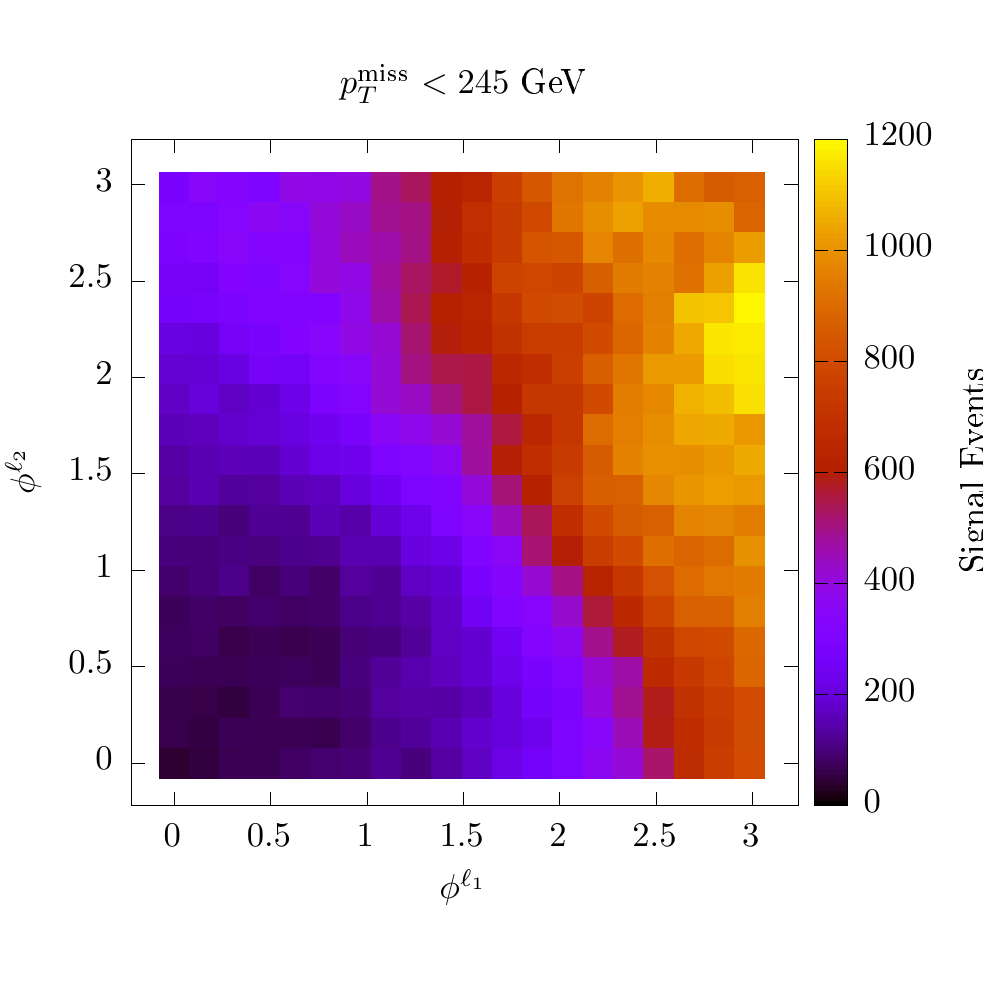}} \scalebox{0.54}{\includegraphics{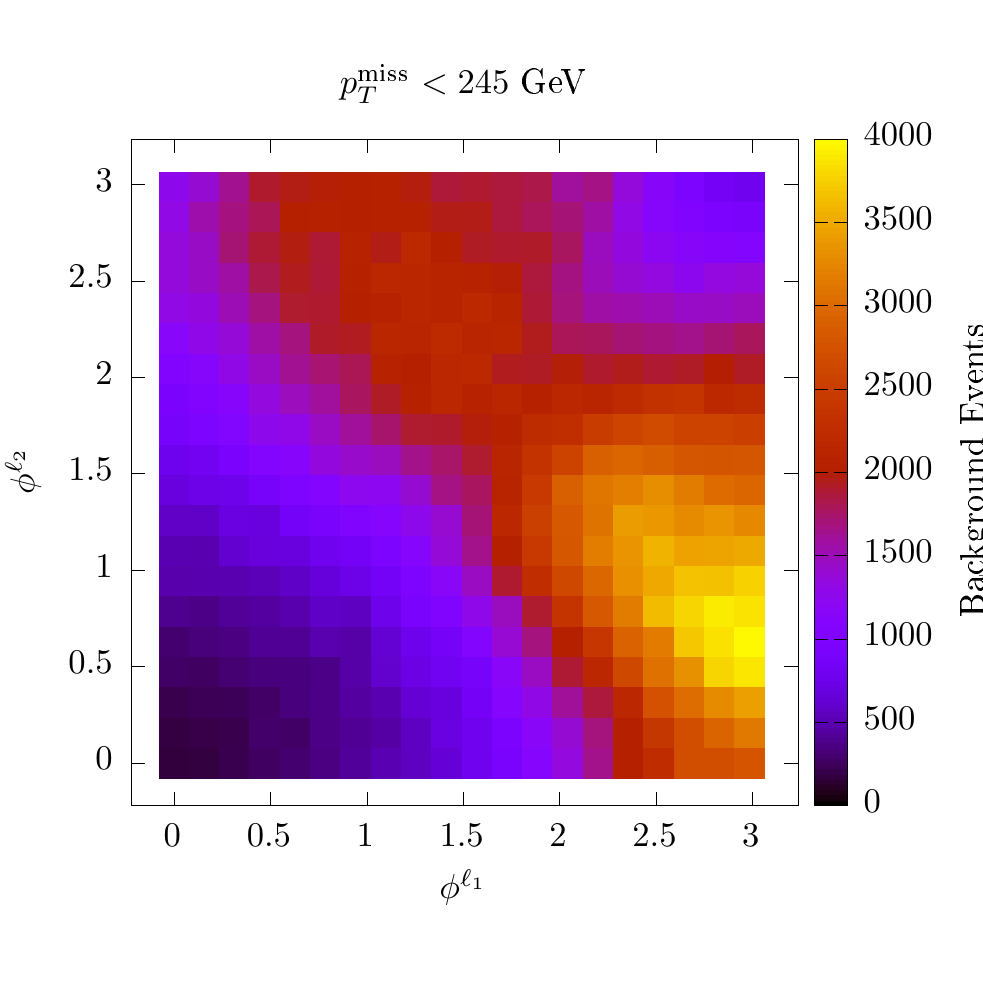}} \scalebox{0.54}{\includegraphics{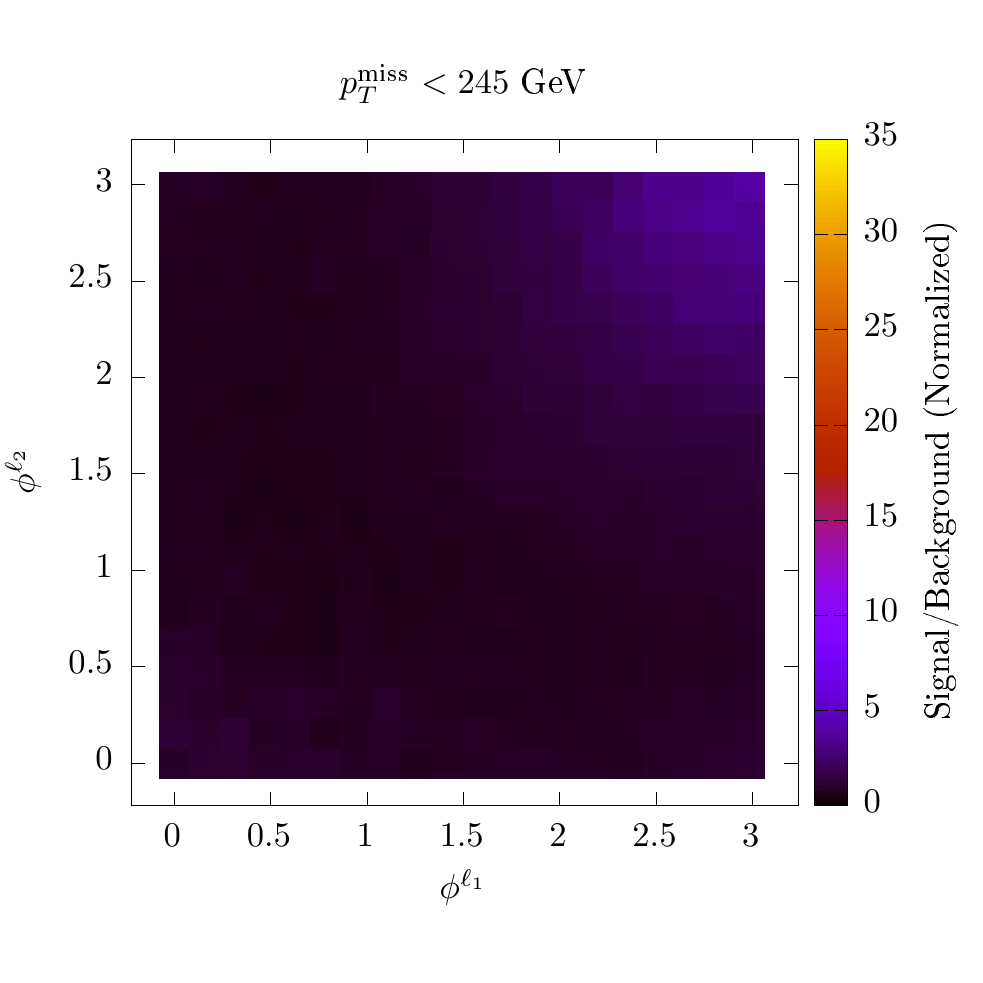}} \\
\scalebox{0.54}{\includegraphics{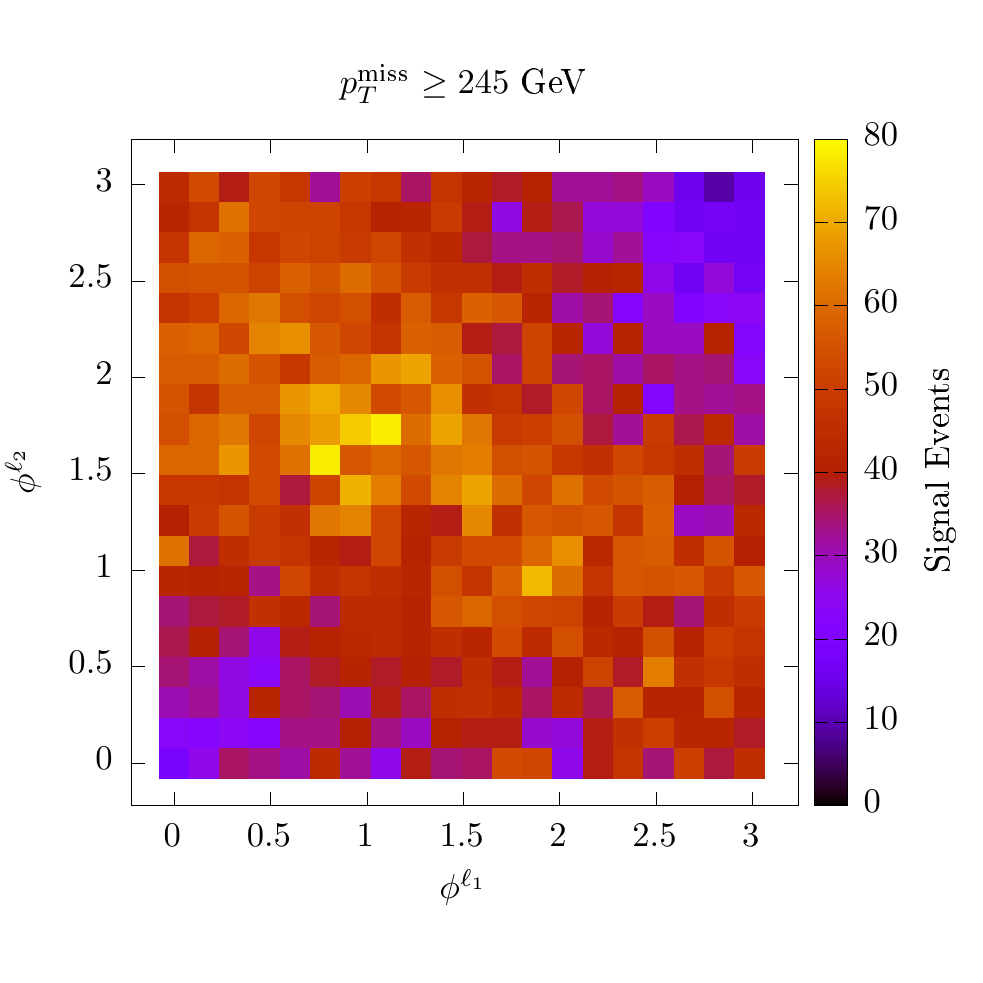}} \scalebox{0.54}{\includegraphics{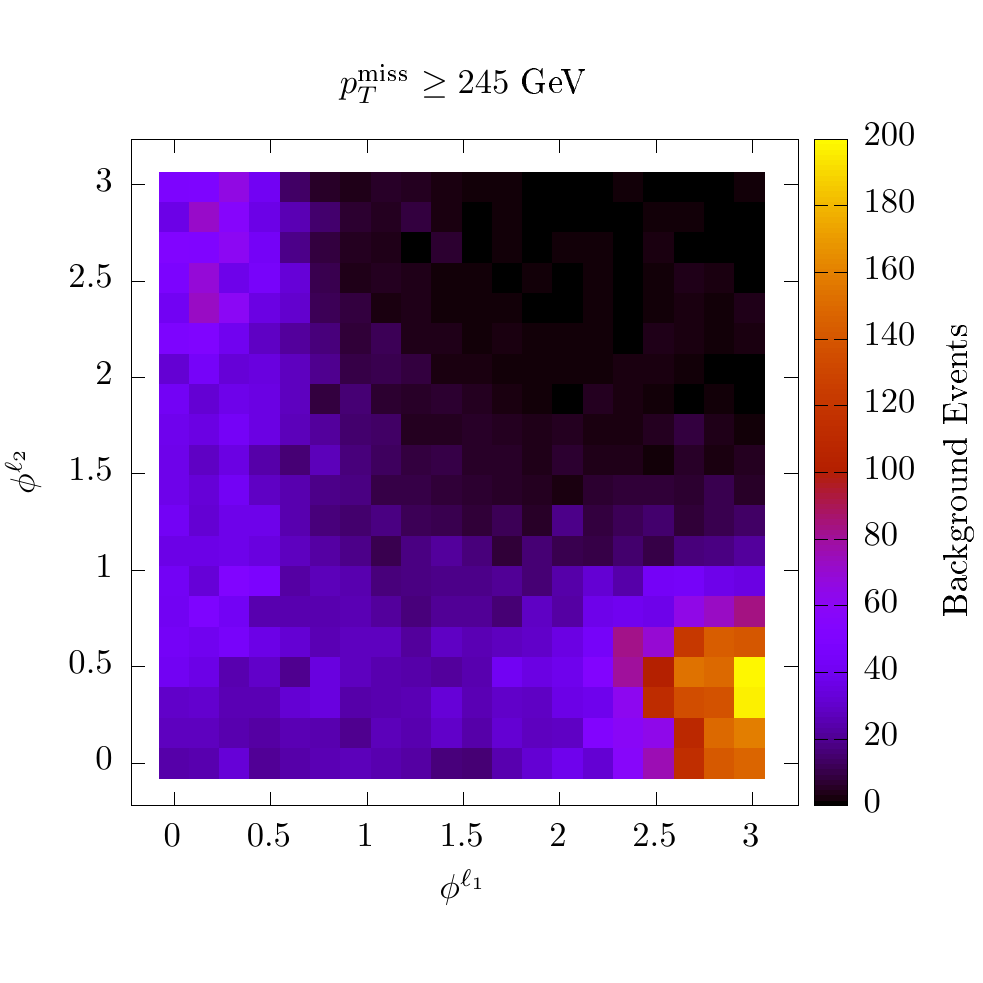}} \scalebox{0.54}{\includegraphics{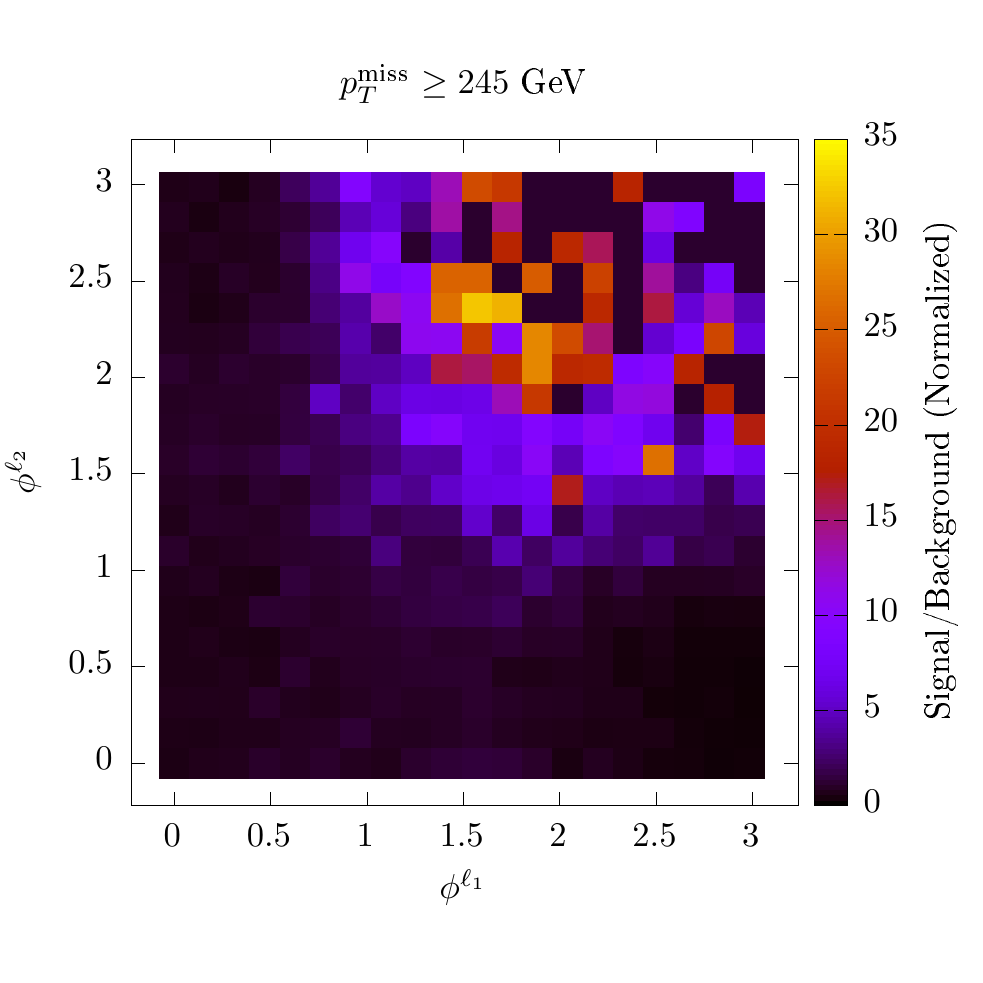}}
\caption{Comparison of the lepton $\phi$ distributions of $(m_{\widetilde t}, m_{\widetilde \chi})$ = $(350, 200)$ data for high and low $p_T^{\text{miss}}$. Top: $p_T^{\text{miss}} < 245$ GeV. Bottom: $p_T^{\text{miss}} \ge 245$ GeV. From left to right: signal, background, and signal over background (after normalizing each histogram individually).}
\label{fig:data_0-11-13}
\end{center}
\end{figure}

We can see this pattern reflected in the simulated data, as well (Figure \ref{fig:data_0-11-13}). We see that, as we go to higher missing energy, both signal and background retreat from the upper-right corner of the plot --- that is, the leptons are less likely to both be directly opposite the missing energy. However, this effect is much more pronounced in the background than the signal, so selecting on events that lie in this area preferentially selects signal events. 

\begin{figure}[!h]
\begin{center}
    \includegraphics[width=1.0\textwidth]{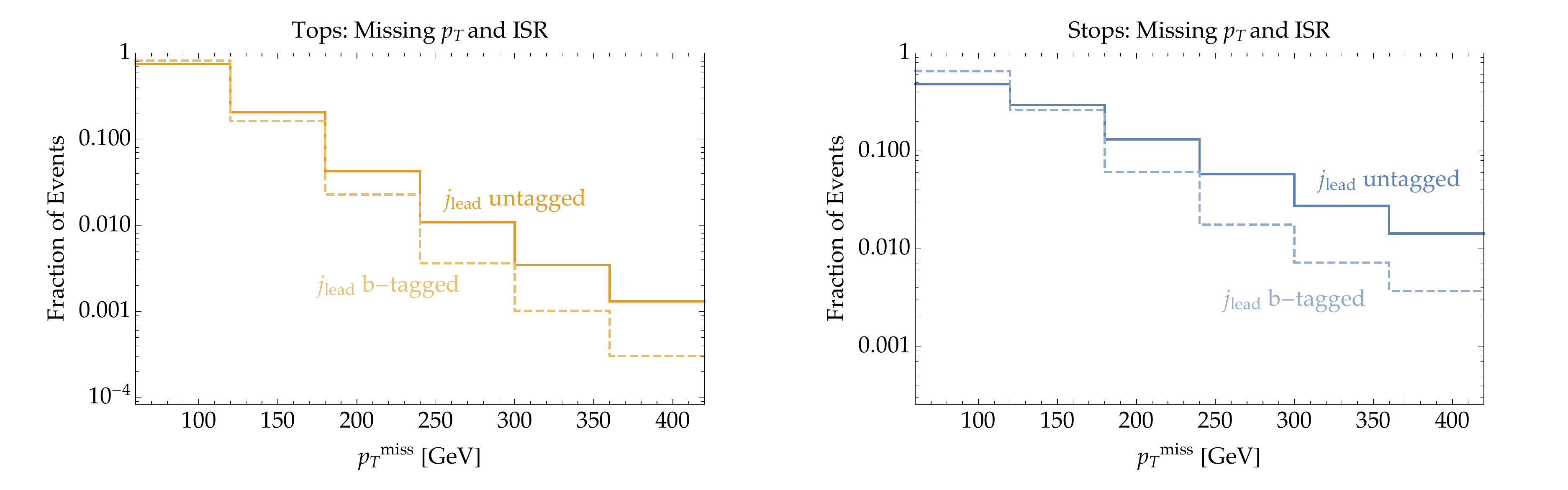}\\
     \includegraphics[width=0.7\textwidth]{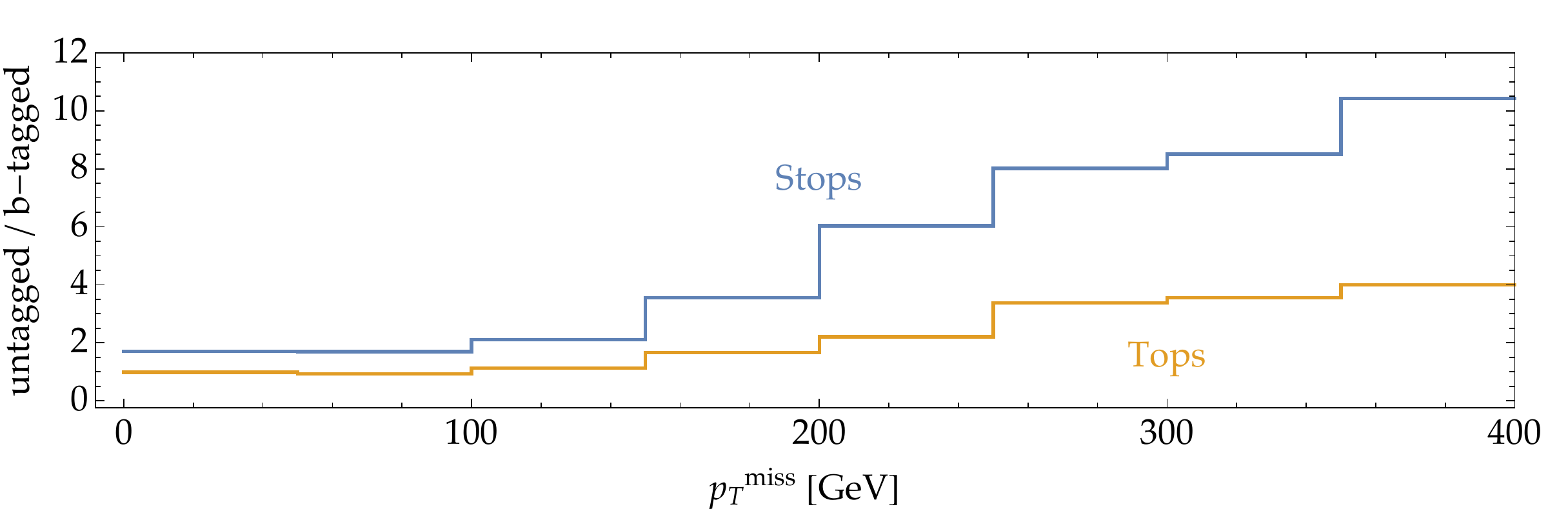}
\end{center}
\caption{{\bf Top panel:} The effect of an ISR jet on the $p_T^{\text{miss}}$ distribution for $t\overline{t}$ events (left) and for ${\widetilde t}{\widetilde t}^*$ events (right) in simulated data, for the parameter point $(m_{\widetilde{t}}, m_{\widetilde{\chi}}) = (350,200)$ GeV. As a proxy for the presence of hard ISR, we plot the case where the highest-$p_T$ jet in the event has no $b$-tag (solid line) versus the case where it has a $b$-tag (lighter dashed line). {\bf Lower panel:} ratio of the number of events with untagged leading jet to events with $b$-tagged leading jet in a given bin of $p_T^{\text{miss}}$. For both stops and tops, the presence of an ISR jet tends to increase $p_T^{\text{miss}}$, but the effect is stronger in signal events than in background events.}
\label{fig:ISRMETeffect}
\end{figure}

Why does the neural network target more complicated patterns in this 3-body case than in the uncompressed 2-body scenario discussed above? Our hypothesis is that the network is sometimes selecting for events with a high-$p_T$ initial state radiation (ISR) jet. As we noted above, relatively little phase space is available to the stop daughters in the rest frame of the stop decay. This tends to suppress the missing momentum in an event, since the missing momentum is determined by the sum
of the momenta of visible particles and these are relatively soft. The exception, as is well-known in the study of compressed SUSY scenarios, is when the entire stop-antistop system is recoiling against a high-momentum ISR jet. In this case all of the decay products are bent toward each other as they recoil against the jet, aligning the invisible particles with each other and increasing the missing momentum. We can test this idea by plotting the $p_T^{\text{miss}}$ distribution
in signal and background events, separating the case where the leading (highest-$p_T$) jet in the event is $b$-tagged (and hence not an ISR jet) from those where it is not (and potentially an ISR jet). As shown in Fig.~\ref{fig:ISRMETeffect}, the presence of ISR tends to enhance $p_T^{\text{miss}}$ in both signal and background, but the effect is much more pronounced in the signal events.

\begin{figure}[!htbp]
\begin{center}
\scalebox{0.65}{\includegraphics{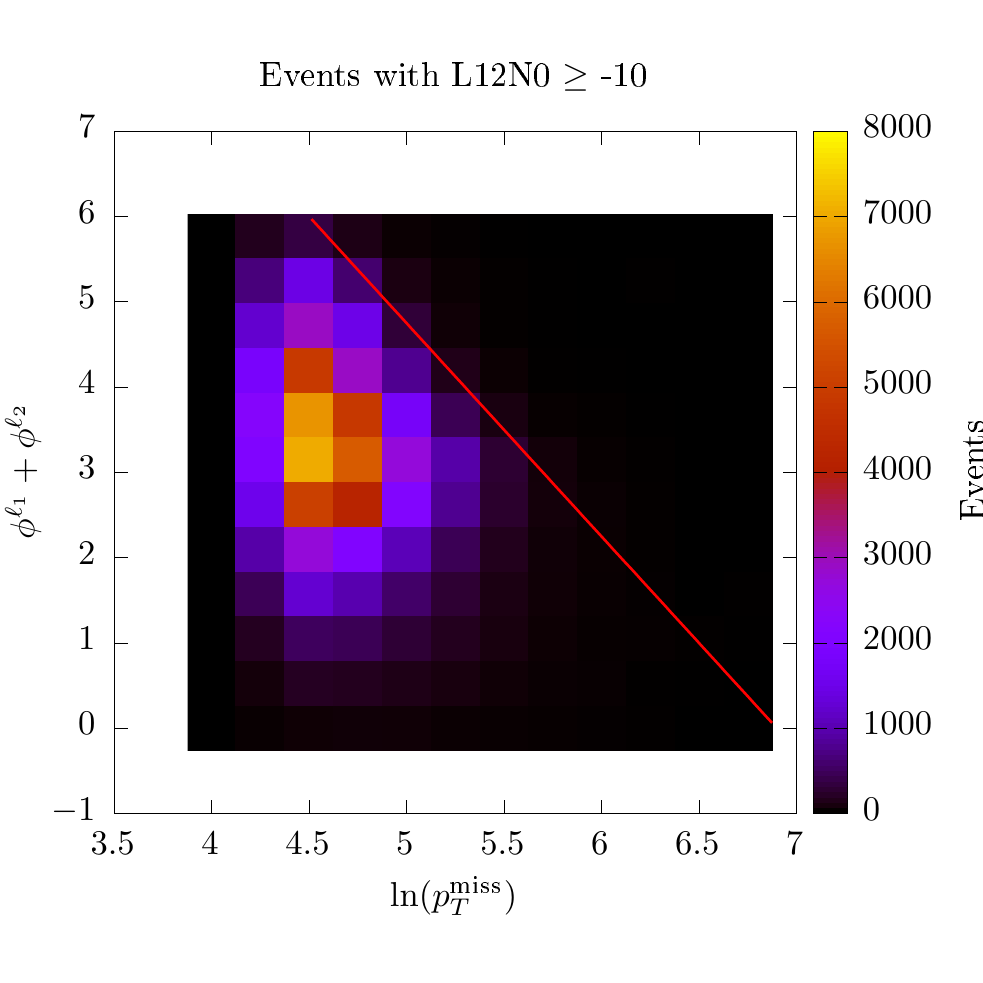}} \scalebox{0.65}{\includegraphics{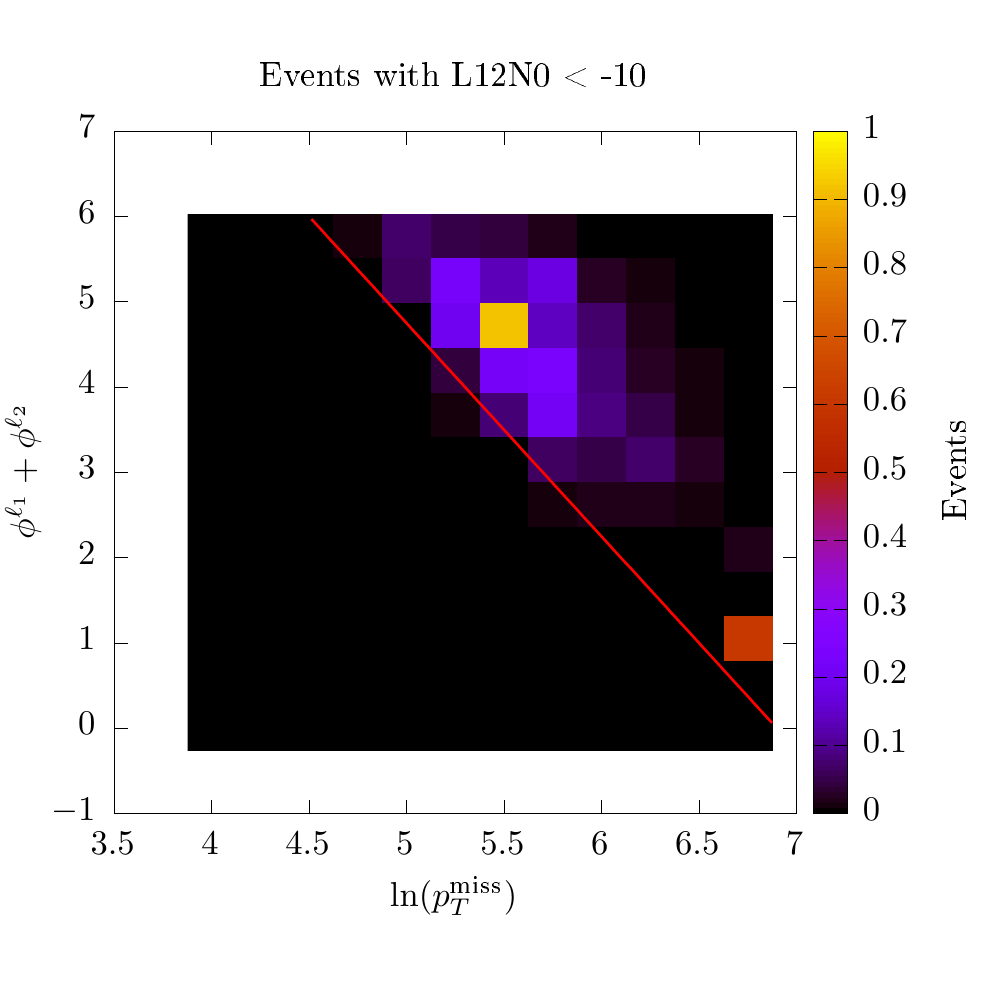}} 
\caption{Data distributions in the $p_T^{\text{miss}}$-$\sum\phi^{\ell}$ plane. We see that, to a reasonable approximation, cutting on a pre-transfer function network output of $-10$ is equivalent to dividing the plane on the displayed line. More signal-like events are concentrated at more negative values of the network output.}
\label{fig:dispEmissPhi}
\end{center}
\end{figure}

\begin{figure}[!htbp]
\begin{center}
\scalebox{0.65}{\includegraphics{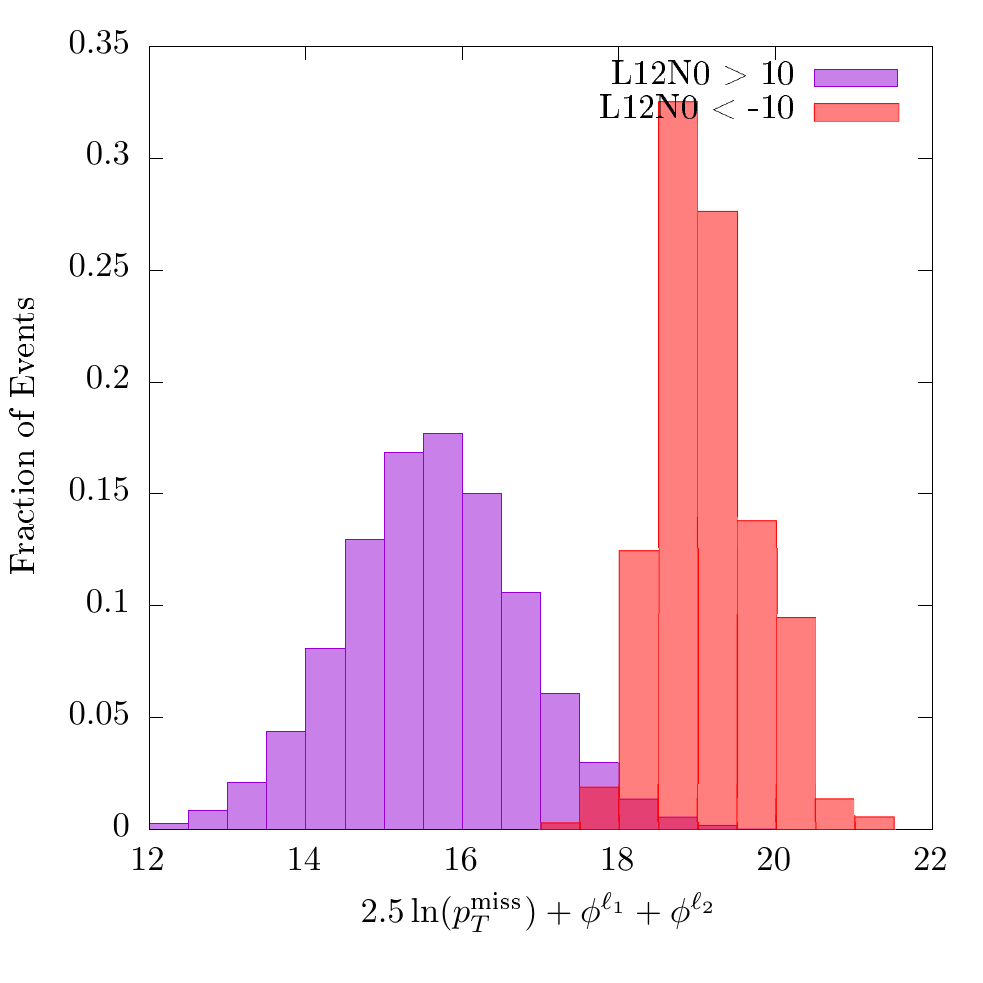}} \scalebox{0.65}{\includegraphics{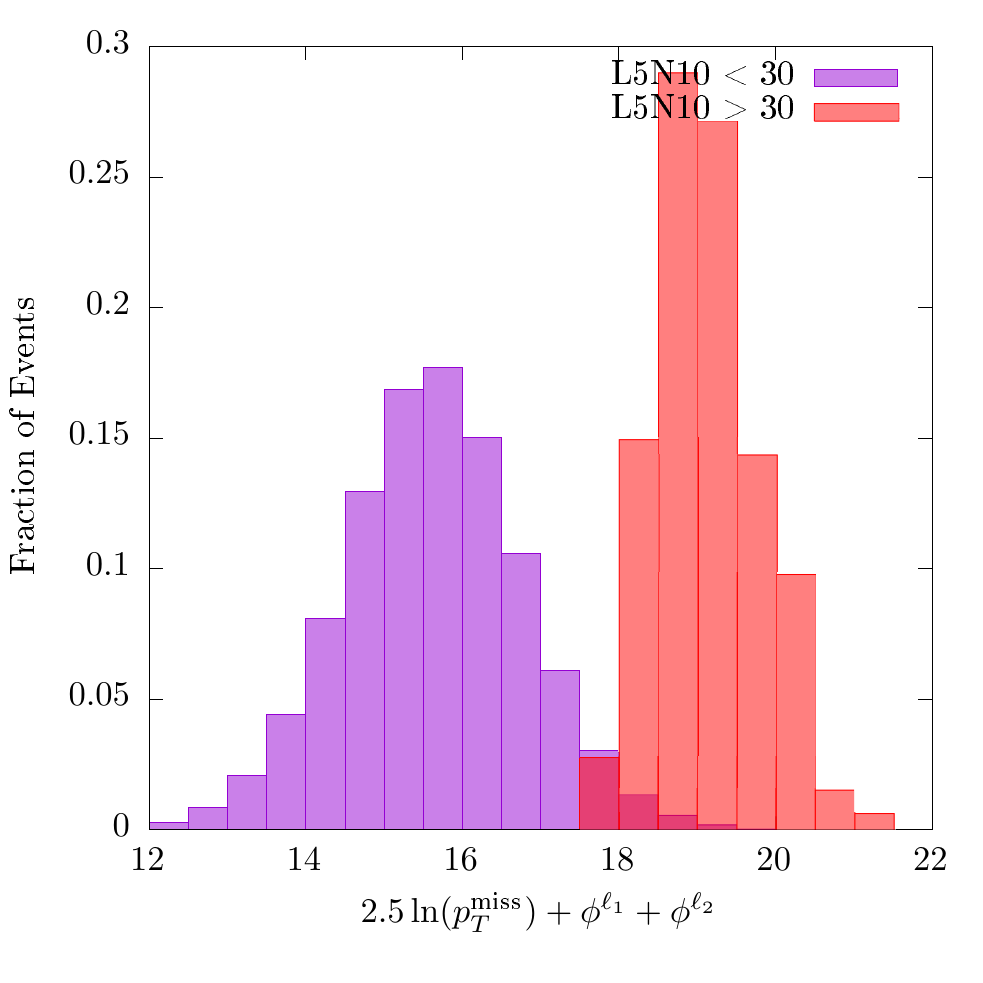}} 
\caption{Comparison of a simple combined variable composed of $p_T^{\text{miss}}$ and the lepton $\phi$s with the network output. We see that this variable is, in the main, able to reproduce the gross behavior of the network, although on its own it is not sufficient to duplicate its discriminative prowess.}
\label{fig:nhist_EmissPhi}
\end{center}
\end{figure}

The role of ISR explains the apparent migration of the signal events at high $p_T^{\text{miss}}$ that we observe in Figure \ref{fig:data_0-11-13}: at relatively low $p_T^{\text{miss}}$, the neural network optimizes rejection of the top background by demanding that leptons point opposite to missing momentum, just as it did for the (750,1) point. At larger $p_T^{\text{miss}}$, the network accounts for the fact that ISR is likely present in the signal and hence the leptons are unlikely to be as well-separated in angle from the missing momentum; it prefers moderate values of lepton $\phi$. This leads us to a reasonable first approximation of the network output: we simply add the lepton $\phi$s to the log of the missing energy. We can see how well this works in Figures \ref{fig:dispEmissPhi} and \ref{fig:nhist_EmissPhi}. These figures show the output of the final neuron $L12N0$, for which more signal-like events correspond to more negative outputs. We conclude that the network requires \emph{either} a very high missing energy \emph{or} the leptons to be facing away from the missing energy, although both phenomena rarely occur at the same time. In this way the network smoothly interpolates between the kinematics of events with large ISR and those without.

\subsubsection{Checking the ISR hypothesis}

In order to verify the hypothesis that the network is selecting events with large ISR jets, we can examine the events it has flagged as the most signal-like and check to see whether they have more energetic ISR than the rest of the population. We thus construct the ``ISR $p_T$'', a variable which (very roughly) approximates the $p_T$ of a hard ISR jet under the assumption that it is relatively energetic compared to the other jets in the
event. It is given by:
\begin{equation}
p_T^{\text{ISR}} = 
\begin{cases}
    p_T^{j_1} & \text{if $j_1$ is not $b$-tagged} \\
    p_T^{j_2} & \text{if $j_2$ is not $b$-tagged} \\
    |\vec{p}_T^{\text{miss}} + \vec{p}_T^{\ell_1} + \vec{p}_T^{\ell_2} + \vec{p}_T^{j_1} + \vec{p}_T^{j_2}| & \text{if both jets are $b$-tagged} \\
\end{cases}
\end{equation}

\begin{figure}[!htbp]
\begin{center}
\scalebox{0.55}{\includegraphics{plots/350-200/nhist_350-200_L12N0_iden.pdf}}\scalebox{0.55}{\includegraphics{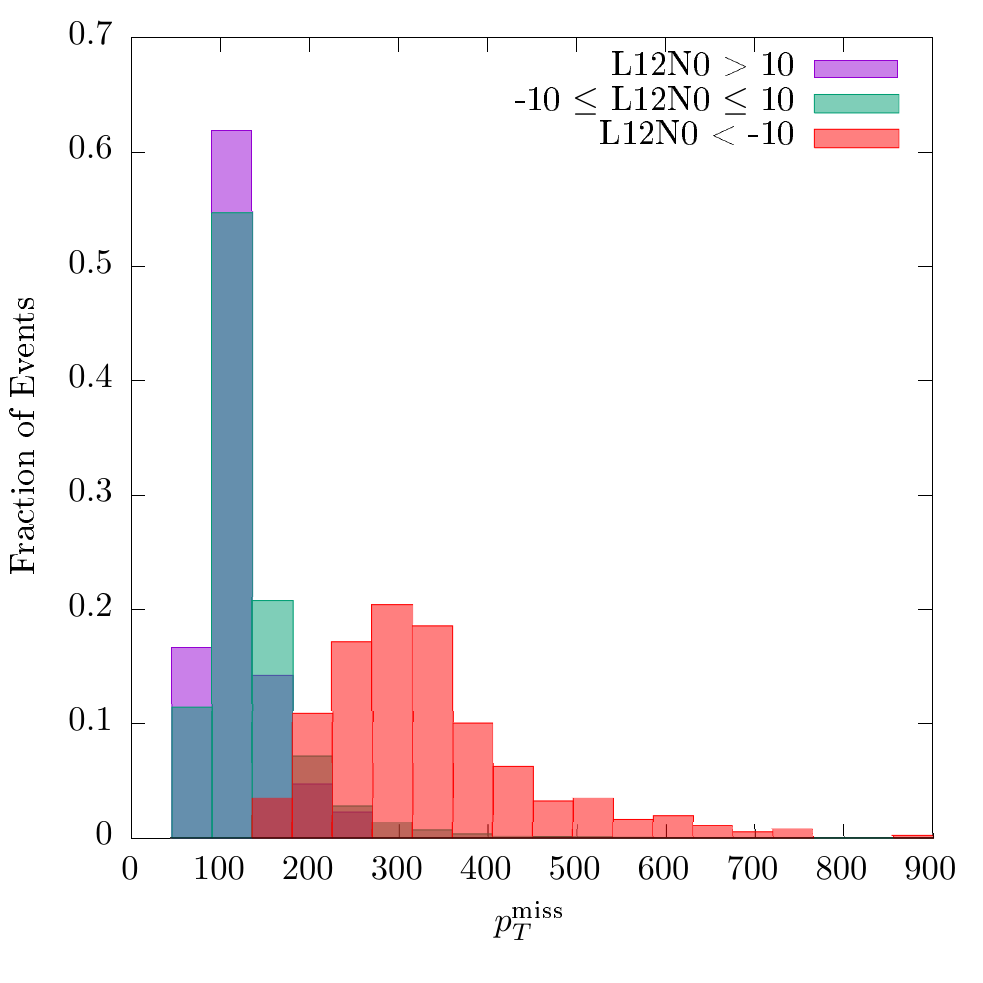}}\scalebox{0.55}{\includegraphics{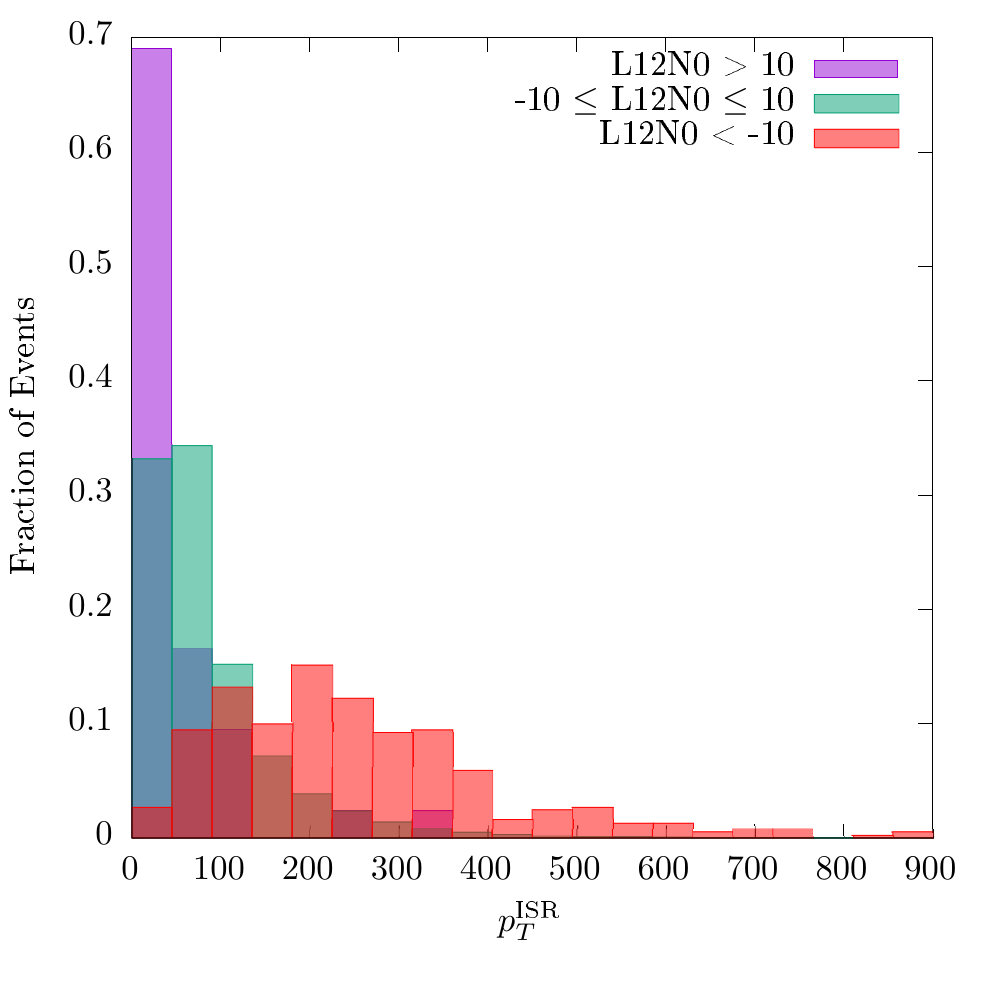}}
\scalebox{0.55}{\includegraphics{plots/350-200/nhist_350-200_L5N10.pdf}}\scalebox{0.55}{\includegraphics{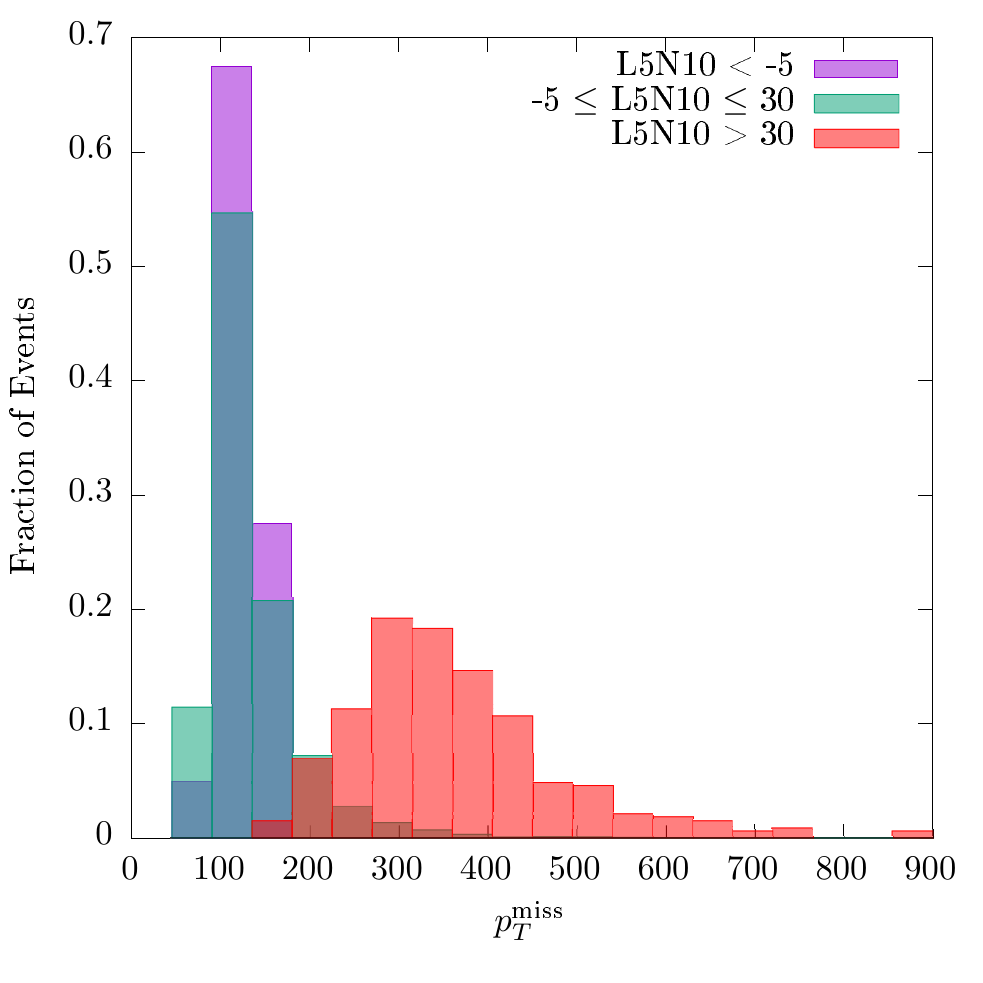}}\scalebox{0.55}{\includegraphics{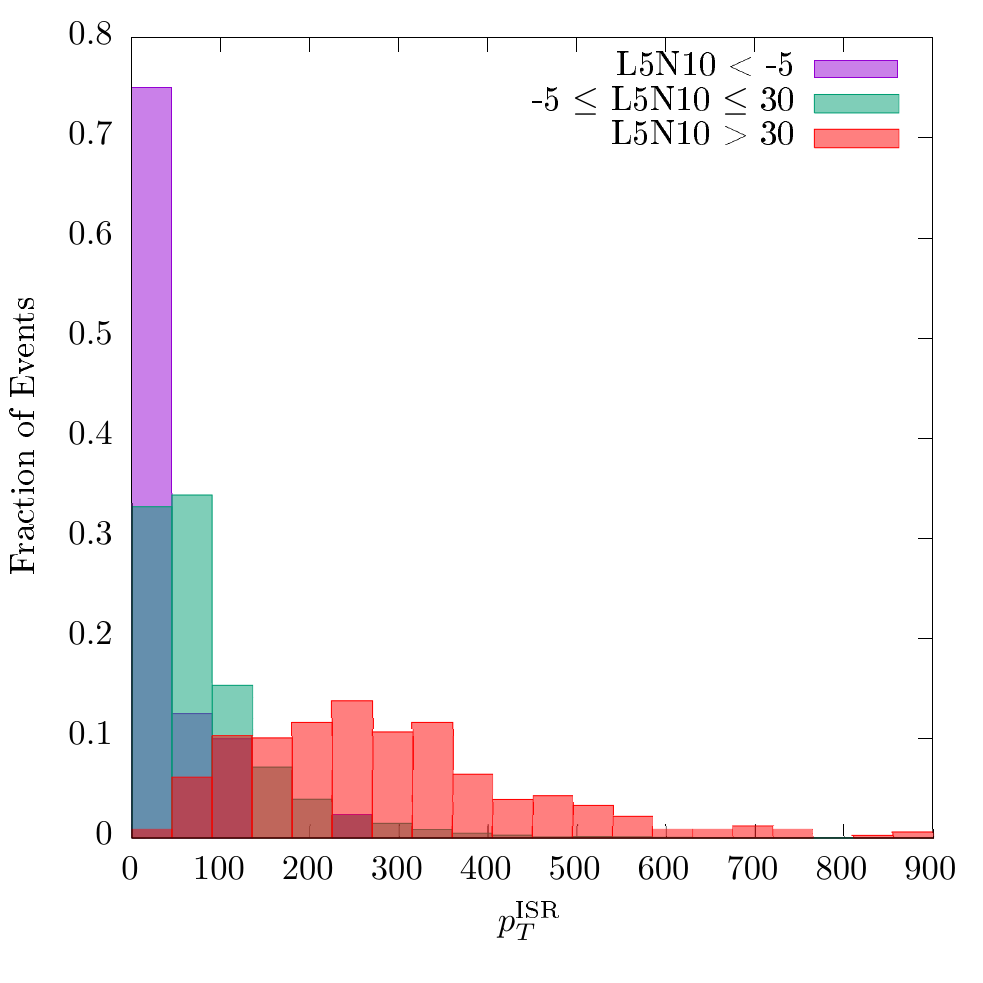}}
\caption{Left: neuron distributions. Right: $p_T^{\text{ISR}}$ distributions for different neuron cuts. Events that the network selects as signal are clearly much more likely to have high $p_T^{\text{ISR}}$, although the network appears to be selecting mostly on $p_T^{\text{miss}}$. Note: signal and background events are mixed together unweighted to produce the plots on the right. This means that the histogram corresponding to middling events contains both signal and background events in approximately a 1:3 ratio.}
\label{fig:nhist_ISR}
\end{center}
\end{figure}

In Figure \ref{fig:nhist_ISR}, we consider the $p_T^{\text{ISR}}$ distributions for very signal-like, average, and very background-like events, as determined by the network. We see that, indeed, events selected by the network are likely to have higher $p_T^{\text{ISR}}$, although the network is clearly not selecting on it directly.

\subsection{Additional correlations}

In order to move towards a more complete understanding of what the network is doing, we must of course add the rest of the input variables. Given the success of the simple variable in figure \ref{fig:nhist_EmissPhi}, the contributions of these variables are clearly subdominant, but we can approach adding them in the same way. Some example correlations are shown in Figure \ref{fig:dispL5N10_variousDisp}. These are complex and the physical interpretation of some of the features is unclear. As in the Hawaii map example of \S\ref{sec:hawaii}, it could be that intermediate neurons of the network are learning parts of a larger, final pattern that are not easily understood on their own. If one were to study the activation patterns of a large number of neurons, it might be possible to develop a better understanding of how the neural network operates. This undertaking is beyond the scope of this paper. 

\begin{figure}[!htbp]
\begin{center}
\scalebox{0.55}{\includegraphics{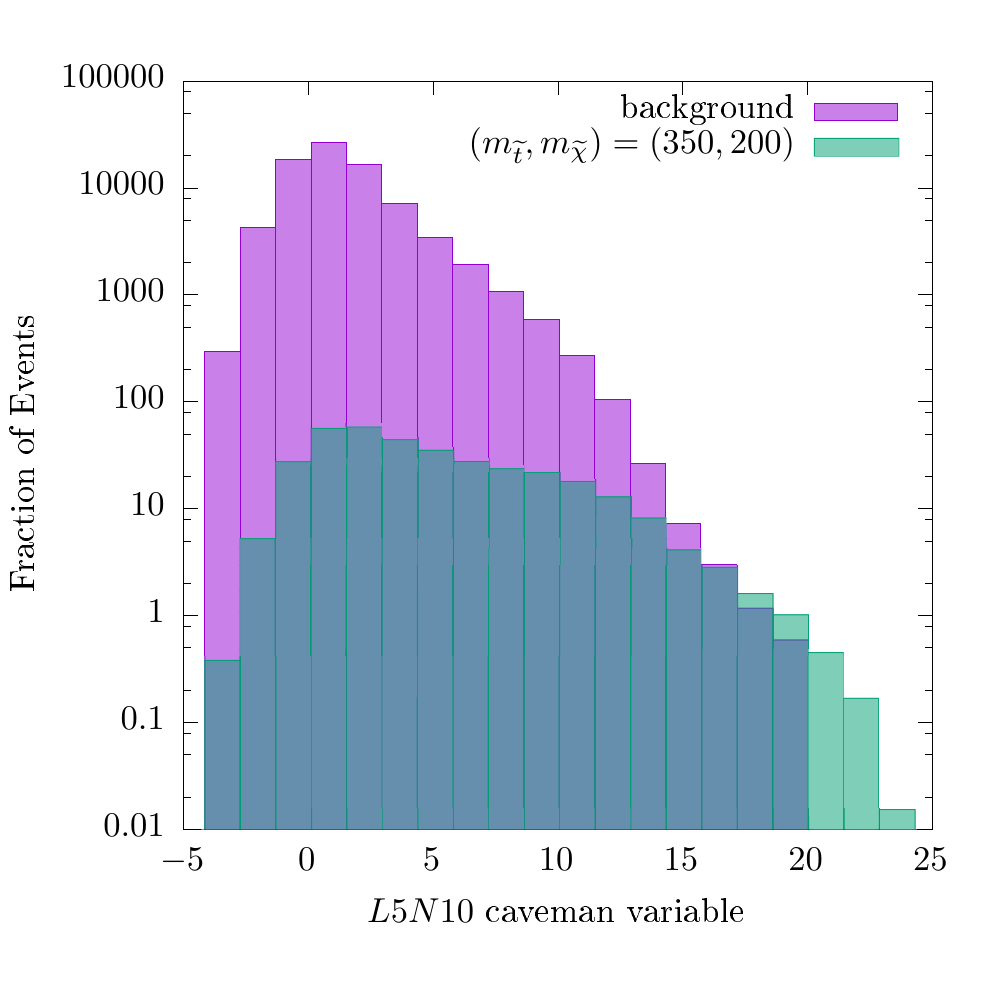}}
\caption{Performance of the caveman variable for neuron $L5N10$ of a network trained on $(350, 200)$ data.}
\label{fig:disp_350-200_L5N10_caveman}
\end{center}
\end{figure}

\begin{figure}[!htbp]
\begin{center}
\scalebox{0.55}{\includegraphics{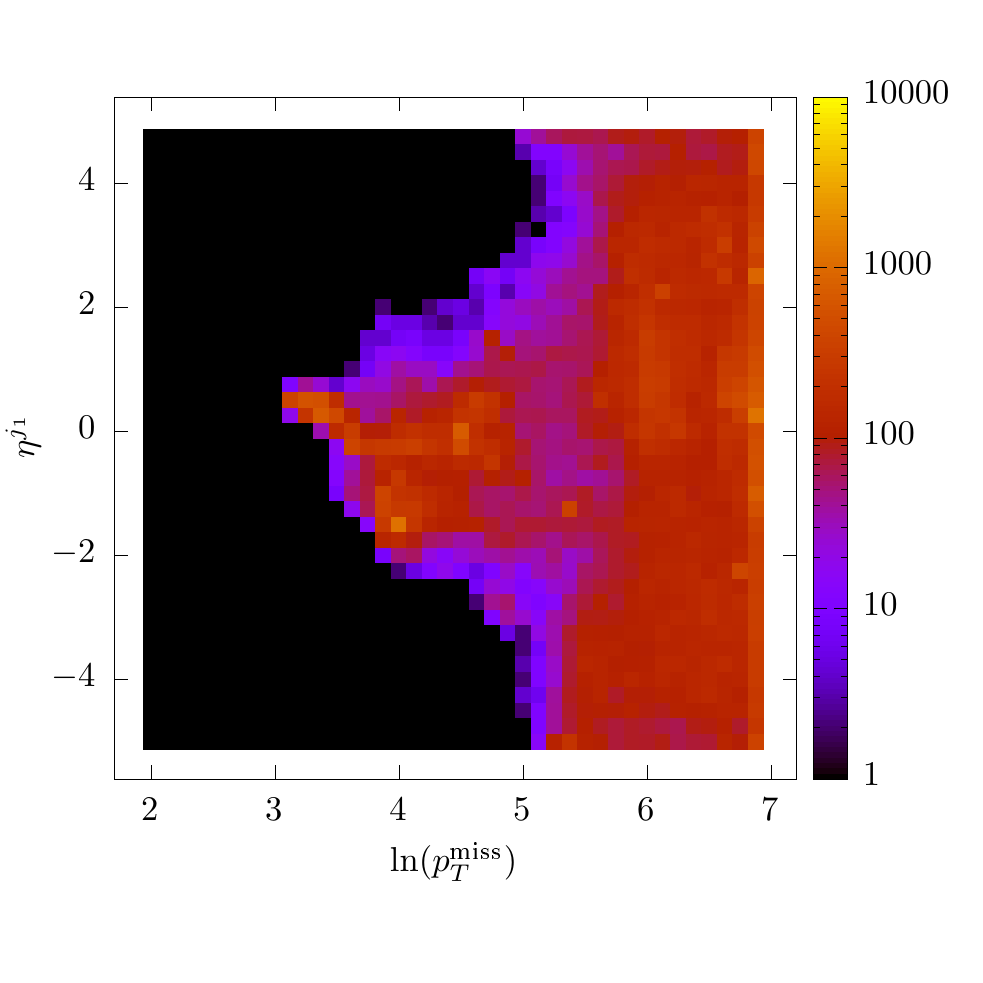}}\scalebox{0.55}{\includegraphics{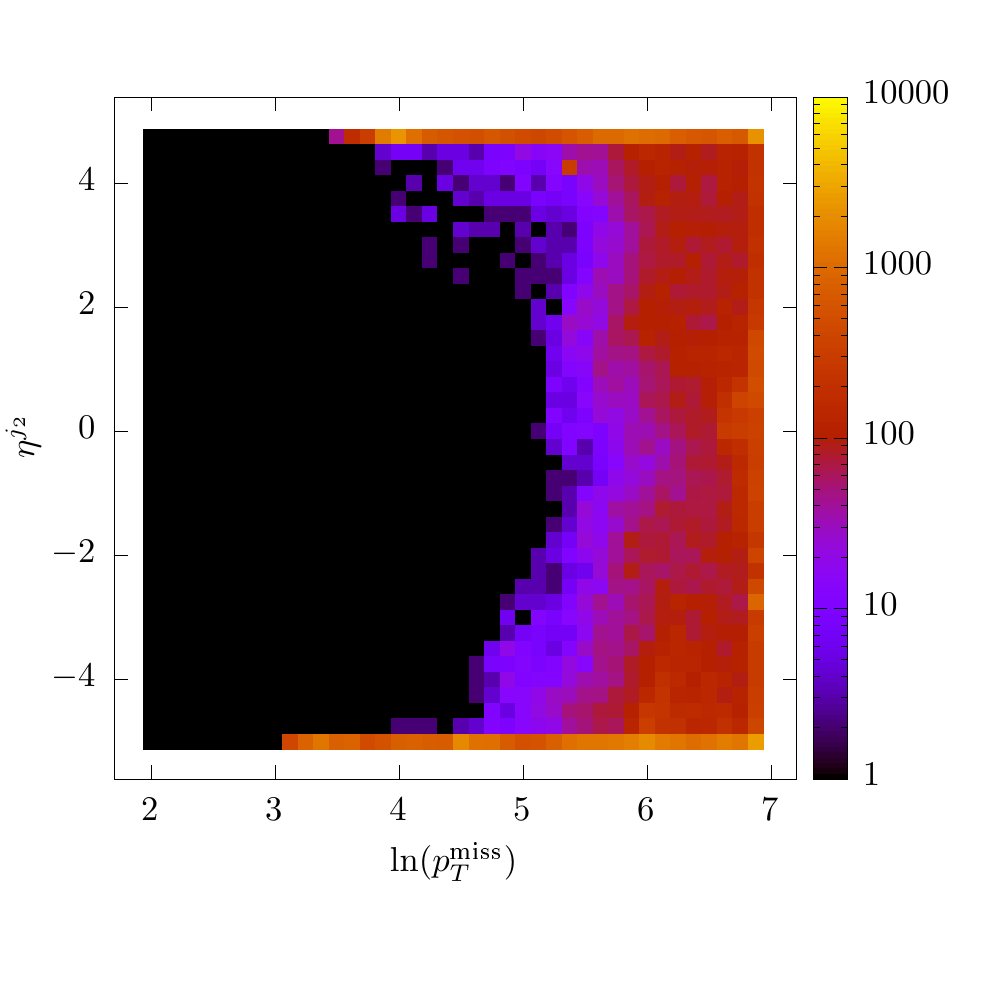}}
\scalebox{0.55}{\includegraphics{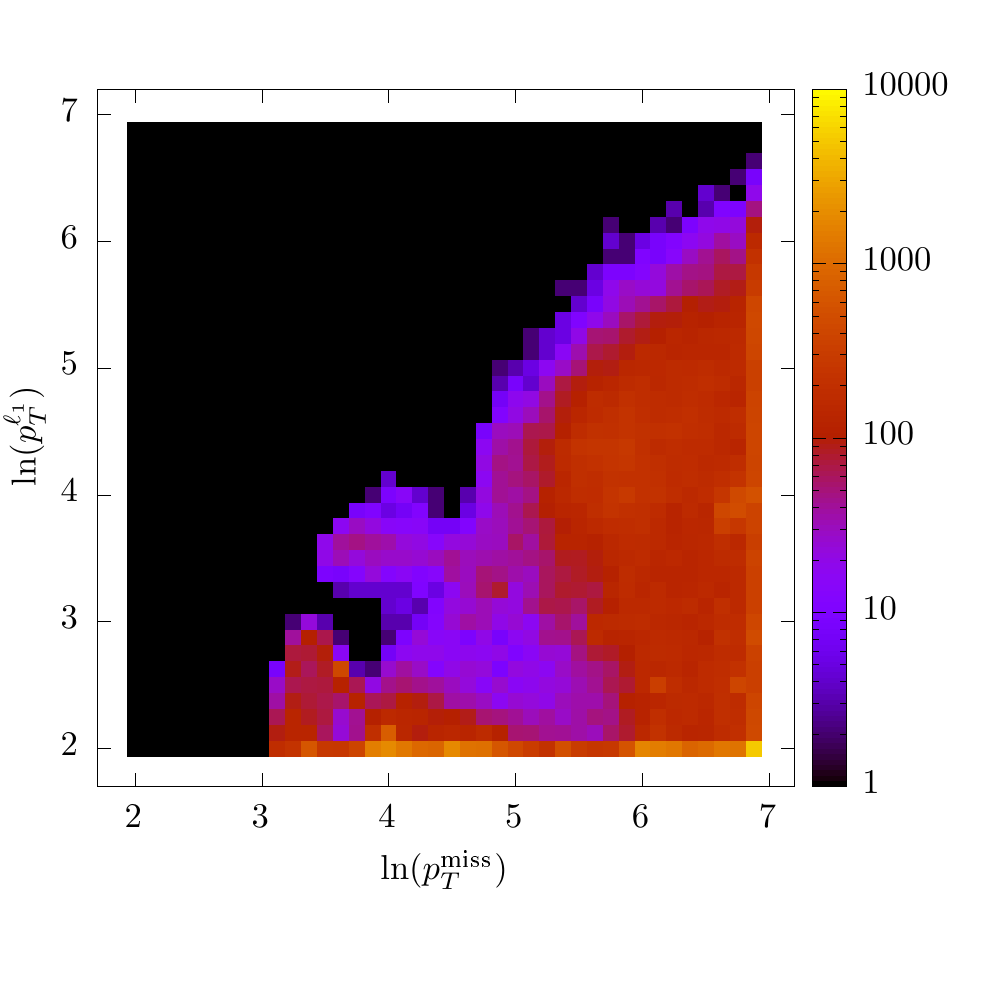}}\scalebox{0.55}{\includegraphics{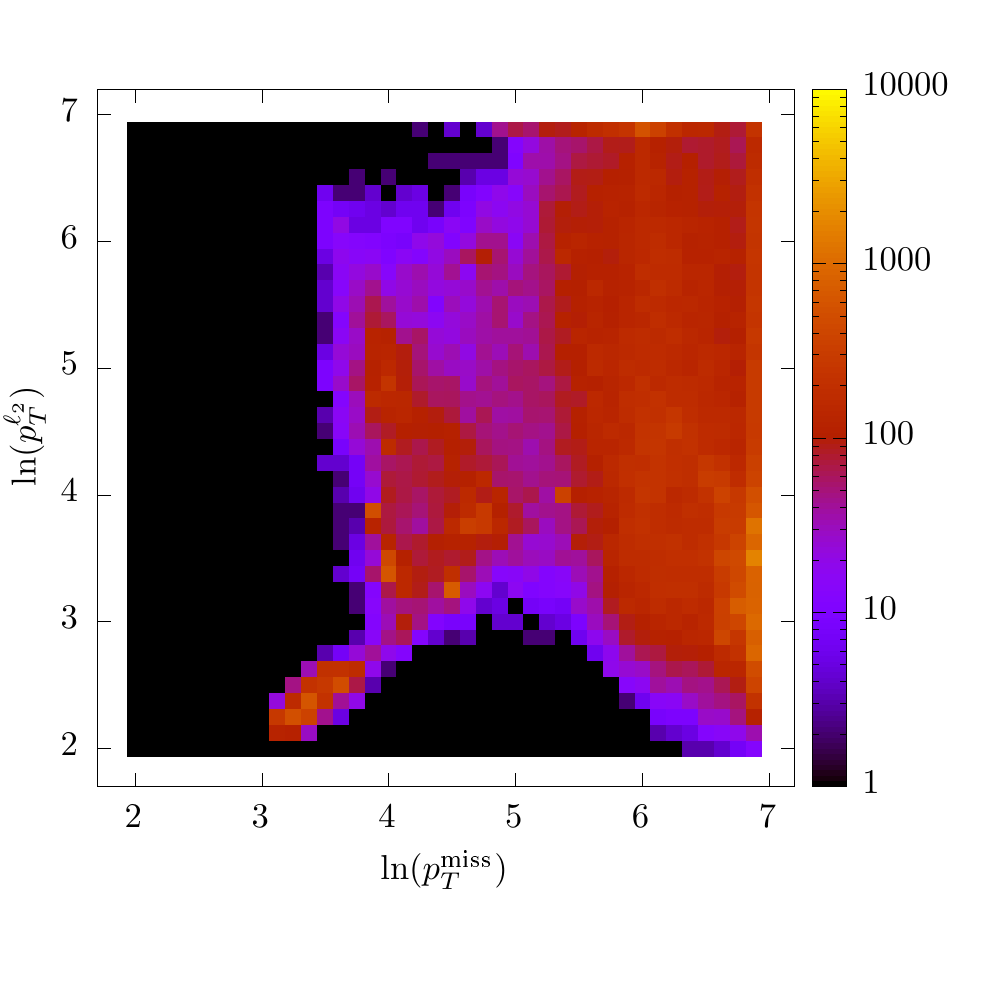}}
\scalebox{0.55}{\includegraphics{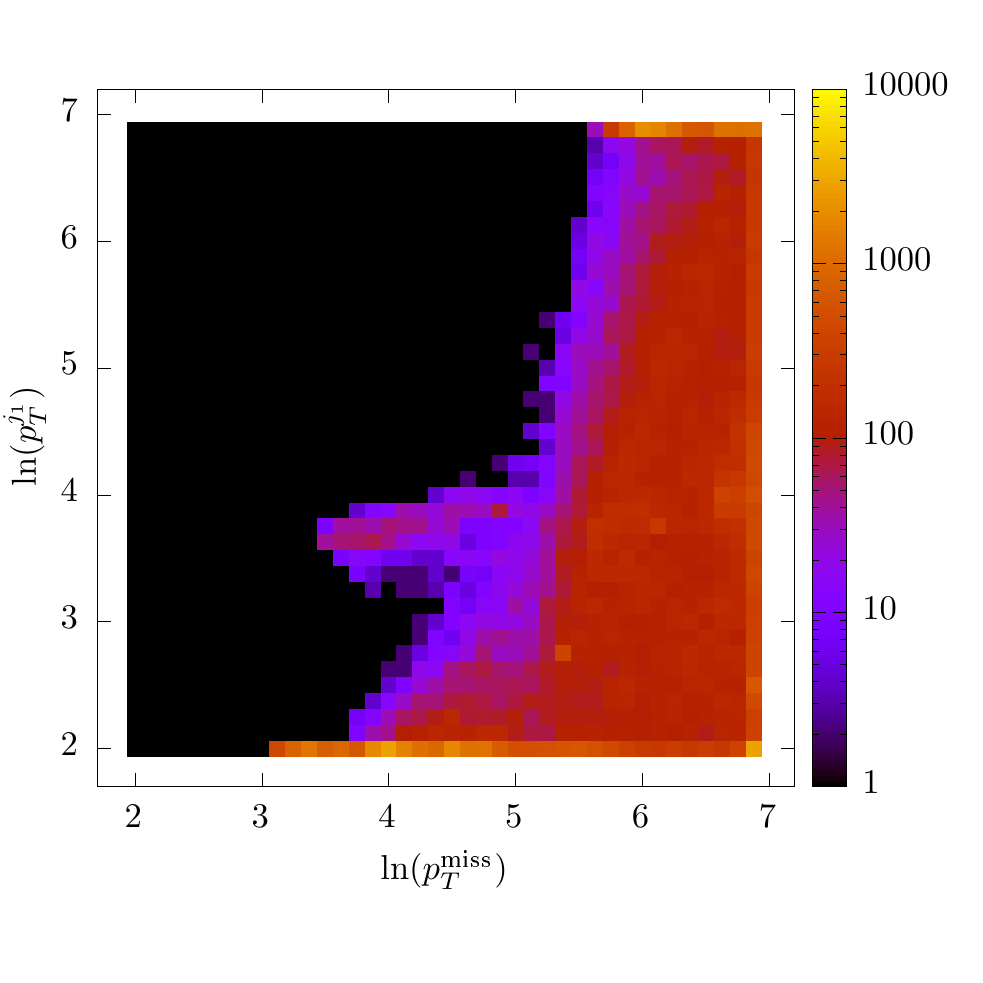}}\scalebox{0.55}{\includegraphics{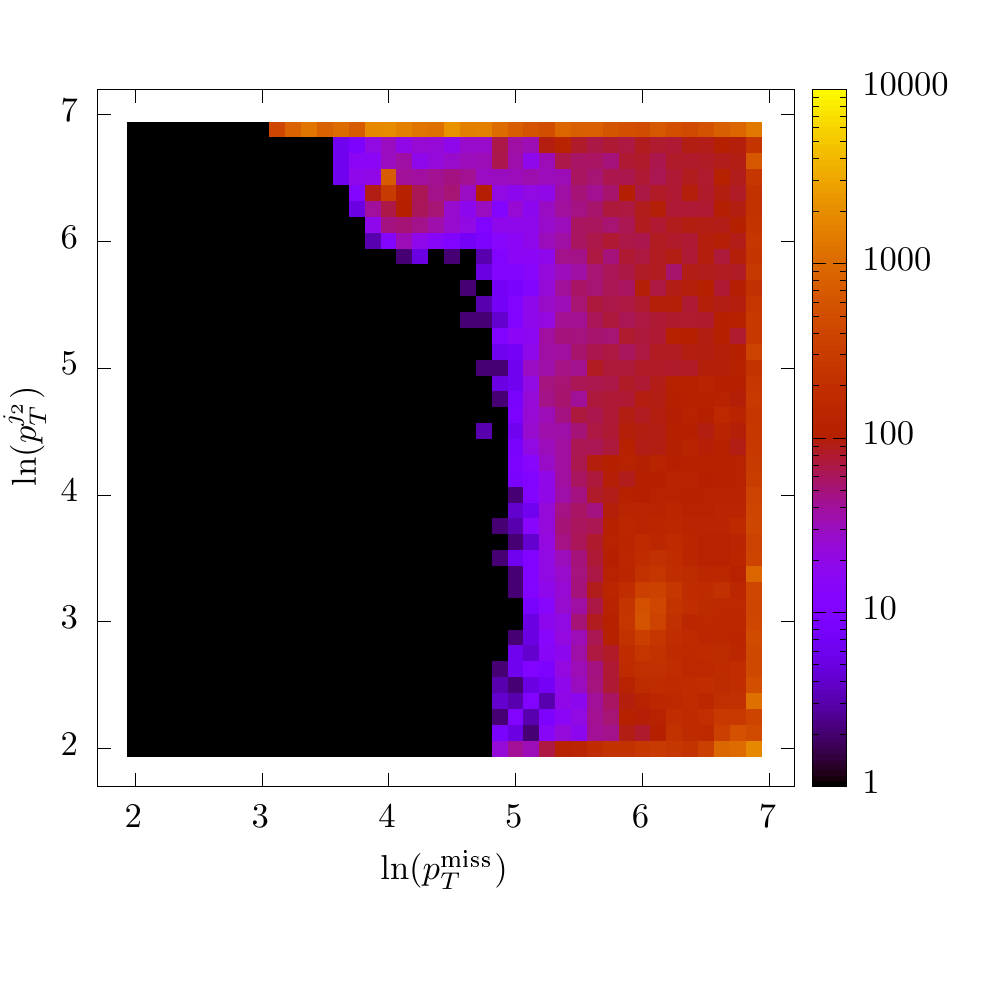}}
\scalebox{0.55}{\includegraphics{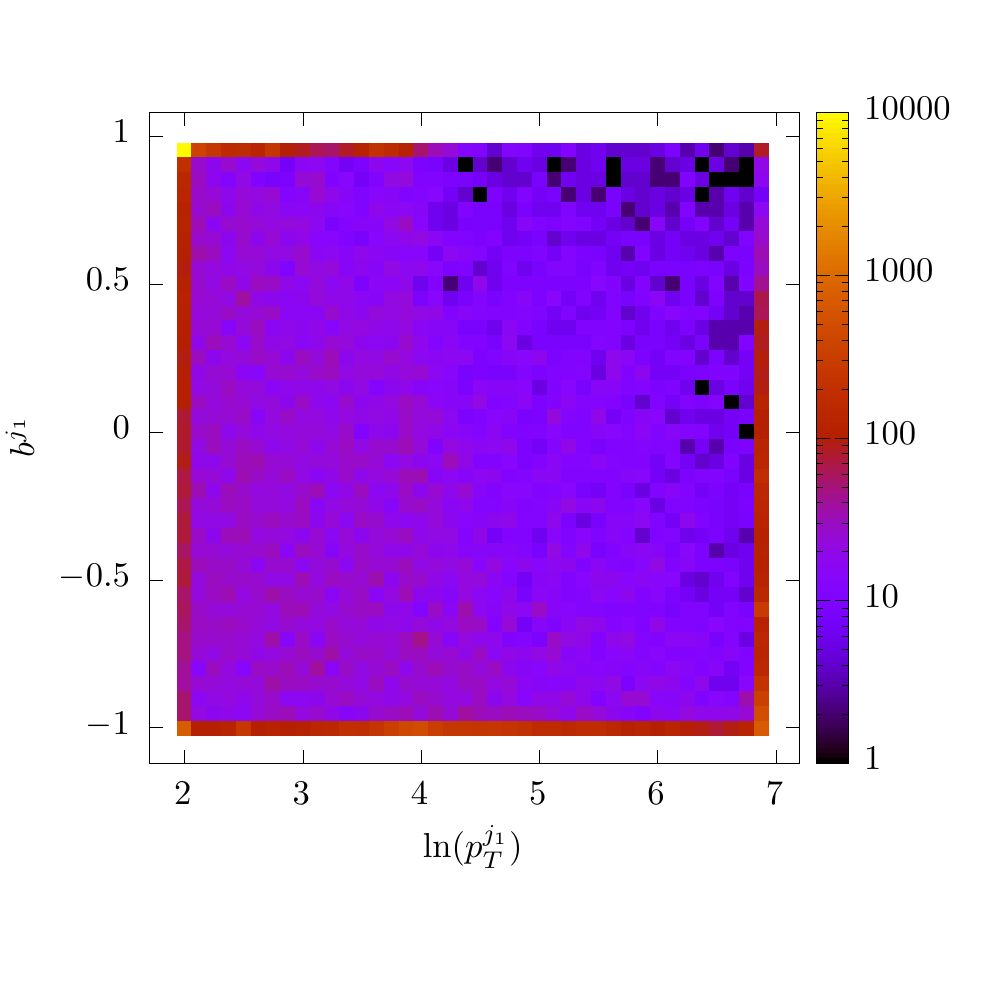}}\scalebox{0.55}{\includegraphics{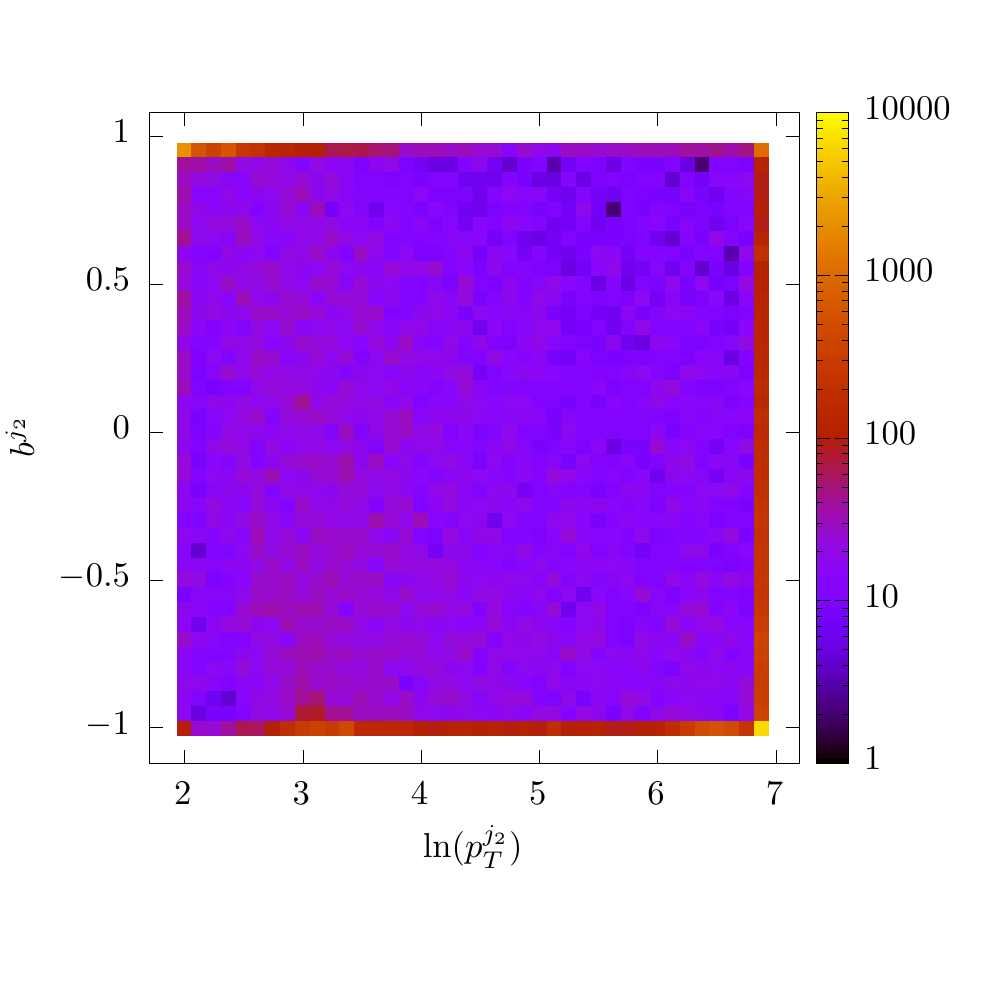}}
\caption{(350, 200) $L5N10$ activation maximization histograms for various variables in combination with $p_T^{\text{miss}}$.}
\label{fig:dispL5N10_variousDisp}
\end{center}
\end{figure}

We can get a very rough idea of how the correlations in these plots perform together by constructing a caveman variable following the procedure in Section \ref{sec:caveman}. The performance of the caveman variable is shown in Figure \ref{fig:disp_350-200_L5N10_caveman}. We see that it performs worse than the neuron itself, as must be expected, but still possesses some discriminative power.

\clearpage

\section{The stealthy region of parameter space}
\label{sec:stealth}

When the neutralino is very light and the stop is only slightly heavier than the top quark, the kinematics of the final-state particles in their decays look very similar, since the stop essentially decays directly to a top with very similar energy and momentum. Nonetheless, by exploiting the fact that the stops, as scalar particles, are unable to transmit to their daughter top quarks the spin correlation that is present when the tops are directly produced from the same process, it is possible
to discern a difference between signal and background distributions. This approach was investigated in \cite{Han:2012fw, Aad:2014mfk}, where it was found that the problem succumbed to selections in a region of parameter space defined by two simple variables: high azimuthal difference between the leptons and small pseudorapidity difference between the tops (and thus presumably their decay products as well). To find out whether we can deduce these correlations, or more effective
ones, using
our methods, we train a network on the mass pair $(m_{\widetilde t}, m_{\widetilde{\chi}^0_1}) = (185, 5)$ GeV. Let's open the black box.

\begin{figure}[!ht]
\begin{center}
\scalebox{0.7}{\includegraphics{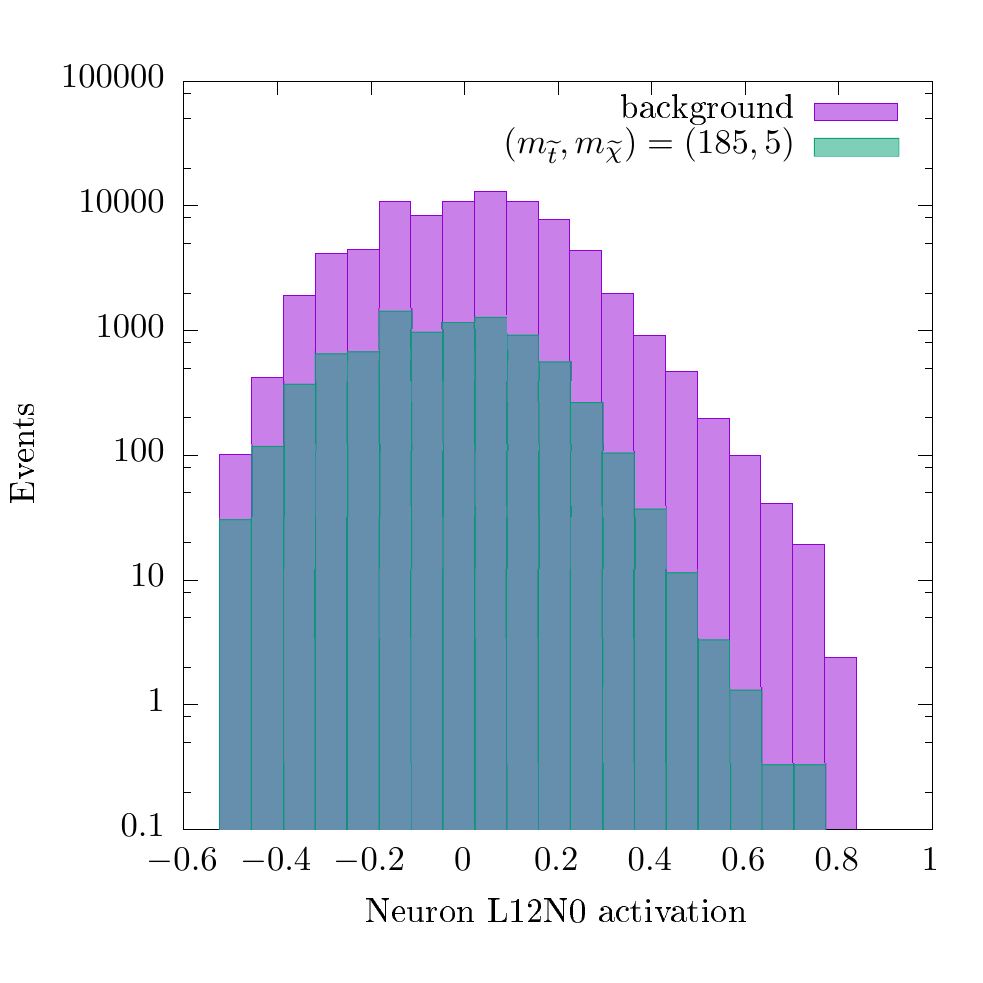}} \\
\caption{Activation histogram of the final neuron in a network trained on $(m_{\widetilde t}, m_{\widetilde{\chi}^0_1}) = (185, 5)$. The network has clearly managed to learn some method of distinguishing signal from background, though even in the most signal-like bins the background rate is higher by nearly a factor of 4.}
\label{fig:nhist_185-5_L12N0}
\end{center}
\end{figure}

Since this is a more difficult problem than the $(750, 1)$ point, we expect the network to be unable to differentiate signal from background nearly as well as before. Indeed, this is what we see in Figure \ref{fig:nhist_185-5_L12N0}: there is no activation threshold beyond which the network selects only stop events. Nonetheless, since the $\widetilde{t}\widetilde{t}^*$ cross-section in this regime is approximately 10\% that of $t\overline{t}$, there are still enough statistics to
meaningfully distinguish a population of $t\overline{t}$ events from one with an injection of $\widetilde{t}\widetilde{t}^*$. In fact, the signal cross-section is large enough to make a brute-force strategy of simply comparing the total cross-sections viable; indeed, if we naively compute the AMS significance value of our network, we find that it is nearly 30 with a threshold that includes over 90\% of events. The problem with this approach has, in the past, been systematic uncertainties in the
$t\overline{t}$ cross-section, although these uncertainties have come down in recent years enough to render it tractable (see refs.~\cite{Czakon:2014fka, Eifert:2014kea, Cohen:2018arg}). We will not attempt to incorporate systematic uncertainties to quote a significance measure for this point, but rather simply seek to qualitatively examine the correlations the network has found, whether they are actually strictly necessary for exclusion or not.

We will mainly confine ourselves to the six most important input variables, listed by activation difference in Table \ref{tab:actDiff_185-5_L12N0}. Other variables, for example the angular positions of the jets, do contribute to the result, but we can draw some interesting conclusions just from considering these six.

\begin{table}
\begin{center}
\begin{tabular}{|c|c|c|c|c|c|c|}
    \hline
    Variable & $\phi^{\ell_2}$ & $\phi^{\ell_1}$ & $\eta^{\ell_2}$ & $p_T^{\text{miss}}$ & $\eta^{\ell_1}$ & $p_T^{\ell_2}$ \\
    \hline
    Activation Difference & 0.55 & 0.51 & 0.51 & 0.45 & 0.44 & 0.44 \\
    \hline
\end{tabular}
\end{center}
\caption{Activation differences of the six most important neurons of a network trained on $(m_{\widetilde t}, m_{\widetilde \chi}) = (185, 5)$ data, with respect to the network output.}
\label{tab:actDiff_185-5_L12N0}
\end{table}

\begin{figure}[!ht]
\begin{center}
\scalebox{0.54}{\includegraphics{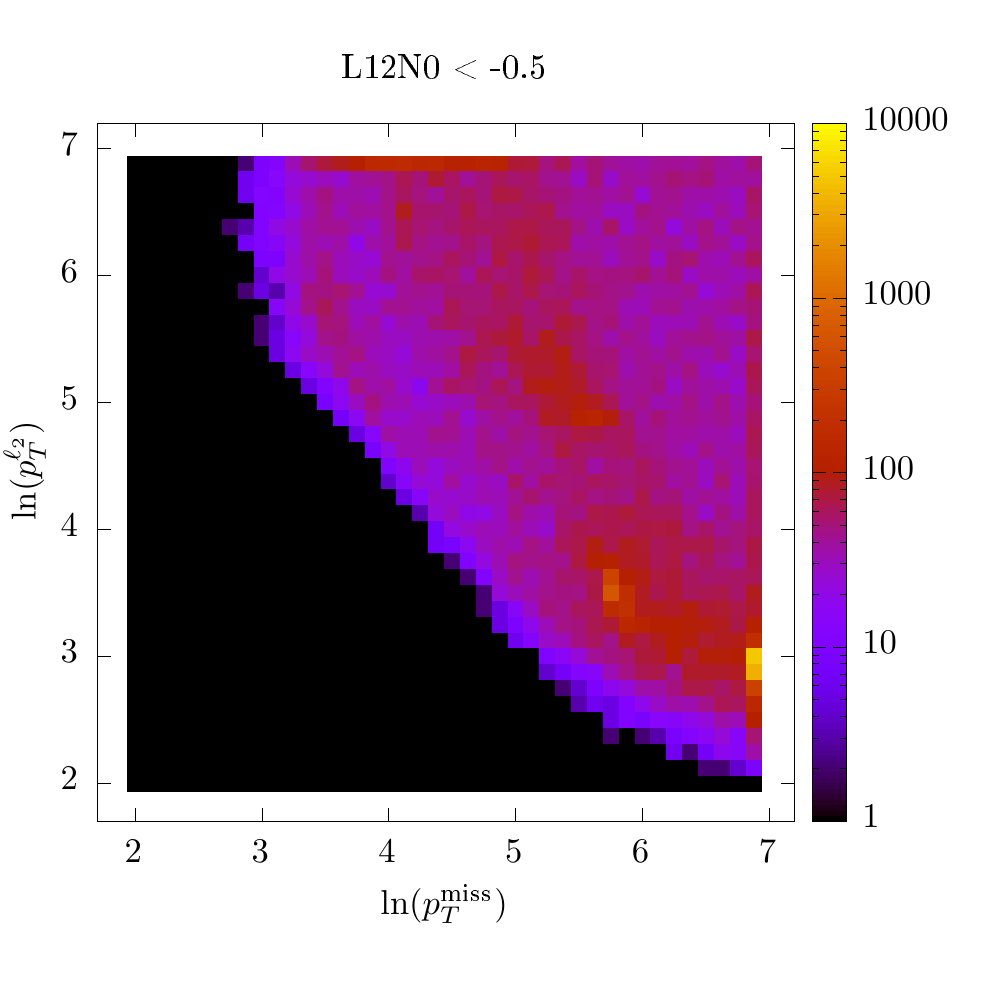}} \scalebox{0.54}{\includegraphics{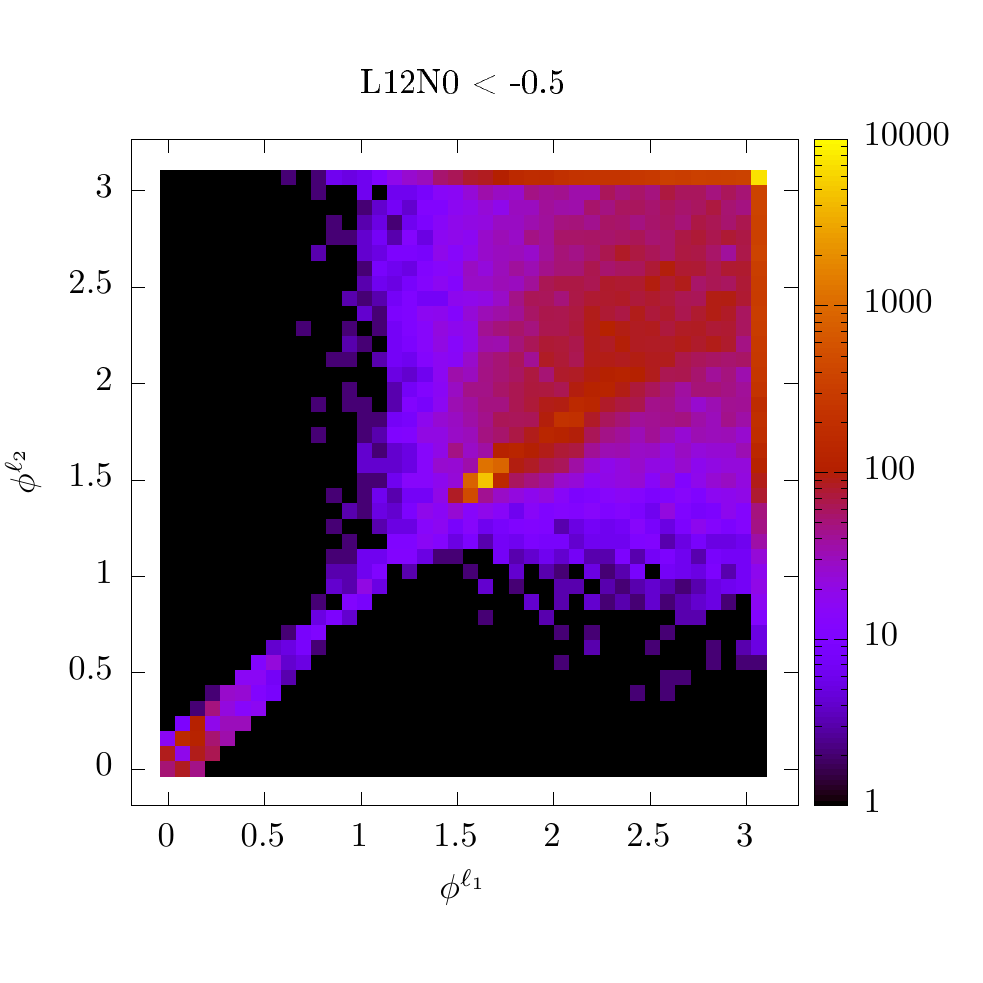}} \scalebox{0.54}{\includegraphics{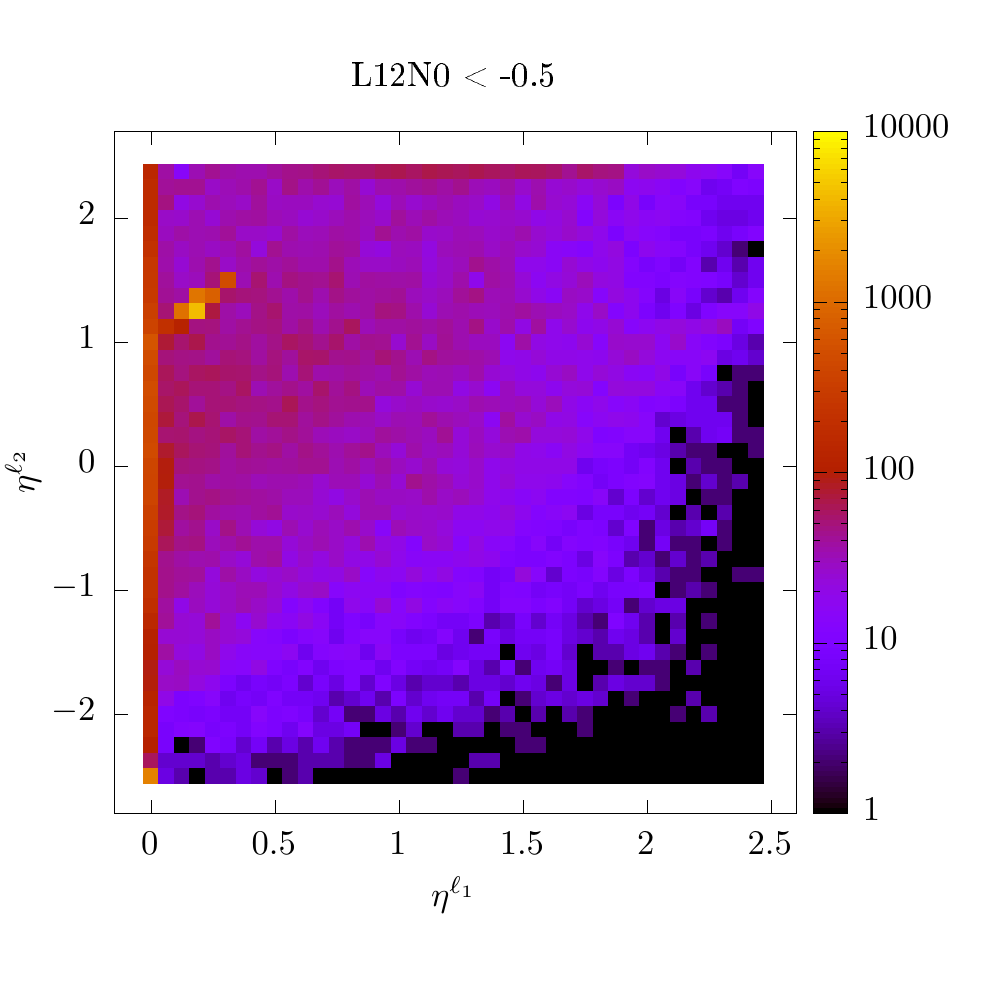}}
\caption{Activation maximization histograms for $L12N0$ involving the 6 most important input variables.}
\label{fig:actMax_185-5_L12N0}
\end{center}
\end{figure}

\begin{figure}[!ht]
\begin{center}
\scalebox{0.54}{\includegraphics{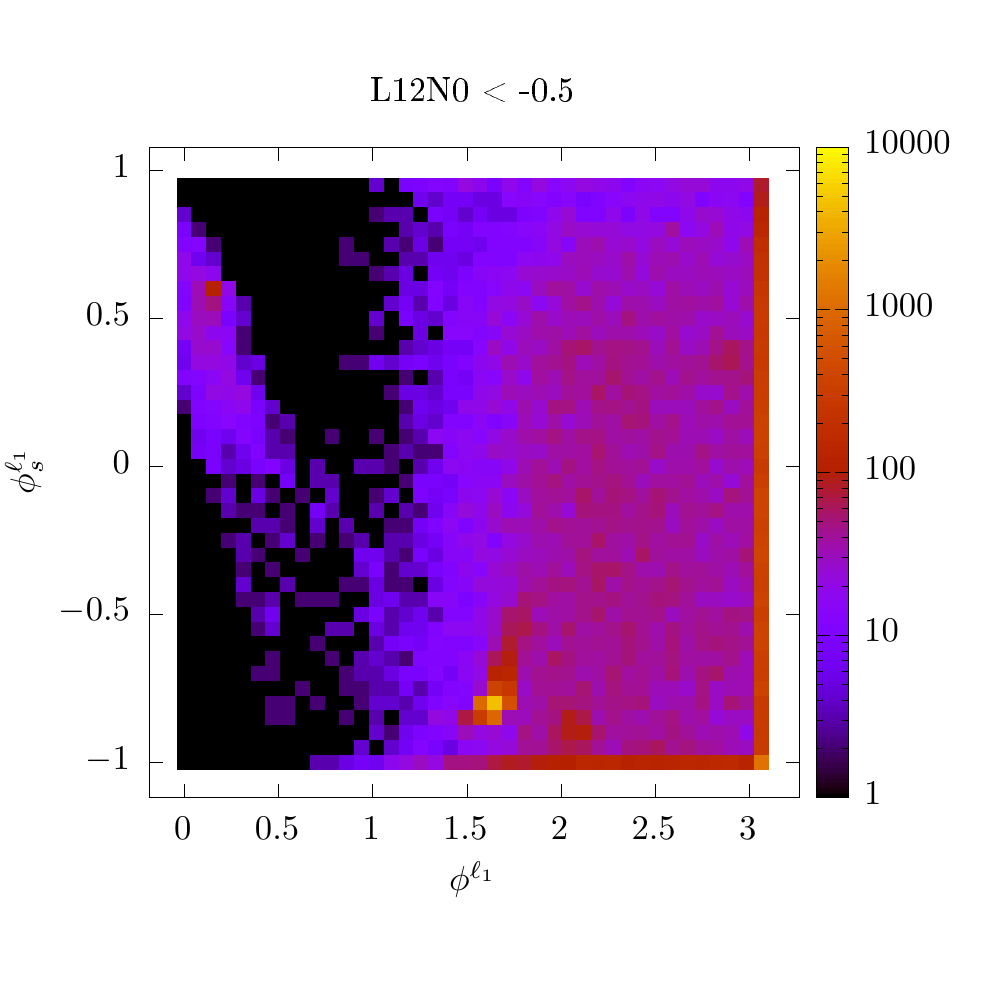}} \scalebox{0.54}{\includegraphics{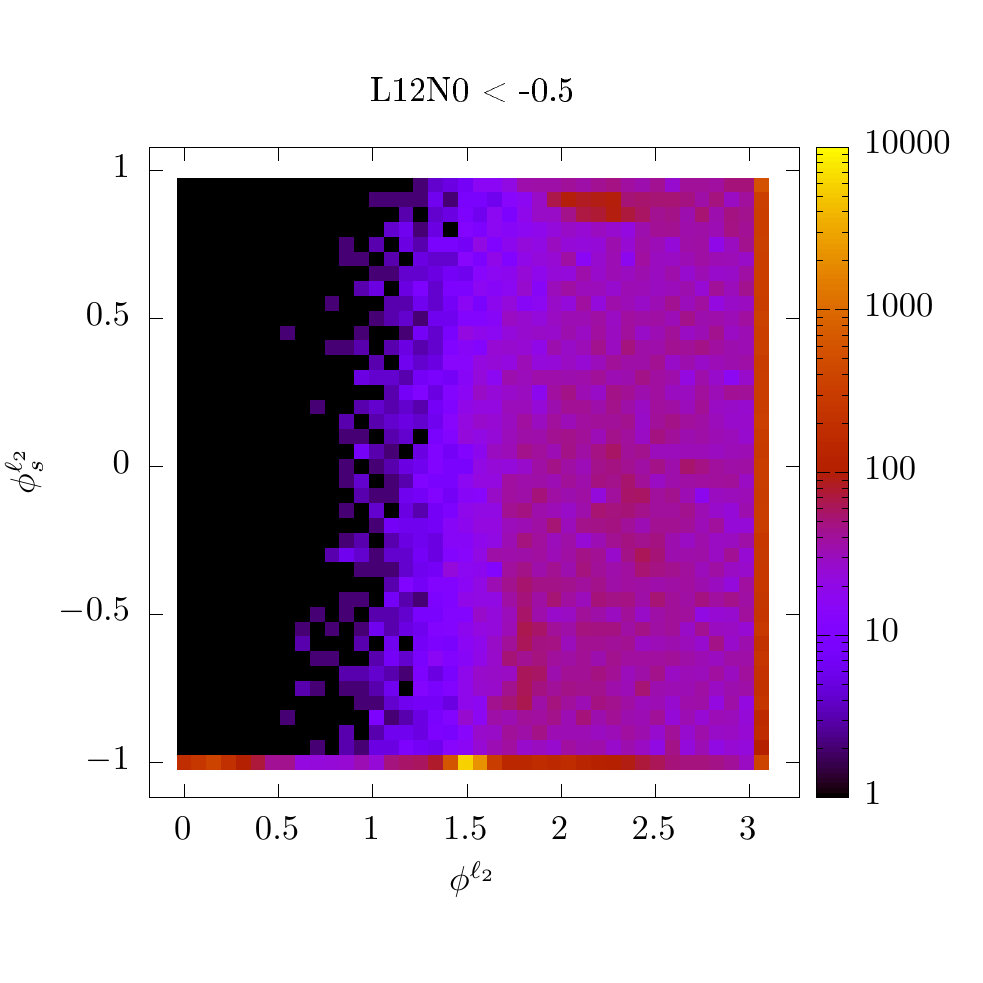}}
\caption{Activation maximization histograms for $L12N0$ between the magnitude of the lepton $\phi$s and their signs.}
\label{fig:actMax_185-5_phiSign}
\end{center}
\end{figure}

Activation maximization histograms for the variables listed in Table \ref{tab:actDiff_185-5_L12N0} are shown in Figure \ref{fig:actMax_185-5_L12N0}. The first plot, showing the correlation between $p_T^{\text{miss}}$ and $p_T^{\ell_2}$, is fairly clear in broad strokes: the network likes high $p_T$ for both the missing energy and the lepton. Indeed, the most important trend from these plots is the same as at our other mass points: high missing energy and leptons facing away from it.
However, there are other aspects which are more mysterious. For example, the second plot seems to suggest that the network accepts low lepton $\phi$s as long as $\phi^{\ell_1} = \phi^{\ell_2}$, but this seems to contradict our \emph{a priori} belief that stops could be distinguished by their high $\Delta \phi^{\ell}$. This puzzle can be resolved by looking at how the network treats the signs of the lepton $\phi$s, which are fed to the network as a separate variable. The correlations between each lepton $\phi$ and its associated sign are shown in Figure \ref{fig:actMax_185-5_phiSign}.

\begin{figure}[!ht]
\begin{center}
\scalebox{0.54}{\includegraphics{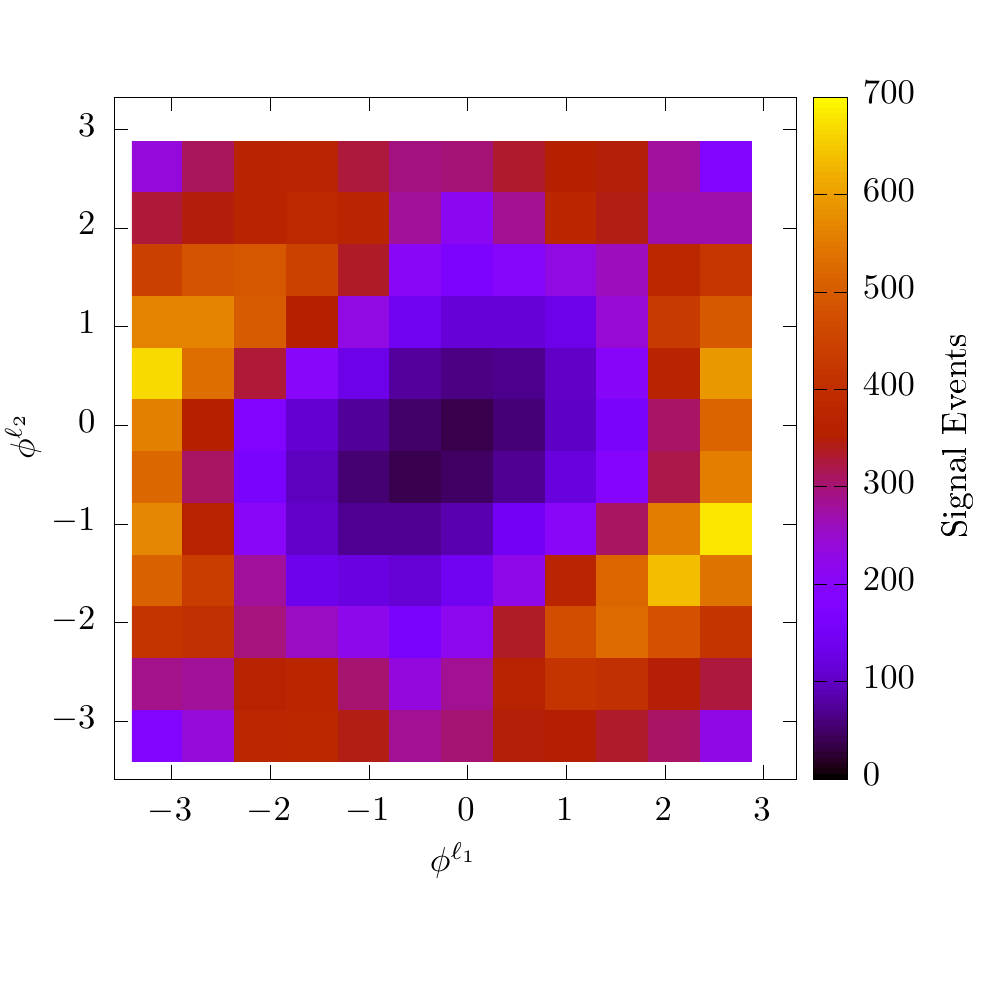}} \scalebox{0.54}{\includegraphics{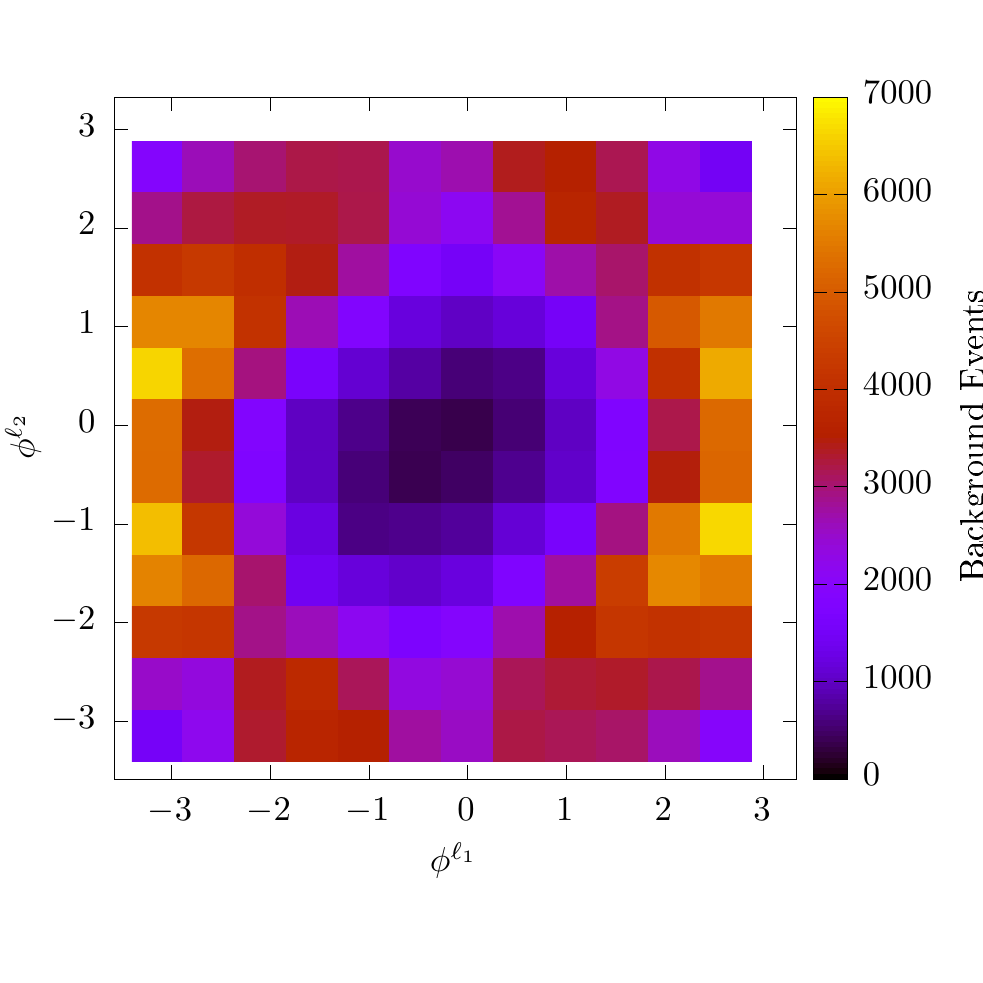}} \scalebox{0.54}{\includegraphics{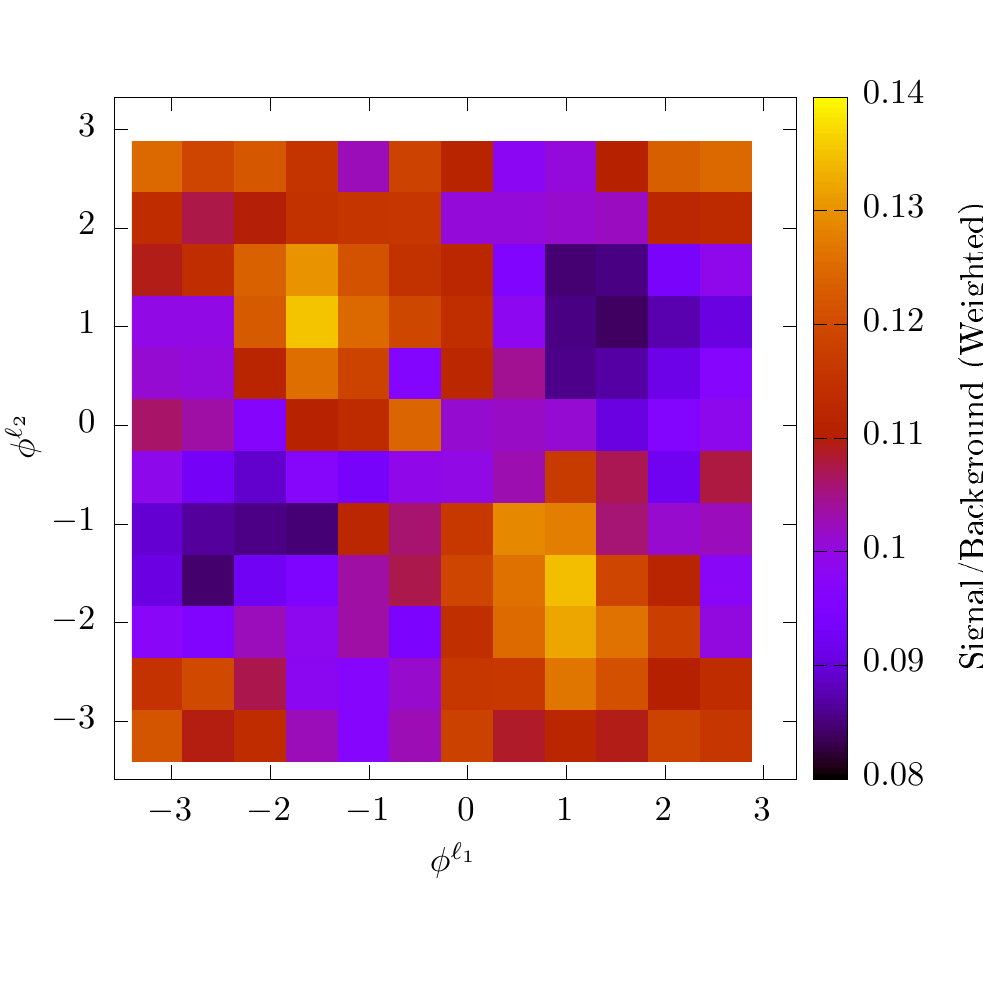}}
\caption{(185, 5) data distributions for $\phi^{\ell_1}$ vs. $\phi^{\ell_2}$.} 
\label{fig:actMax_185-5_L4N11_data}
\end{center}
\end{figure}

We see that there are at least two distinct clusters: at very high $\phi$, the leptons are together anyway, and the $\phi$ signs are spread out across the full range. However, at lower $\phi$, the neuron prefers $\phi_s^{\ell_1} > 0$ and $\phi_s^{\ell_2} < 0$, implying that the leptons are on opposite sides of the detector, and $\Delta
\phi \sim \pi$, as expected. Thus, the network has learned a pattern more specific than simply high $\Delta \phi$: it likes the leptons to not only be across the detector from each other, but also roughly equidistant from the missing energy. We did not anticipate this feature from prior work on spin correlations, but the neural network output motivates us to inspect the distributions of $(\phi^{\ell_1}, \phi^{\ell_2})$ defined relative to the missing $p_T$ angle. These distributions (Figure \ref{fig:actMax_185-5_L4N11_data}) clearly show, as in earlier work, that the signal to background ratio is largest at high $\Delta \phi(\ell_1, \ell_2)$, but they convey the additional information that the signal peaks where both leptons are far from the $p_T^{\text{miss}}$. In this way, the neural network has taught us a lesson about the preferred azimuthal angular distribution that we previously did not know. On the other hand, there is additional information in the output of the neural network that remains confusing. For instance, there is a bright cluster where both leptons are approximately orthogonal to $p_T^{\text{miss}}$ on the {\em same} side, which does not appear to be a region of high signal-to-background ratio in the simulation unless the network is exploiting more subtle correlations with additional variables that we have not yet identified.

\begin{figure}[!ht]
\begin{center}
\scalebox{0.7}{\includegraphics{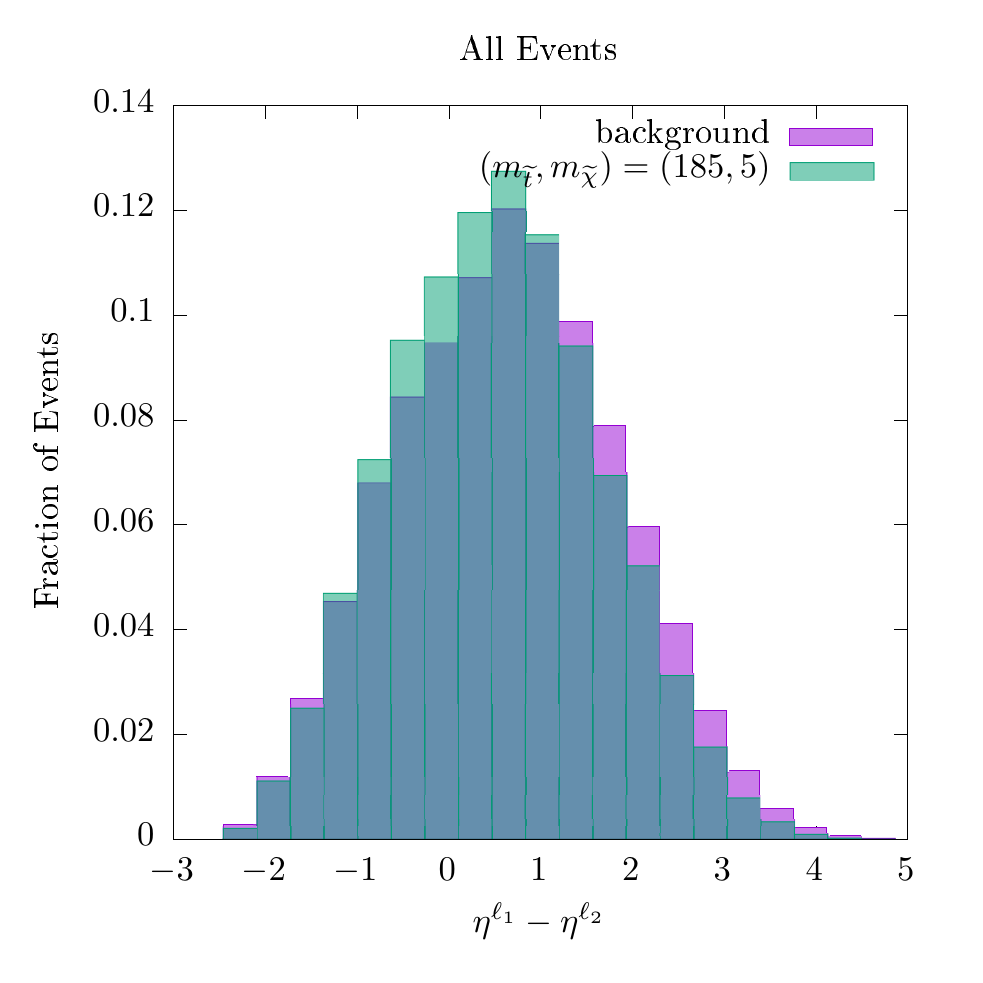}} \scalebox{0.7}{\includegraphics{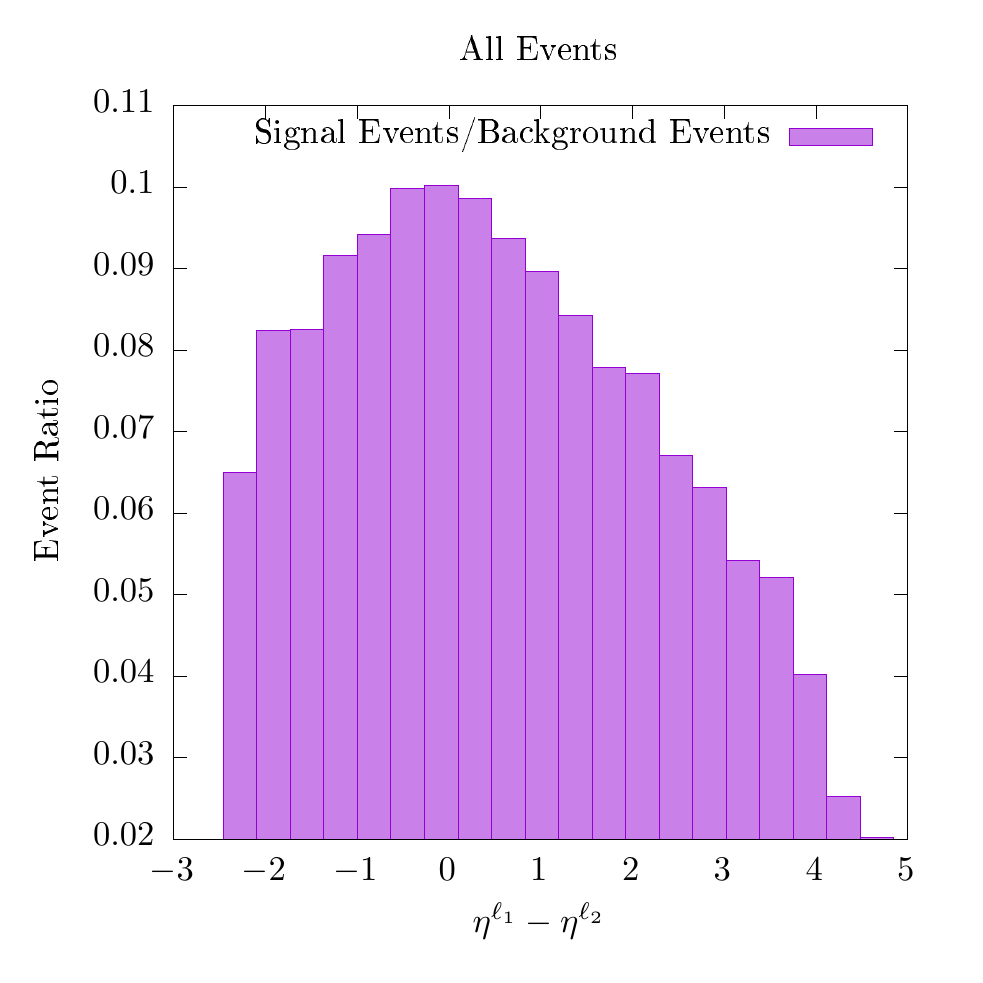}}    \\
\scalebox{0.7}{\includegraphics{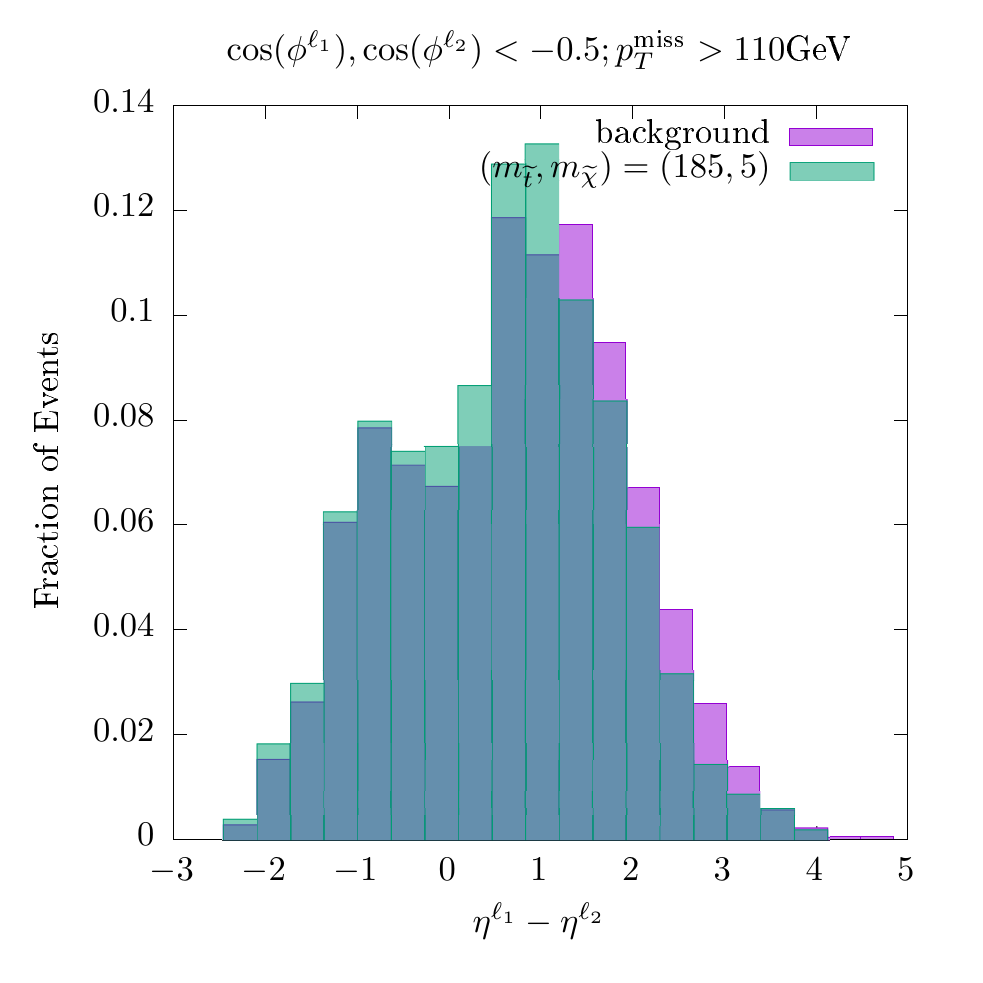}} \scalebox{0.7}{\includegraphics{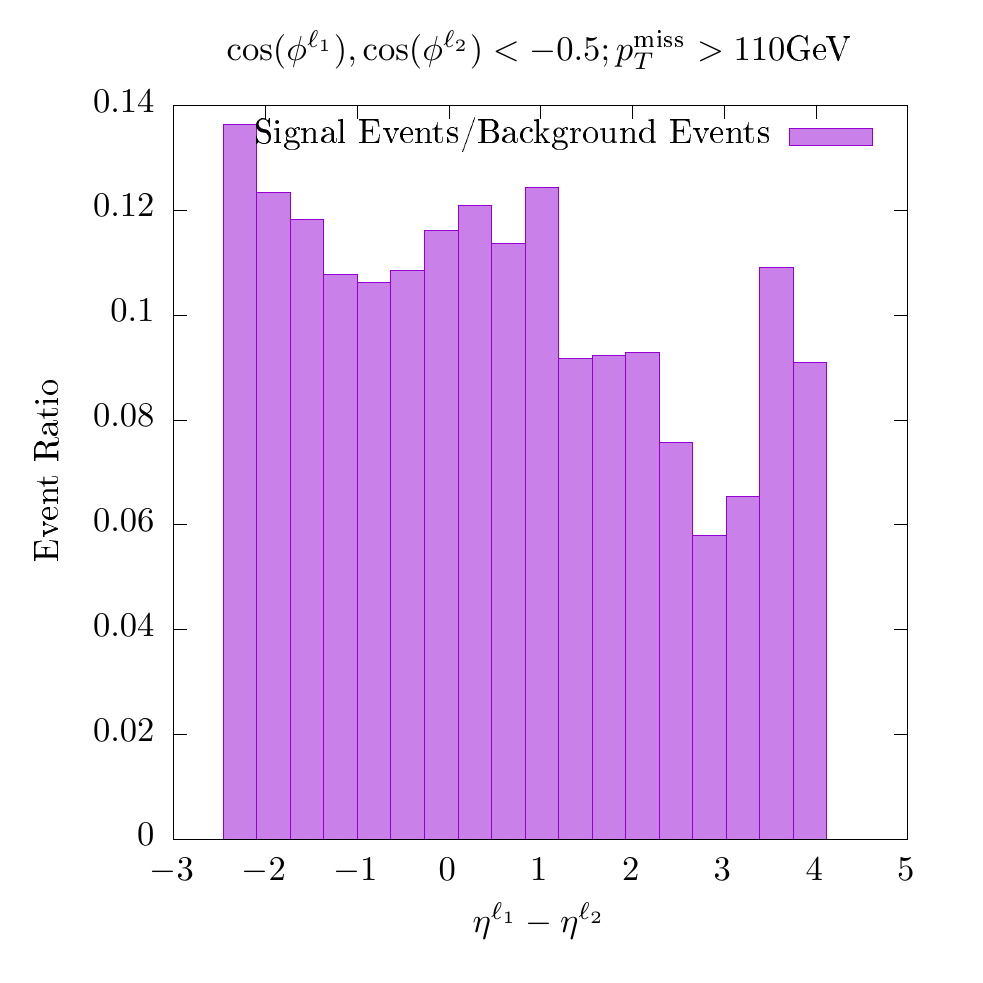}}
\caption{$\eta^{\ell_1}$-$\eta^{\ell_2}$ correlations in the (185, 5) data set. Top: all events. Bottom: Only events which pass the cuts $\cos(\phi^{\ell_1}) < -0.5$, $\cos(\phi^{\ell_2}) < -0.5$, and $p_T^{\text{miss}} > 110 \text{ Gev}$. We see that, not only do events which pass these cuts have an elevated signal-to-background ratio, this increase is significantly exacerbated for very negative $\eta^{\ell_1}-\eta^{\ell_2}$.}
\label{fig:185-5_deltaEtaData}
\end{center}
\end{figure}

The third plot in Figure \ref{fig:actMax_185-5_L12N0} is also a bit confusing at first, since it suggests that the network is looking for a negative value of $\eta^{\ell_1}-\eta^{\ell_2}$, while we expected to be looking for low $|\eta^{\ell_1}-\eta^{\ell_2}|$. However, if we take the other preferences of the network into account, the reason for this discrepancy becomes somewhat more clear. In the right panel of Fig.~\ref{fig:actMax_185-5_L4N11_data}, observe that the signal-to-background ratio is high not only in the region discussed in the previous paragraph but also in the four corners, where both leptons are azimuthally opposite the $p_T^{\text{miss}}$. In Figure \ref{fig:185-5_deltaEtaData}, we show how the $\eta^{\ell_1} - \eta^{\ell_2}$ distribution behaves in the full simulated datasets compared to those where we isolate the corner region we identified in Fig.~\ref{fig:actMax_185-5_L4N11_data}. If we consider the entire dataset, the highest signal-to-background ratios come only from the region where the two lepton $\eta$s are close together, as expected from earlier literature on spin effects on the cross section. Background events are more concentrated relative to signal events for both significantly negative and positive $\eta^{\ell_1} - \eta^{\ell_2}$. However, when we consider only events with relatively high $p_T^{\text{miss}}$ and lepton $\phi$s opposite the $p_T^{\text{miss}}$, background events continue to be favored at large positive $\eta^{\ell_1} - \eta^{\ell_2}$ but signal events are favored at large negative $\eta^{\ell_1} - \eta^{\ell_2}$. Because the sign of $\eta$ is defined relative to the leading lepton, this corresponds to a relatively central hardest lepton with the second lepton more forward in the same direction. This is the pattern the network has found. As we saw in \S\ref{sec:threebody}, this suggests that the neural network may be continuously interpolating between different strategies. For instance, the signal-to-background plot of lepton $\phi$ distributions in Fig.~\ref{fig:actMax_185-5_L4N11_data} showed two distinct peaks, one with leptons approximately orthogonal to missing $p_T$ and one with them opposite missing $p_T$. The network's strategy targets both of these regions.

\section{Conclusions}
\label{sec:conclusions}

We have proposed a method for efficiently extracting qualitative information about the behavior of a neural network, and demonstrated that it is possible to glean through it non-trivial physical insight about the underlying distributions upon which the network was trained. In some cases (the uncompressed point), we found that the network had learned seemingly simpler, but no less effective, approximations of the techniques already used by physicists. In others (the stealth point), we
found that the network appeared to have discovered useful multivariable correlations not present in the literature to our knowledge. In all cases, we were able to obtain a fairly good qualitative picture of at least the first-order behavior of the network, and link our findings to demonstrable, understandable aspects of the kinematics under study.

In constructing the activation maximization histograms, we make no use of any actual data, but merely consider the network themselves in isolation. This is, in our view, both a boon and a drawback of the technique. It is advantageous because it means that any correlations we observe must necessarily arise from preferences of the network itself --- and the network is not likely to learn patterns which do not aid in the separation of signal and background. On the other hand, it
means that the path of gradient ascent is not limited to physically possible configurations. This can make analysis of the resulting histograms more an art than a science at times, especially since neural networks can behave unpredictably when taken out of the input domain of their training set. It would be interesting to repeat our analysis with some sort of constraints or guidelines on the allowed input spaces, to see if the results were more or less understandable. For example,
perhaps a generative adversarial network \cite{2014arXiv1406.2661G} could be used to constrain the input space to only physically plausible events.

As the techniques of artificial intelligence become more refined and more powerful, it is inevitable that they begin to creep into areas of endeavor once thought the unassailable domain of human creativity. It is our hope that the data-driven techniques we have outlined can be useful in collider physics as a complement to more traditional techniques, since they allow a direct evaluation of the raw data to inform our understanding of the important physics rather than the reverse.

\section*{Acknowledgments}

This work has been supported in part by the NSF Grant PHY-1415548 and the DOE Grant DE-SC0013607. Some computations in this paper were run on the Odyssey cluster supported by the FAS Division of Science, Research Computing Group at Harvard University.

\appendix

\section{Details of ``caveman variables''}
\label{app:caveman}

Here we will outline the basic procedure for constructing the caveman variables introduced in \S\ref{sec:caveman}. We must emphasize that, due to their obvious crudeness, these variables are not meant to replace the output of the neural network in any real capacity. Instead, they are intended to serve two purposes: first, to demonstrate that the correlations visible in our activation maximization histograms really do have discriminative power, and, second, to give us a sense of the
correlations most important to the network with more precision than the activation differences. An outline of the procedure we use is as follows:

\begin{itemize}
    \item Choose a neuron for which to construct the caveman variable, and an earlier layer of the network to provide the inputs to the variable. This input layer can also be the set of actual inputs to the network; the ultimate goal is to construct a variable that stretches from the networks inputs to the final layer. However, for more complicated networks, it may be necessary to do this in several jumps by constructing a ``super-caveman'' variable out of a number of intermediate caveman
    variables. Even in these cases, though, much of the interesting computation can go on in the early layers, and examining early neurons can provide insight. The caveman variables we present below only consider the input layer, however.
    \item Construct the activation maximization set for the chosen neuron. That is, given a set of inputs randomly distributed over the space of possible inputs, perform gradient ascent on each input until it produces a neuron activation above a predetermined threshold. This results in a set of artificial events which highly activate the neuron, produced completely independently from any actual data.
    \item For each possible combination of 1, 2, or 3 input variables, use the artificial events in this activation maximization set to fill a histogram with axes corresponding to the chosen variables. To save on computation time, we restrict ourselves to the 20 histograms of each dimension with the highest mutual informations. These represent combinations of inputs which are correlated with each other; that is, for which the network requires a pattern involving multiple inputs to activate.
    \item In each histogram, find clusters or ``islands'' of points which represent common patterns sought by the network; we use a variant of the DBSCAN algorithm \cite{ester1996density} for this task. Each cluster will represent a single term in the caveman variable. We distinguish between input tuples which contain one or more inputs along the boundary of the histogram and tuples which are confined to the interior, dividing them into separate clusters. The reason for this is that points on the boundary of the
    histogram often represent ``runaway'' directions in the classification space --- that is, they indicate that the network is looking for an input or combination of inputs to be as high or as low as possible, rather than any specific value. (Recall for instance the toy problem examples in Figs.~\ref{fig:toydata_pos_L2N36} and \ref{fig:toydata_pos_L2N11}, where neurons have learned to favor particular corners but activation maximization populates mostly diagonal lines.) We therefore treat these clusters differently.
    \item Model the clusters. We do this by fitting functions via gradient descent to be 1 within the area of the cluster and 0 elsewhere. For interior clusters, we use second degree polynomials passed through a sigmoidal transfer function to clamp the value between 0 and 1. For boundary clusters, we use first degree polynomials with cross-terms (ie. $x$, $y$, and $x y$ but not $x^2$) with no clamping.
    \item Combine the cluster terms together linearly, fitting the coefficients to best match the neuron activation through gradient descent. We apply an $L1$ regularization term to the utility function to suppress the less important terms, leaving only a manageable and configurable number of terms to consider.
\end{itemize}

\begin{figure}[!ht]
\begin{center}
\scalebox{0.5}{\includegraphics{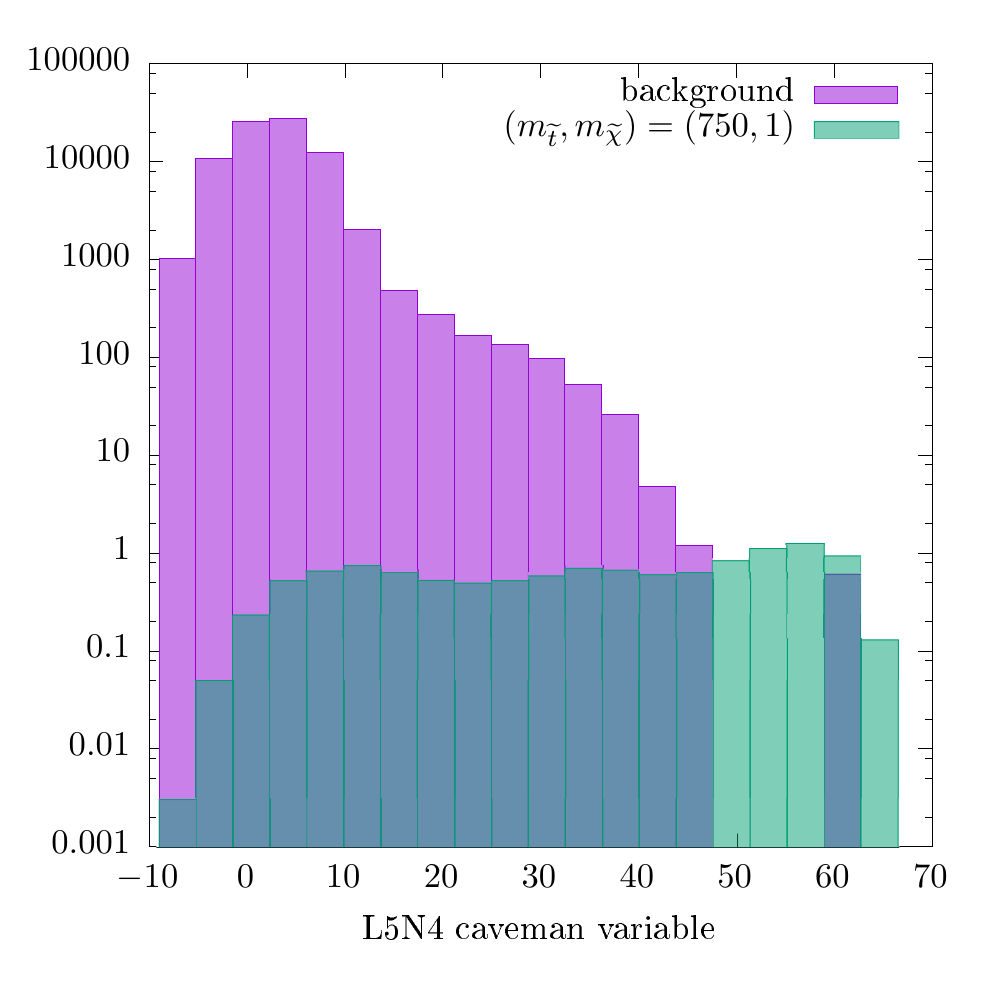}} \scalebox{0.5}{\includegraphics{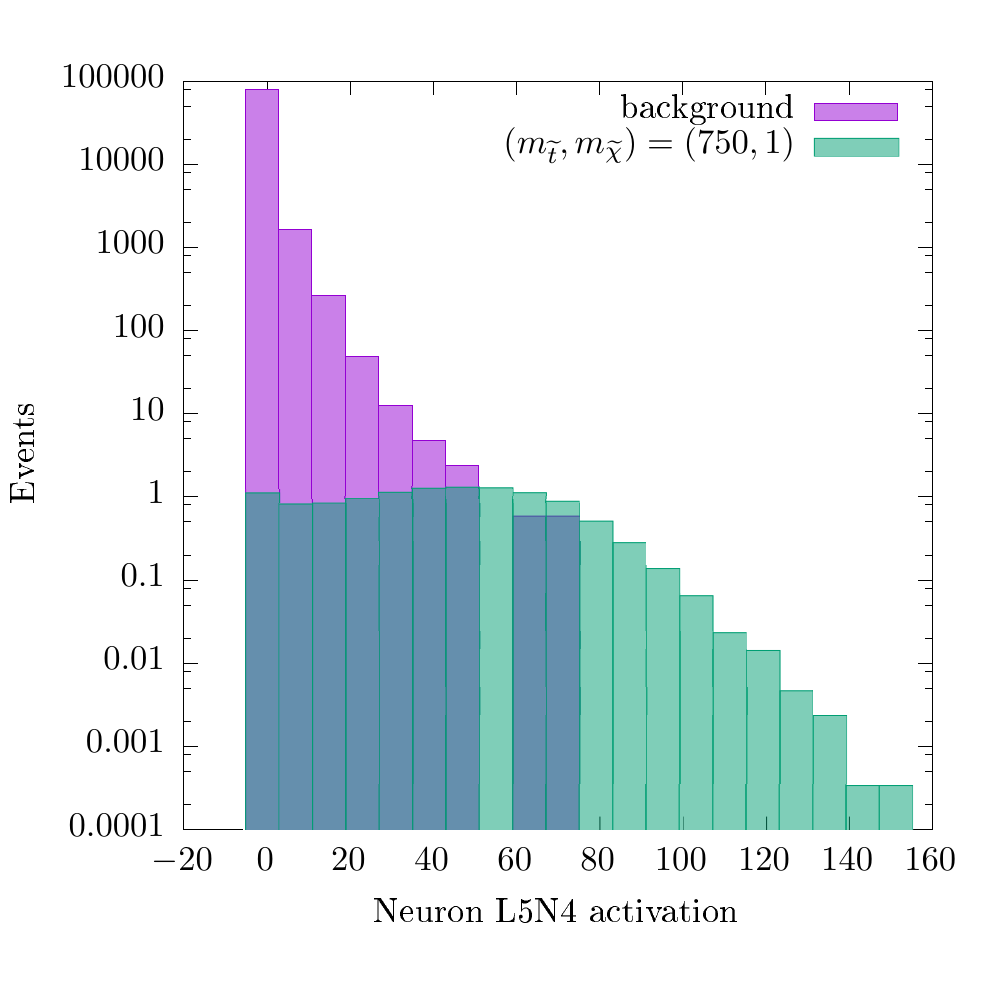}}    \\
\scalebox{0.5}{\includegraphics{plots/350-200/nhist_350-200_CaveL5N10_45.pdf}} \scalebox{0.5}{\includegraphics{plots/350-200/nhist_350-200_L5N10.pdf}}    \\
\scalebox{0.5}{\includegraphics{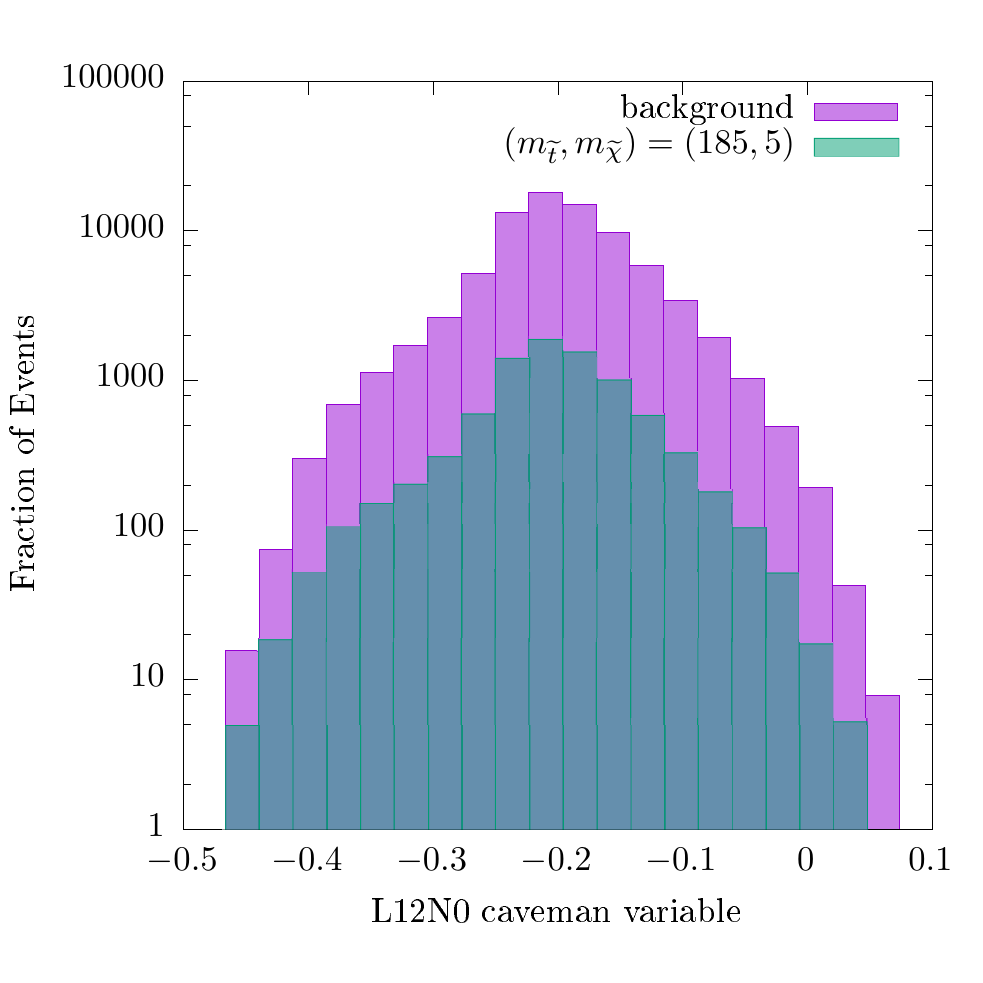}} \scalebox{0.5}{\includegraphics{plots/185-5/nhist_185-5_L12N0.pdf}}    \\
\caption{Comparison of caveman variables (left) with their model neurons (right). The caveman variables clearly exhibit some of the same discriminative power as their parent neurons, albeit with somewhat lesser efficacy.}
\label{fig:cavemancomparison}
\end{center}
\end{figure}

Some comparisons of the discriminative power of several caveman variables with their parent neurons are shown in Figure \ref{fig:cavemancomparison}.

\bibliographystyle{utphys}
\bibliography{stopsearch}

\end{document}